\documentclass[a4paper,12pt]{article}
\pdfoutput=1 

\usepackage{jheppub} 

\usepackage[T1]{fontenc}
\usepackage[english]{babel}
\usepackage{hyperref}
\usepackage{ifpdf}
\usepackage{subfigure}
\usepackage{slashed}
\usepackage{amssymb}
\usepackage{amsfonts}
\usepackage{pdfpages}
\usepackage{epsf}
\usepackage{float}
\usepackage{rotating}
\usepackage{graphicx}
\usepackage{amsmath}
\usepackage{fancyhdr}
\usepackage{pbox}
\usepackage[usestackEOL]{stackengine}
\usepackage{makecell,multirow}
\usepackage{babel}
\usepackage{graphics}
\usepackage{color}
\usepackage{multirow}

\newcommand{\nn}{\nonumber}
\newcommand{\lsim}{\mathrel{\mathop{\kern 0pt \rlap
			{\raise.2ex\hbox{$<$}}}
		\lower.9ex\hbox{\kern-.190em $\sim$}}}
\newcommand{\gsim}{\mathrel{\mathop{\kern 0pt \rlap
			{\raise.2ex\hbox{$>$}}}
		\lower.9ex\hbox{\kern-.190em $\sim$}}}

\newcommand{\be}{\begin{equation}}
\newcommand{\ee}{\end{equation}}
\newcommand{\bea}{\begin{eqnarray}}
\newcommand{\eea}{\end{eqnarray}}

\title{\boldmath Scrutinizing Vacuum Stability in IDM with Type-III Inverse seesaw}
\preprint{ IITH-PH-0006/20\\\hspace*{12cm}IP/BBSR/2020-4}
\author[a]{Priyotosh Bandyopadhyay,}
\author[a]{Shilpa Jangid,}
\author[b,c]{Manimala Mitra}

\affiliation[a]{
	Indian Institute of Technology Hyderabad, Kandi,  Sangareddy-502287, Telangana, India}
\affiliation[b]{Institute of Physics, Sachivalaya Marg, Bhubaneswar 751005, India}
\affiliation[c]{Homi Bhabha National Institute, BARC Training School Complex, Anushakti Nagar, Mumbai 400094, India }

\emailAdd{bpriyo@phy.iith.ac.in, ph19resch02006@iith.ac.in, manimala@iopb.res.in}

\abstract{We consider the extension of the Standard Model (SM) with an inert Higgs doublet that also contains two or three sets of $SU(2)_L$ triplet fermions with hypercharge zero and analyze the stability of electroweak vacuum for the scenarios. The model represents a  Type-III inverse seesaw mechanism for neutrino mass generation with a Dark matter candidate. 
An effective potential approach calculation with two-loop beta function have been carried out in deciding the fate of the electroweak vacuum. Weak gauge coupling $g_2$ shows a different behaviour as compared to the Standard Model. The modified 
running of  $g_2$, along with the Higgs quartic coupling and Type-III Yukawa couplings become crucial in determining the stability of electroweak vacuum. The interplay between two and three generations of such triplet fermions reveals that extensions with two generations is favoured if we aspire for Planck scale stability. Bounds on the Higgs quartic couplings, Type-III Yukawa and  number of triplet fermion generations are drawn for different mass scale of Type-III fermions.  The phenomenologies of inert doublet  and Type-III fermions  at the LHC and other experiments are commented upon. }


\begin{document}

\maketitle
\flushbottom
\section{Introduction}
The last element of the Standard Model (SM)  was the Higgs boson which was discovered at the CMS and ATLAS detectors of the Large Hadron Collider (LHC)~\cite{Aad:2012tfa, Chatrchyan:2012xdj}. The spin, parity measurements and the combined analysis show the SM-like behaviour of the Higgs boson \cite{Aad:2013xqa,Sirunyan:2018koj,ATLAS:2018doi}. However it has been shown that Standard Model electroweak (EW)  vacuum on its own can run into metastability due to quantum corrections \cite{Isidori:2001bm, Bezrukov:2012sa, Degrassi:2012ry, Buttazzo:2013uya}. It is well known that the addition of scalars enhance the stability of the EW vacuum via positive loop contributions to the Higgs quartic coupling. Various models, which  include scalars from different gauge representations have been proposed \cite{singletex,Bandyopadhyay:2016oif,Haba:2015rha,2HDMs,2HIDM,Belyaev:2016lok,PBSJ,PlascenciaIDM,Tripletex,LR,Cai1,Cai2} to enhance the stability of EW vacuum. On the contrary an extension with fermion often gives negative contributions to the Higgs quartic couplings that it couples to. Such negative contributions then tend to pull such Higgs quartic couplings toward instability much faster. Thus models with extra fermions and where  Majorana masses of the fermions are spontaneously generated are constrained from the vacuum stability \cite{PBBDSJ,VSLG,exwfermion,Coriano:2015sea,Rose:2015fua,Garg:2017iva,Casas:1999cd, EliasMiro:2011aa, Rodejohann:2012px, Masina:2012tz, Farina:2013mla, Ng:2015eia, Bambhaniya:2016rbb,Khan:2012zw,Das:2019pua,Baek:2012uj, Lindner:2016kqk, DuttaBanik:2018emv, Wang:2018lhk,Xiao:2014kba, Gopalakrishna:2018uxn,Mohapatra:2014qva, Dev:2015vjd}. 

Apart from the problem of vacuum metastability in SM which depends on the top quark and Higgs boson masses \cite{Tanabashi:2018oca, Markkanen:2018pdo,espinoza}, the theory also fails to provide a stable dark matter (DM) candidate, as well as to give successful  explanation for the very tiny eV scale  neutrino masses, and their mixings. In this work,  we focus on these two aspects by extending the SM with  $SU(2)_L$ triplet fermions, and $SU(2)_L$ inert doublet scalar. The triplet fermion generates  the eV light neutrino mass via Type-III seesaw mechanism, while  the inert Higgs doublet provides a  dark matter candidate, as well as stabilizes the EW vacuum. 

The minimal Type-III extensions have one to three generations of $SU(2)_L$ fermions with hypercharge zero, which mix with the SM charged and neutral fermions, and also generates tiny eV neutrino mass via electroweak symmetry breaking\cite{TypeIII,TypeIII2,TypeIII3}. Different extensions of Type-III seesaw and their collider signatures have been studied in \cite{TypeII4}-\cite{TypeIII9}, including their spin measurement at the LHC \cite{PBSD}. The stability of EW vacuum in some of these scenarios are studied in \cite{Gogoladze:2008ak, Chen:2012faa, Lindner:2015qva, Goswami:2018jar}. In this article,  we consider the inverse seesaw mechanism of neutrino mass generation with two generations of triplet fermions. One among them couples with the SM Higgs boson via Type-III Yukawa coupling,  and generates the Dirac  mass term. The  other triplet fermion generates the Majorana mass term for the triplet fermion. 

As discussed earlier an extension with scalar enhances stability of EW vacuum and if the scalar is in the form of $SU(2)_L$ inert ($Z_2$-odd ) doublet then it also provides the much needed dark matter candidate \cite{2HIDM}. SM extension with such inert doublet in the context of vacuum stability have been studied extensively\cite{2HIDM,Belyaev:2016lok,PBBDSJ,PlascenciaIDM,PBSJ}. Fields in inert doublet have very interesting phenomenology due to their compressed spectrum and the possibility of real and pseudoscalar dark matter particle \cite{Choubey:2017hsq}-\cite{Banerjee:2019luv}. In this article we will investigate the effect of inert doublet in the context of Type-III fermions.

In our model we have both $SU(2)_L$ triplet fermion and $SU(2)_L$ doublet scalar. Being in the triplet representation of $SU(2)_L$ the new fermions contribute in the evolution of weak gauge coupling $g_2$ such that $g_2$ now increases with running scale. This behaviour substantially changes the dynamics of couplings responsible for the EW vacuum stability. We shall see how an enhanced $g_2$ causes a much lower perturbative scale compared to Type-I \cite{PBBDSJ,VSLG} or only IDM case \cite{PBSJ}.
Specially we see that with three generations of Type-III fermions it is difficult to attain Planck scale perturbativity and thus two generations are more favoured.

The paper is organised as follows. In section~\ref{model},  we describe the model, and present the  EW symmetry breaking conditions for this model. We discuss the perturbativity and the interplay of two and three generations of $SU(2)_L$ triplet fermions in section~\ref{pert}. The EW vacuum stability with all three posibilites are covered in section~\ref{stability}.  In section~\ref{concl},  we discuss the phenomenological consequences and present our conclusion. The expressions of the  two-loop beta functions used in our analysis are presented in Appendix~\ref{betaf1}. 

\section{The Model}\label{model}
The SM is augmented with an inert doublet (ID) and three Type-III fermions with ISS mechanism. We first discuss the scalar sector of the model in Section~\ref{sec:2.1}. We consider  two different scenarios, viz., a canonical Type-III seesaw with small Yukawa couplings and an inverse seesaw (ISS) with large Yukawa couplings to study effects in perturbativity of dimensionless couplings and vacuum stability. The fermionic sector with Type-III seesaw and inverse seesaw is discussed in Section~\ref{sec:2.2}. 

\subsection{The Scalar Sector} \label{sec:2.1}
The scalar part  consists of two $SU(2)_L$-doublets $\Phi_1$ and $\Phi_2$,  both with hypercharge $1/2$: 
\begin{align}
\Phi_1
	\ = \ \left(\begin{array}{c}
	G^+   \\
    h+ i G^0   \end{array}\right) \, , \qquad
		\Phi_2
	\ = \ \left(\begin{array}{c}
	H^+   \\
	H+ i A   \end{array}\right) \,.
\end{align}


The tree-level SM gauge invariant scalar potential i.e., invariant under $SU(2)_L \times U(1)_Y$ is given by~\cite{Branco:2011iw,Plascencia:2015xwa} 
\begin{align}
  V_{\rm scalar} \ &= \ m_{11}^2\Phi_1^\dagger \Phi_1 + m_{22}^2\Phi_2^\dagger\Phi_2-(m_{12}^2\Phi_1^\dagger \Phi_2+{\rm H.c}) \nonumber \\
  & \qquad +\lambda_1(\Phi_1^\dagger \Phi_1)^2+\lambda_2(\Phi_2^\dagger \Phi_2)^2+
 \lambda_3(\Phi_1^\dagger \Phi_1)(\Phi_2^\dagger \Phi_2) +\lambda_4(\Phi_1^\dagger \Phi_2)(\Phi_2^\dagger \Phi_1) \nonumber \\ 
 & \qquad +\big[\lambda_5(\Phi_1^\dagger \Phi_2)^2+\lambda_6(\Phi_1^\dagger \Phi_1)(\Phi_1^\dagger \Phi_2) +   \lambda_7(\Phi_2^\dagger \Phi_2)(\Phi_1^\dagger \Phi_2) +{\rm H.c}\big], \label{eq:2.2}
\end{align}
where we have chosen all the Higgs quartic couplings $\lambda_{1,2,3,4}$ and mass terms $ m_{11}^2, m_{22}^2 $ as real. While $m_{12}^2$ and the $\lambda_{5,6,7}$ couplings are in general complex but for this study we also have taken them as real number. A $Z_2$ symmetry is imposed to prohibit the  flavor changing neutral currents at tree-level. Under this  $Z_2$ symmetry  $\Phi_2$ is odd and $\Phi_1$ is even. This choice also make $\Phi_2$ as inert which can be DM candidate. The $\lambda_6$, $\lambda_7$ and  $m_{12}$ also get removed and Eq.~\eqref{eq:2.2} reduces to 
\begin{align}
V_{\rm scalar} & \ =  \ m_{11}^2\Phi_1^\dagger \Phi_1 + m_{22}^2\Phi_2^\dagger\Phi_2 + \lambda_1(\Phi_1^\dagger \Phi_1)^2 + \lambda_2(\Phi_2^\dagger \Phi_2)^2  \nonumber \\ 
& \qquad +
\lambda_3(\Phi_1^\dagger \Phi_1)(\Phi_2^\dagger \Phi_2)+ \lambda_4(\Phi_1^\dagger \Phi_2)(\Phi_2^\dagger \Phi_1) + \big[\lambda_5(\Phi_1^\dagger \Phi_2)^2 + {\rm H.c}\big].  
\label{eq:2.3}
\end{align}

The $\Phi_1$ takes a real vacuum expectation value (VEV) which breaks the EW symmetry  as follows
\begin{equation}
	\langle \Phi _1 \rangle \ = \ \frac{1}{\sqrt 2}\left(
\begin{array}{c}
	0 \\
	v \\
\end{array}
\right) \, ,
\end{equation}
here we choose $v\simeq 246$ GeV guided by the masses of gauge bosons and SM-like Higgs boson. $\Phi_2$, being $Z_2$-odd, does not have any part in breaking the EW symmetry, so it behaves like a  `inert' Higgs doublet. Such model is often named as inert doublet model (IDM).  We replace $m^2_{11}$ by the minimzation condition follows: 
\begin{align}
m_{11}^2 \ = \ -\lambda _1v^2 \, ,
\end{align}
where the respective physical mass eigenvalues can be written as 
\begin{eqnarray}\label{mass}
M_{h}^2 & \ = \ & 2\lambda_1 v^2  \, , \nn\\
M_{H_0}^2 &\ = \ & \frac{1}{2}[2m_{22}^2+v^2(\lambda_3+ \lambda_4+2\lambda_5)] \, , \nn\\
M_{A}^2 &\ = \ & \frac{1}{2}[2m_{22}^2+v^2(\lambda_3+\lambda_4-2\lambda_5)] \, , \nn\\
M_{H^\pm}^2 &\ = \ & m_{22}^2+\frac{1}{2}v^2 \lambda_3 \, .
\end{eqnarray} 
This is to note that, being $Z_2$ odd $\Phi_2$ is inert, which prohibits any mixing between $\Phi_2$ and $\Phi_1$. This also implies that the gauge eigenstates and the mass eigenstates  are the same for the Higgs bosons from both $Z_2$ odd or even multiplets.  In this scenario, $\Phi_2$ being $Z_2$-odd does not talk to the fermions. Moreover, we get two CP even neutral Higgs bosons $h$ and $H_0$. Here we choose $h$ as the SM-like Higgs boson with mass around 125 GeV discovered at the LHC.  The spectrum has one pseudoscalar Higgs boson $A$ and a pair of charged Higgs bosons $H^\pm$. It is evident from Eq.~\eqref{mass} that the heavy Higgs bosons $H_0$, $A$ and $H^\pm$ are from $\Phi_2$ so they are nearly degenerate in mass spectrum with possible splitting by the help of $\lambda_5$. The sign of $\lambda_5$ is crucial in making one of the scalars between $A$ and $H_0$ as the lightest, and possible DM candidate. The $Z_2$-odd symmetry prohibits some of the decays of $\Phi_2$-type Higgs bosons..

	\subsection{The Type-III and Inverse Seesaw Lagrangians } \label{sec:2.2}
In addition to the SM particle contents, the Type-III seesaw  model contains  $SU(2)_L$ fermionic triplets $\Sigma$ with zero hypercharge. Being in the adjoint representation of the $SU(2)_L$ group, the Majorana mass term $M_N$ of such triplets is gauge invariant. In terms of the usual two-by two notation for triplets, the beyond SM interactions are described by the Lagrangian:
\begin{eqnarray}
{\cal L}_{\rm III} \ = \ Tr[\overline{\Sigma}i  \slashed{D}\Sigma] -\frac{1}{2}Tr[\overline{\Sigma} M_{N}\Sigma^c + \overline{\Sigma^c}M^*_{N}\Sigma]- \sqrt{2}(\widetilde{\Phi}^{\dagger}_1\overline{\Sigma}Y_{N}L+\overline{L}Y_{N}^{\dagger}\Sigma \widetilde{\Phi_1}),
\end{eqnarray}
where  $L\equiv \left(\nu, \ell \right)_{L}$ corresponds to the SM lepton doublet, while $\widetilde{\Phi}_1=i \sigma_2 \Phi_1^\star$ (with $\sigma_2$ is the second Pauli matrix), $\Sigma^c \equiv C\overline{\Sigma}^{T}$ for each fermionic triplet as shown below. 
\begin{center}
	
	$\Sigma = \left(
	\begin{array}{cc}
	\Sigma^0/\sqrt{2} & \Sigma^+ \\
	\Sigma^- & -	\Sigma^0/\sqrt{2} \\
	\end{array}
	\right)$ , \qquad
	$\Sigma^c = \left(
	\begin{array}{cc}
	\Sigma^{0c}/\sqrt{2} & \Sigma^{-c} \\
	\Sigma^{+c} & -	\Sigma^{0c}/\sqrt{2}\\
	\end{array}
	\right)$\end{center}
We drop the generation indices here, but this is to remind that there are three set of fermionic triplets for three leptonic doublets. The covariant derivative generates the coupling between the $W$ bosons and the triplet fermions and they are proportional to $g_2$ as shown below,
\begin{center}
	$\slashed{D}_{\mu}=\slashed{\partial}_{\mu}-i \sqrt{2}g_2\left(
	\begin{array}{cc}
	W^3_{\mu}/\sqrt{2} & W^+_{\mu} \\
	W^-_{\mu} & -		W^3_{\mu}/\sqrt{2}\\
	\end{array}
	\right)$.
\end{center}
Without loss of generality, we start from the basis, where $M_{N}$ is real and diagonal. In order to consider the mixing of fermionic triplets with the charged leptons, it is convenient to express the four degrees of freedom of each charged triplet in terms of a single Dirac spinor:
\begin{eqnarray}
\psi=\Sigma^{+c}_R + \Sigma^-_R.
\end{eqnarray}
On the other hand the neutral fermionic triplet components can be left in two-component notation, since they have only two degrees of freedom and mix with neutrinos, which are also described by two-component fields. This leads to the Lagrangian as follows
\begin{eqnarray}
{\cal L}_{\rm III} & = & \bar{\psi}i\slashed{\partial}\psi + \overline{\Sigma^0_R}i\slashed{\partial}\Sigma^0_R-\overline{\psi}M_{N}\psi-\left(\overline{\Sigma^0_R}\frac{M_{N}}{2}\Sigma^{0c}_R + h.c. \right)\nn\\
&&	+g\left(W^+_{\mu}\overline{\Sigma^0_R}\gamma_{\mu}P_R \psi+W^+_{\mu}\overline{\Sigma^{0c}_R}\gamma_{\mu}P_L \psi  + h.c.\right)-gW^3_{\mu}\overline{\psi}\gamma_{\mu}\psi\nonumber \\
&&	-\left( \Phi^0\overline{\Sigma^0_R}Y_{N}\nu_L+\Phi^+\overline{\Sigma^0_R}Y_{N}\ell_L +\sqrt{2}(\Phi^0\overline{\psi}Y_{N}\ell_L -\Phi^+ \overline{\nu^c_L}Y^T_{N}\psi)+ h.c.\right).
\end{eqnarray}
The mass term of the charged sector shows the usual aspect for Dirac particles:
\begin{center}
	${\cal L} \ni - \left(
	\begin{array}{cc}
	\overline{l_R} & \overline{\psi_R}\\
	\end{array}
	\right)$
	$\left(
	\begin{array}{cc}
	m_l & 0 \\
	Y_{N}v & M_{N}\\
	\end{array}
	\right)$    $	\left(
	\begin{array}{cc}
	l_L  \\
	\psi_L\\
	\end{array}
	\right)$
	$-\left(
	\begin{array}{cc}
	\overline{l_L} & \overline{\psi_L}\\
	\end{array}
	\right)$
	$ \left(
	\begin{array}{cc}
	m_l & 	Y^{\dagger}_{N}v \\
	0 & M_{N}\\
	\end{array}
	\right)$	$ \left(
	\begin{array}{cc}
	l_R  \\
	\psi_R\\
	\end{array}
	\right)$,
\end{center}
where $m_\ell=y_\ell <\phi^0>$ with $y_\ell$ is leptonic Yukawa coupling and $v=\sqrt{2} <\phi^0>=246$ GeV. On the other hand the symmetric mass matrix for the neutral states is given by
\begin{center}
	${\cal L} \ni - \left(
	\begin{array}{cc}
	\overline{\nu_L} & \overline{\Sigma^{0c}}\\
	\end{array}
	\right)$
	$\left(
	\begin{array}{cc}
	0 & Y^{\dagger}_{N}v/2\sqrt{2} \\
	Y^*_{N}v/2\sqrt{2}& M_{N}/2\\
	\end{array}
	\right)$    $	\left(
	\begin{array}{cc}
	\nu^c_L  \\
	\Sigma^0\\
	\end{array}
	\right)$\\
	$-\left(
	\begin{array}{cc}
	\overline{\nu^c_L} & \overline{\Sigma^0}\\
	\end{array}
	\right)$
	$ \left(
	\begin{array}{cc}
	0 & Y^T_{N}v/2\sqrt{2} \\
	Y_{N}v/2\sqrt{2} & M_{N}/2\\
	\end{array}
	\right)$	$ \left(
	\begin{array}{cc}
	\nu_L  \\
	\Sigma^{0c}\\
	\end{array}
	\right)$.
\end{center}
The neutrino mass matrix in this case can be written as 
	\begin{equation}\label{numass}
\mathcal{M}_\nu \ = 
\ 
\left( {\begin{array}{cc}
	0 & M_D \\
	M_D^\intercal & M_{N} \\
	\end{array} } \right) \, \quad {\rm{where}} \quad M_D \ = \ \frac{v}{\sqrt 2}Y_{N} \, .
\end{equation}
Thus the light  neutrino mass can be written as 
\begin{eqnarray}
m_{\nu}=-\frac{v^2}{2}Y^T_{N}\frac{1}{M_{N}}Y_{N},
\end{eqnarray}
which mixes the left-handed neutrinos and the neutral fermionic triplet components (right-handed neutrinos). This leads to the full  mass matrix for the neutral states as:
	
	\begin{equation}\label{numass}
	\mathcal{M}_\nu \ = 
	\ 
	\left( {\begin{array}{cc}
		0 & M_D \\
		M_D^\intercal & M_{N} \\
		\end{array} } \right) \, .
	\end{equation}
Diagonalizing the mass matrix we obtain one small neutrino eigenvalue in the limit of $||M_D||\ll ||M_{N}||$ as : 
	\begin{eqnarray}\label{megst}
	m_{\nu} & \ \simeq & -M_D M_{N}^{-1} M_D^\intercal \,  ,
	\end{eqnarray}
where the triplet fermions take the mass values  around $M_{N}$. From Eq.~\eqref{megst}, 
	it is evident that to have the  light neutrino mass $m_\nu\lesssim 0.1$ eV, we need $Y_{N} \lesssim \mathcal{O}(10^{-6})$ for  $M_{N}\sim {\cal O}(100)~{\rm GeV}$.
However, such choices of small couplings cannot affect the RG evolution of other couplings \cite{PBBDSJ}, a coupling motivated by inverse seesaw can be relevant here. 
	
	The  collider signatures of heavy neutrinos and leptons rely upon larger Yukawa couplings. These are further restricted from  electroweak precision data \cite{Atre:2009rg, Deppisch:2015qwa,delAguila:2008pw,Akhmedov:2013hec,deBlas:2013gla}.  In the  inverse seesaw frame work ~\cite{Mohapatra:1986aw, Mohapatra:1986bd,ISS,ISS2}, we introduce another set of fermions which are  $SU(2)_L$  triplet, $\Sigma_{2i}$ (with $i=1,2,3$) accompanying the $\Sigma_{1i}$. The inverse seesaw Lagrangian is  given by,
	\begin{align}\label{lmass}
	{\cal L}_{\rm ISS} \ = \  Tr[\overline{\Sigma_{1i}} \slashed{D}\Sigma_{1i}]+ Tr[\overline{\Sigma_{2i}} \slashed{D}\Sigma_{2j}] -\frac{1}{2}Tr[\overline{\Sigma_{2i}} \mu_{\Sigma_{ij}}\Sigma^c_{2j} + \overline{\Sigma^c_{2i}}\mu^*_{\Sigma_{ij}}\Sigma_{2j}] \nonumber \\
	-\left(\widetilde{\Phi}^{\dagger}_1\overline{\Sigma_{1i}}\sqrt{2}Y_{N_{ij}}L_j + Tr[\overline{\Sigma}_{1i} M_{N_{ij}}\Sigma_{2j}]   +  {\rm H.c.}   \right),
	\end{align}
	where $M_{D}$ is a 3$\times$3 Dirac mass mixing matrix, $M_{N}$ is the mixing mass term between the two triplets and $\mu_\Sigma$ is the small lepton number violating mass term for the $\Sigma_{2}$-fields. In the basis of $\{\nu^c_L, \Sigma^0_1, \Sigma^0_2\}$, the full $9 \times 9$ neutral components mass matrix can be written as
	\be\label{massm}
	\mathcal{M}_{\nu} \ = \ \begin{pmatrix}
		0 \, \quad M_D\, \quad 0 \\
		M^\intercal_D\, \quad 0\, \quad M_{N}\\
		0\,\quad   M^\intercal_{N}\, \quad  \mu_{\Sigma} 
	\end{pmatrix}.
	\ee
	one obtains the three light neutrinos  diagonalizing the mass matrix Eq.~\eqref{massm}  as described below
	\bea\label{neueigen}
	m_{\nu} \ \simeq \ M_D M^{-1}_{N}\mu_{\Sigma} (M^\intercal_{N})^{-1}M^\intercal_D \, , 
	\eea
other neutrino masses are heavy and they are given by $M_{N}\pm \mu_{\Sigma}/2$. Here the small $\mu_{\Sigma}$ share the load of seesaw making $Y_N \sim \mathcal{O}(0.1)$. Such large Yukawa 
couplings give significant negative contributions in the stability of EW vacuum \cite{Ipek:2018sai}. 
	\section{Perturbativity}\label{pert}
	To illustrate the theoretical bounds from perturbativity behaviour of the dimensionless couplings, we impose the following conditions on the dimensionless couplings as perturbative limit at a given scale $\mu$,
	\begin{align}
	\left|\lambda_i\right|  \ \leq \ 4 \pi, \qquad
	\left|g_j\right| \ \leq \ 4 \pi, \qquad \left|Y_k\right|  \ \leq \ \sqrt{4\pi} \, ,
	\end{align}
	where $g_j$ with $j=1,2$ are the EW gauge couplings\footnote{The running of $g_3$ remains the as in the SM, as the new fields do not have $SU(3)$ charges.},  $\lambda_i$ with $i=1,2,3,4,5$ are the quartic couplings corresponding to the scalars and $Y_k$ with $k=u,d,\ell$ are the Yukawa couplings for the quarks and leptons.  The  extension of SM with a $SU(2)_L$ inert doublet as well as by $SU(2)_L$ triplet fermions can change the running $g_2(\mu)$,  which in turn affects the progression of other couplings namely the $\lambda_i$ relevant for the vacuum stability and perturbativity. Below we discuss that how $g_2(\mu)$  gets affected via the extra scalar and fermions of this model. 
	
\subsection{Running of Gauge couplings:}\label{gaugecoup}
\begin{figure}[t]
	\begin{center}
		\mbox{\subfigure[]{\includegraphics[width=0.5\linewidth,angle=-0]{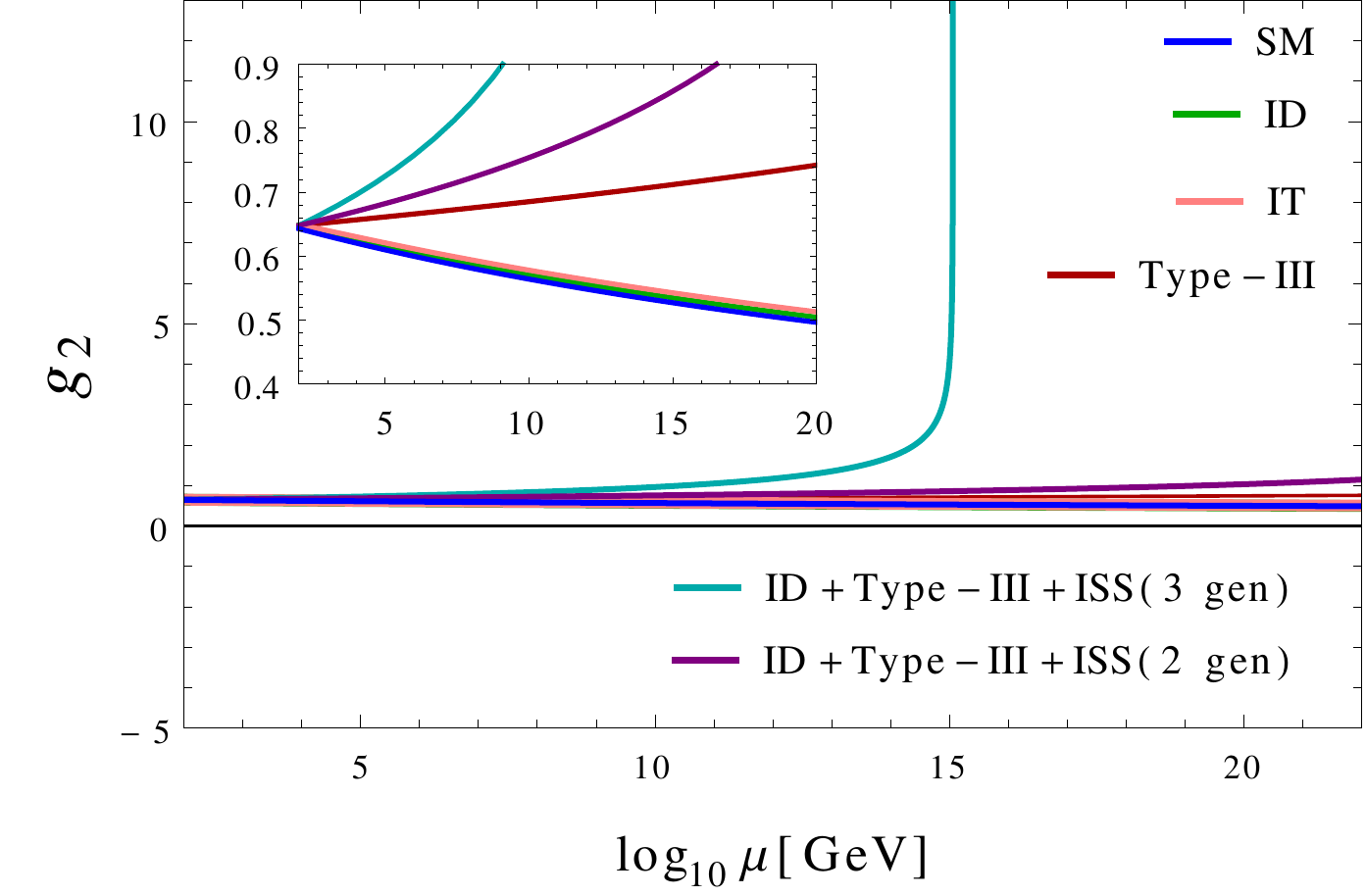}\label{f1a}}
			\subfigure[]{\includegraphics[width=0.5\linewidth,angle=-0]{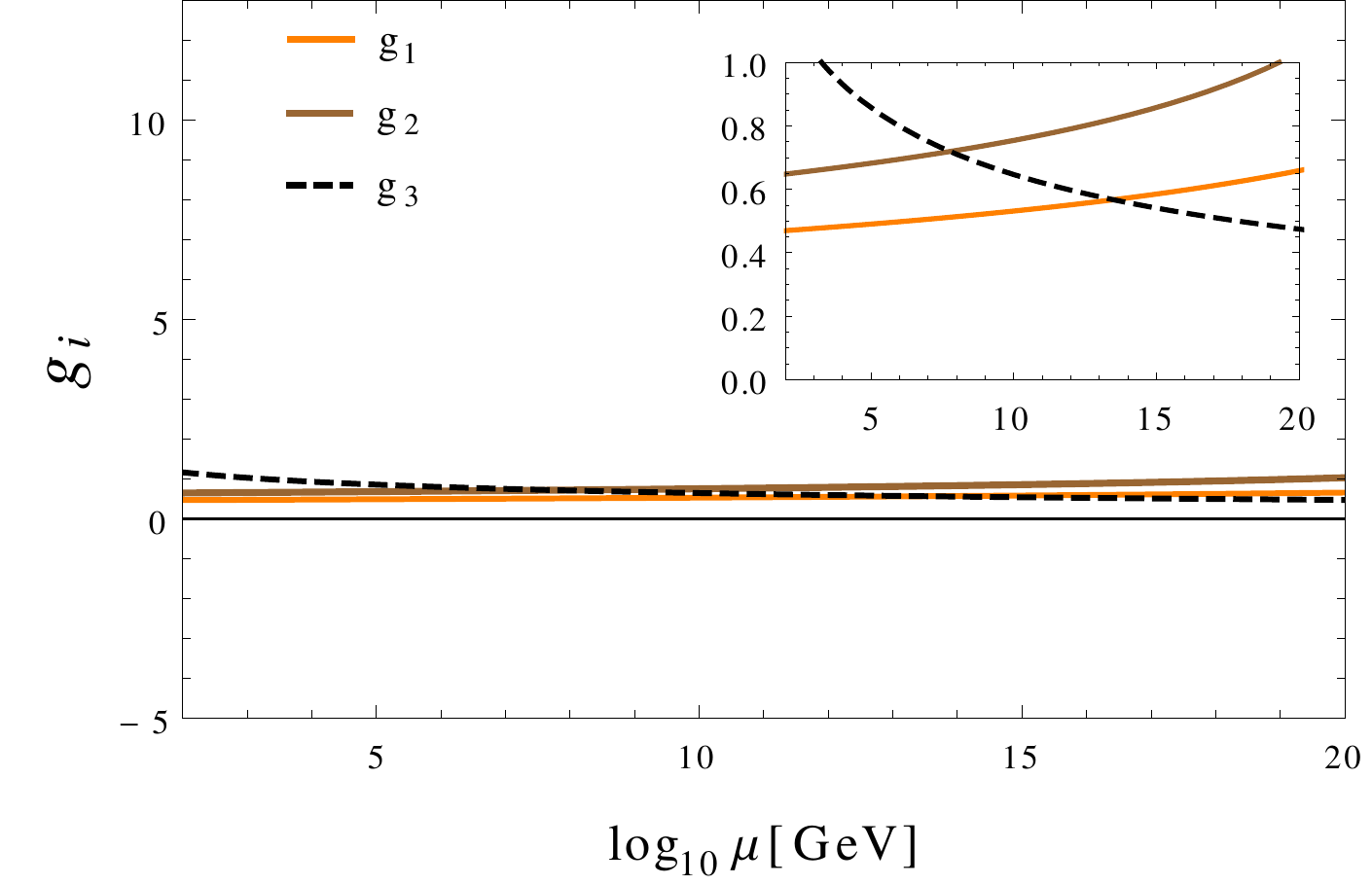}\label{f3a}}}
			\mbox{\subfigure[]{\includegraphics[width=0.5\linewidth,angle=-0]{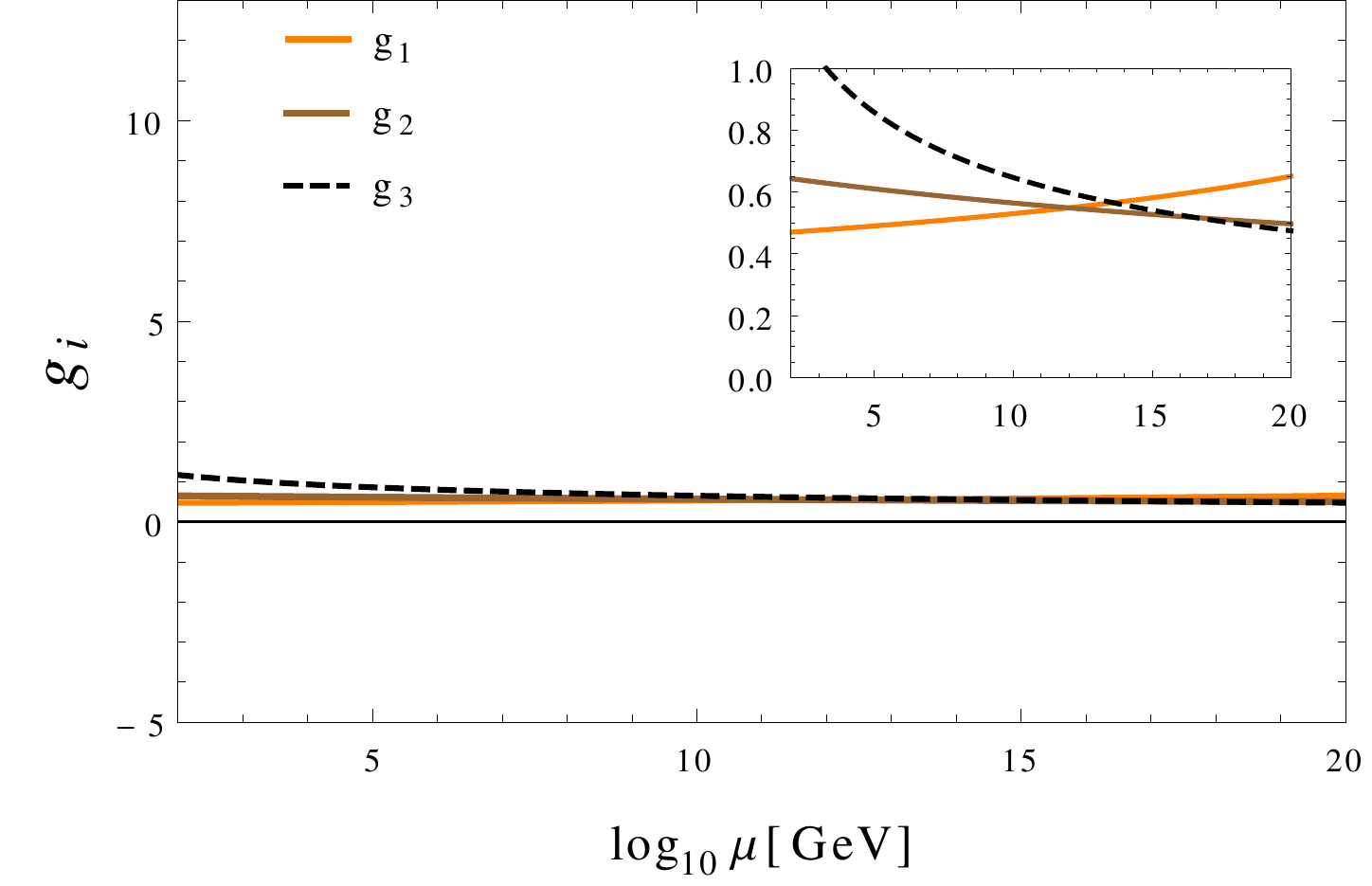}\label{f2a}}}
		\caption{(a) Running of $g_2$ with the scale $\mu$ for the SM in blue, SM + ID in green, SM + ITM in pink, SM + three generations of Type-III fermions in red, SM+ID +  three generations of Type-III fermions in sky-blue and SM+ ID+  two generations of Type-III fermions in purple in two-loop. (b) Running of all three gauge couplings $g_i$ at two-loop for the SM+ ID+ two generations of Type-III fermions.  (c) Running of all three gauge couplings $g_i$ at two-loop for the SM. }\label{gaugec}
	\end{center}
\end{figure}

Eqs.~\ref{SMgauge}-~\ref{SMgauge3} and Figure~\ref{f2a} describe the evolution of the SM gauge couplings at a given scale $\mu$ (not explicitly mentioned in those Eqs.) at two-loop level. Both $g_2$ and $g_3$ decrease with the increase in the running scale $\mu$  and  remain perturbative in the high scale limit. However, such behaviour can change substantially with the inclusion of other $SU(2)_L$ fields such as ID and Type-III Seesaw fermions, which are in the triplet of $SU(2)_L$ representation.

\bea\label{SMgauge}
\beta_{g_1} &= & 
\frac{1}{16\pi^2}\Bigg[\frac{21}{5} g_{1}^{3}\Bigg]
+\frac{1}{(16\pi^2)^2}\Bigg[\frac{1}{50} g_{1}^{3} \Big(180 g_{2}^{2}  + 208 g_{1}^{2}  -25 \mbox{Tr}\Big({Y_d  Y_{d}^{\dagger}}\Big) \nonumber \\ 
& +& 440 g_{3}^{2} -45 \mbox{Tr}\Big({Y_{N}  Y_{N}^{\dagger}}\Big)  -75 \mbox{Tr}\Big({Y_e  Y_{e}^{\dagger}}\Big)  -85 \mbox{Tr}\Big({Y_u  Y_{u}^{\dagger}}\Big) \Big)\Bigg], \,  \\\label{SMgauge2}
\beta_{g_2}  &=  & 
\frac{1}{16\pi^2}\Bigg[-\frac{19}{6} g_{2}^{3}\Bigg]+\frac{1}{(16\pi^2)^2}\Bigg[
\frac{1}{30} g_{2}^{3} \Big(-15 \mbox{Tr}\Big({Y_e  Y_{e}^{\dagger}}\Big)  + 175 g_{2}^{2}  + 27 g_{1}^{2}   \nonumber \\
 & +& 360 g_{3}^{2} - 45 \mbox{Tr}\Big({Y_d  Y_{d}^{\dagger}}\Big) -45 \mbox{Tr}\Big({Y_u  Y_{u}^{\dagger}}\Big) \Big)\Bigg], \,   \\\label{SMgauge3}
\beta_{g_3} &= &
\frac{1}{16\pi^2}\Bigg[-7 g_{3}^{3}\Bigg]+
\frac{1}{(16\pi^2)^2}\Bigg[-\frac{1}{10} g_{3}^{3} \Big(-11 g_{1}^{2}  + 20 \mbox{Tr}\Big({Y_d  Y_{d}^{\dagger}}\Big)  \nonumber \\
&+& 20 \mbox{Tr}\Big({Y_u  Y_{u}^{\dagger}}\Big) + 260 g_{3}^{2}  -45 g_{2}^{2} \Big)\Bigg].
\eea

In Eq.~\ref{g2IDM}-Eq.~\ref{g2t2}, we show the modified evolution of the gauge couplings in the presence of the ID and triplet fermions. Eq.~\ref{g2IDM} shows that an inclusion of inert $SU(2)_L$ doublet makes $\beta_{g_2}$ less negative at one-loop and so as at two-loop. The behaviours can be verified from  the Figure~\ref{f1a}, where the blue curve implies SM and green line represents  SM with ID. For the comparison we also show that an addition of a $Y=0$ $SU(2)_L$ triplet scalar also reduces the negative impact in $\beta_{g_2}$, as can be read from Eq.~\ref{g2ITM},  which is also evident from Figure~\ref{f1a} pink curve. However, from Eq.~\ref{g2tIII} we can see that even at one-loop $\beta_{g_2}$ has become positive with the factor changed to $\frac{5g^3_2}{6}$ if we extend the SM with three generations of Type-III fermions. This   behaviour continues even at two-loop. The running of $g_2$ in this case has been depicted by red curve of Figure~\ref{f1a}. Unlike the SM, or ID scenario, the coupling $g_2$ in this case increases with increasing $\mu$. In addition to the three generations of Type-III fermions if we add a $SU(2)_L$ inert doublet which gives the much needed DM candidate,  the factor at one-loop is  enhanced to $5 g^3_2$  with further enhancement at two-loop and also visible by the sky-blue curve in Figure~\ref{f1a}. Certainly we can see that $g_2$ coupling remians perturbative till Planck scale for all the scenarios except ID + Type-III + ISS scenario with three generations of fermionic triplet. It is evident that with three generations of fermionic triplet $g_2$ coupling loses perturbativity around $10^{15}$ GeV.  However, restricting it only to two generations of triplet fermions, and one ID, this perturbativity scale pushes till the Planck scale. This prompt us to choose only two generations of Type-III fermions along with ID, rather than the three generations. The running of the gauge coupling  $g_2$ for the two generation scenario is evident from the purple curve of Figure~\ref{f1a},  and also from Eq.~\ref{g2t2}. Figure.~\ref{f3a} represents  running of all the three gauge couplings for the two generation scenario. Therefore, from EW scale point of view, SM extension with an ID and two generations of Type-III fermions with ISS mechanism are more motivated, as this accommodates  perturbativity of the gauge couplings  till Planck scale, as well as, can explain two small  light neutrino masses, and provide a dark matter candidate from the ID. However, one lives with the different  asymptotic behaviour of $g_2$ as compared to SM.

\bea\label{g2IDM}
\beta_{g^{IDM}_2}  & =   &
\frac{1}{16\pi^2}\Bigg[-3 g_{2}^{3}\Bigg]+\frac{1}{(16\pi^2)^2}\Bigg[
\frac{1}{10} g_{2}^{3} \Big(120 g_{3}^{2}  + 12 g_{1}^{2}  -15 \mbox{Tr}\Big({Y_d  Y_{d}^{\dagger}}\Big)  \nonumber\\
&-&15 \mbox{Tr}\Big({Y_u  Y_{u}^{\dagger}}\Big)  -5 \mbox{Tr}\Big({Y_e  Y_{e}^{\dagger}}\Big)  + 80 g_{2}^{2} \Big)\Bigg] \, .  \\\label{g2ITM}
	\beta_{g^{ITM}_2}  &=  &
\frac{1}{16\pi^2}\Bigg[-\frac{17}{6} g_{2}^{3}\Bigg]+\frac{1}{(16\pi^2)^2}\Bigg[
\frac{1}{30} g_{2}^{3} \Big(-15 \mbox{Tr}\Big({Y_e  Y_{e}^{\dagger}}\Big)  + 27 g_{1}^{2}  + 360 g_{3}^{2}  \nonumber\\
&+ &455 g_{2}^{2}  -45 \mbox{Tr}\Big({Y_d  Y_{d}^{\dagger}}\Big)  -45 \mbox{Tr}\Big({Y_u  Y_{u}^{\dagger}}\Big) \Big)\Bigg] \, .  \\\label{g2tIII}
\beta^{Type-III }_{g_2, \,3gen}  & =&   
\frac{1}{16\pi^2}\Bigg[\frac{5}{6} g_{2}^{3}\Bigg]+\frac{1}{(16\pi^2)^2}\Bigg[
\frac{1}{60} g_{2}^{3} \Big(-165 \mbox{Tr}\Big({Y_e  Y_{e}^{\dagger}}\Big)  -30 \mbox{Tr}\Big({Y_e  Y_{e}^{\dagger}}\Big)  \nonumber  \\
&+& 4190 g_{2}^{2} + 54 g_{1}^{2}  + 720 g_{3}^{2}  -90 \mbox{Tr}\Big({Y_d  Y_{d}^{\dagger}}\Big)  -90 \mbox{Tr}\Big({Y_u  Y_{u}^{\dagger}}\Big) \Big)\Bigg], \,  \\ \label{g2t3}
\beta_{g_2,\, 3gen} & =&  
\frac{1}{16\pi^2}\Bigg[5 g_{2}^{3}\Bigg]+\frac{1}{(16\pi^2)^2}\Bigg[
\frac{1}{10} g_{2}^{3} \Big(120 g_{3}^{2}  + 12 g_{1}^{2}  + 1360 g_{2}^{2}  -15 \mbox{Tr}\Big({Y_d  Y_{d}^{\dagger}}\Big)  \nonumber \\
& -& 15 \mbox{Tr}\Big({Y_u  Y_{u}^{\dagger}}\Big) - 55 \mbox{Tr}\Big({Y_{N}  Y_{N}^{\dagger}}\Big)  -5 \mbox{Tr}\Big({Y_e  Y_{e}^{\dagger}}\Big) \Big)\Bigg], \,   \\ \label{g2t2}
\beta_{g_2,\, 2gen}  & =&   
\frac{1}{16\pi^2}\Bigg[\frac{7}{3} g_{2}^{3}\Bigg]+\frac{1}{(16\pi^2)^2}\Bigg[
\frac{1}{30} g_{2}^{3} \Big(-15 \mbox{Tr}\Big({Y_e  Y_{e}^{\dagger}}\Big)  -165 \mbox{Tr}\Big({Y_{N}  Y_{N}^{\dagger}}\Big) \nonumber \\
 &+& 2800 g_{2}^{2}  + 360 g_{3}^{2}  + 36 g_{1}^{2}  -45 \mbox{Tr}\Big({Y_d  Y_{d}^{\dagger}}\Big)  -45 \mbox{Tr}\Big({Y_u  Y_{u}^{\dagger}}\Big) \Big)\Bigg].
\eea

	\begin{figure}[t]
		\begin{center}
			\includegraphics[width=0.6\linewidth,angle=-0]{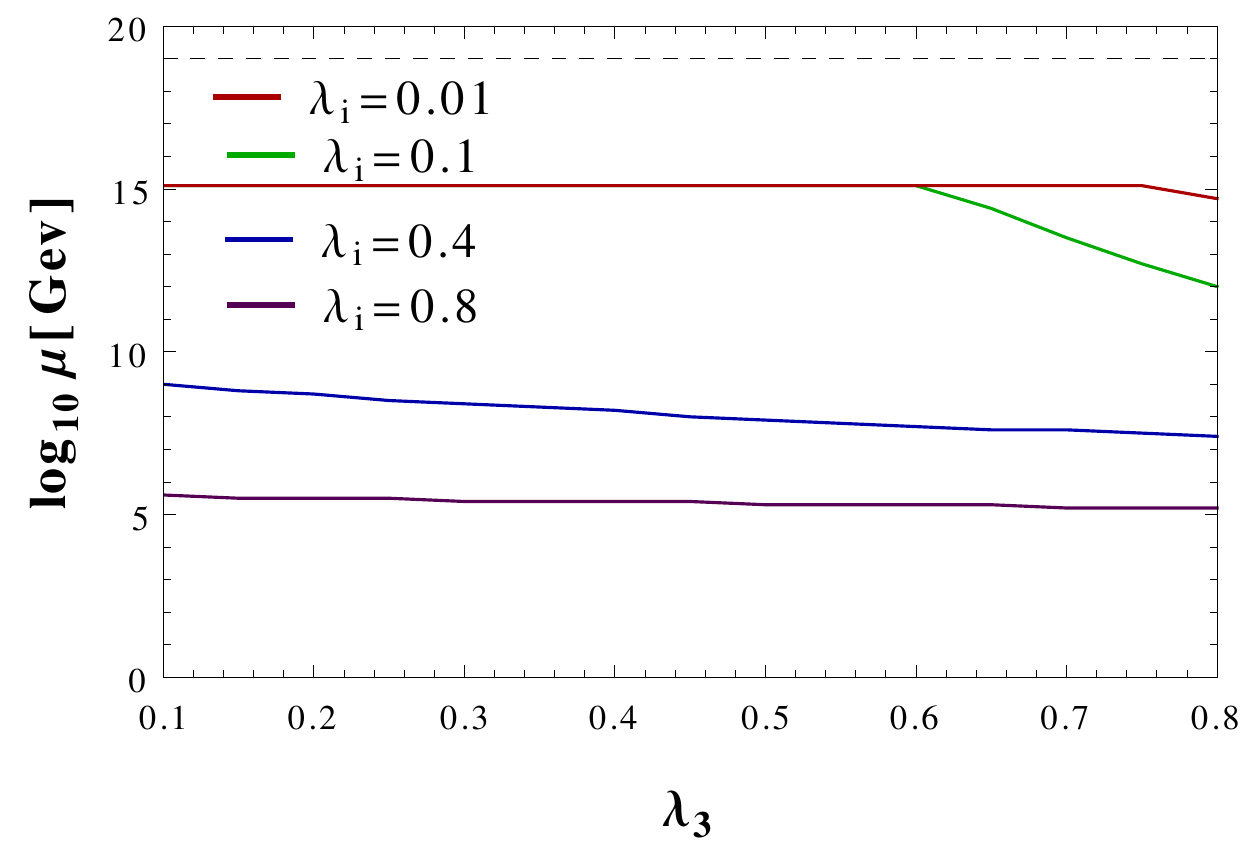}
			\caption{Bounds on Perturbative  scale of the scalar quartic couplings verses $\lambda_3$ at two-loop for Yukawa coupling $Y_N=0.02$, considering three generations of fermionic triplet. Different initial values are designated as red, green, blue and purple curves for $\lambda_i = 0.01, 0.1, 0.4, 0.8$ respectively at the EW scale.}\label{fig1l}
		\end{center}
	\end{figure}
The perturbative nature of the scalar quartic couplings are also modified  by the inclusion of ID and triplet fermions. Below, we first focus on the  analysis  of  the perturbative behaviour of the scalar quartic couplings with respect to $\lambda_{3}$  in the scenario with SM associated with an  ID, and three  generations of fermionic triplet. The perturbative limit is calculated if at least one of the coupling crosses the perturbativity or hits Landau pole. Figure~\ref{fig1l}  describes the perturbative behaviour of $\lambda_{3}$ with the scale $\mu$ for $Y_N=0.02$, where the other quartic couplings $\lambda_{(i=2,4,5)}$ are kept at different values at the electroweak scale. For the coupling $\lambda_1$, we choose $\lambda_1=0.1264$, which gives  SM like Higgs boson mass around $125.5$ GeV. Here $\lambda_i, i=2,4,5$ at the EW scale are chosen in very weak ($\lambda_i = 0.01$ in red ), weak ($\lambda_i = 0.10$ in green ), moderate ($\lambda_i = 0.40$ in blue ) and strong ($\lambda_i = 0.80$ in purple) coupling limits respectively.

	In ID+Type-III+ ISS scenario we have three generations of fermionic triplet $\Sigma_{1i}$ which generates the Dirac term and three generations of additional fermionic triplet $\Sigma_{2j}$ instrumental for inverse seesaw mechanism. Having $ \rm SU(2)_L$ charge they contribute to the beta functions of the $\rm SU(2)_L$ gauge coupling, i.e. $\beta_{g_2}$ positively; which is somewhat different than normal Type-I + ISS case \cite{PBBDSJ}. Additionally $\rm Y_N$ also contributes positively to the beta functions of $\lambda_{3,4,5}$ at one-loop and negative effects only comes at two-loop. 
	Both $\rm Y_N$  and $\beta_{g_2}$ push the $\lambda_{3,4,5}$ towards non-perturbative limit. The Higgs quartic couplings  $\lambda_1$ get negative corrections from the Yukawa coupling $Y_N$,  pushing the Higgs potential toward instability at one-loop and two-loop. The detailed two-loop beta functions are given in  Appendix~\ref{betaf1},  where only two generation effects are shown.  However,  in Figure~\ref{fig1l}  we have considered all three generations of $\rm SU(2)_L$ triplet fermions along with ID which makes the theory more stable,  but simultaneously becomes non-perturbative below Planck scale for all corresponding values of $\lambda_i$ and $\rm Y_N$. Thus for the Planck scale perturbativity  we should restrict ourselves  to two generations of fermionic triplets. For $\lambda_i=0.01, \, 0.10$ which is less than $\lambda_1=0.1264$, $\lambda_1$ hits Landau pole before going into instability around $10^{15}$ GeV (red and green lines). This happens due to positive effect of $g_2$, which is different than the Type-I case \cite{PBBDSJ}.  The bending happens due to further positive effects of $\rm \lambda_1 Tr(Y^\dagger_N Y_N)$ and other quadratic terms involving $\lambda_i$s. Such bending effects grows from $\lambda_i=0.01 $ (brown line)  to $\lambda_i=0.10 $ (green line). For $\rm \lambda_i \geq 0.2$ the perturbative limits come at much smaller case as other $\lambda_i$s hit the Landau pole before $\lambda_1$.
	
\begin{figure}
		\begin{center}
			\mbox{\subfigure[$Y_N=0.01$]{\includegraphics[width=0.5\linewidth,angle=-0]{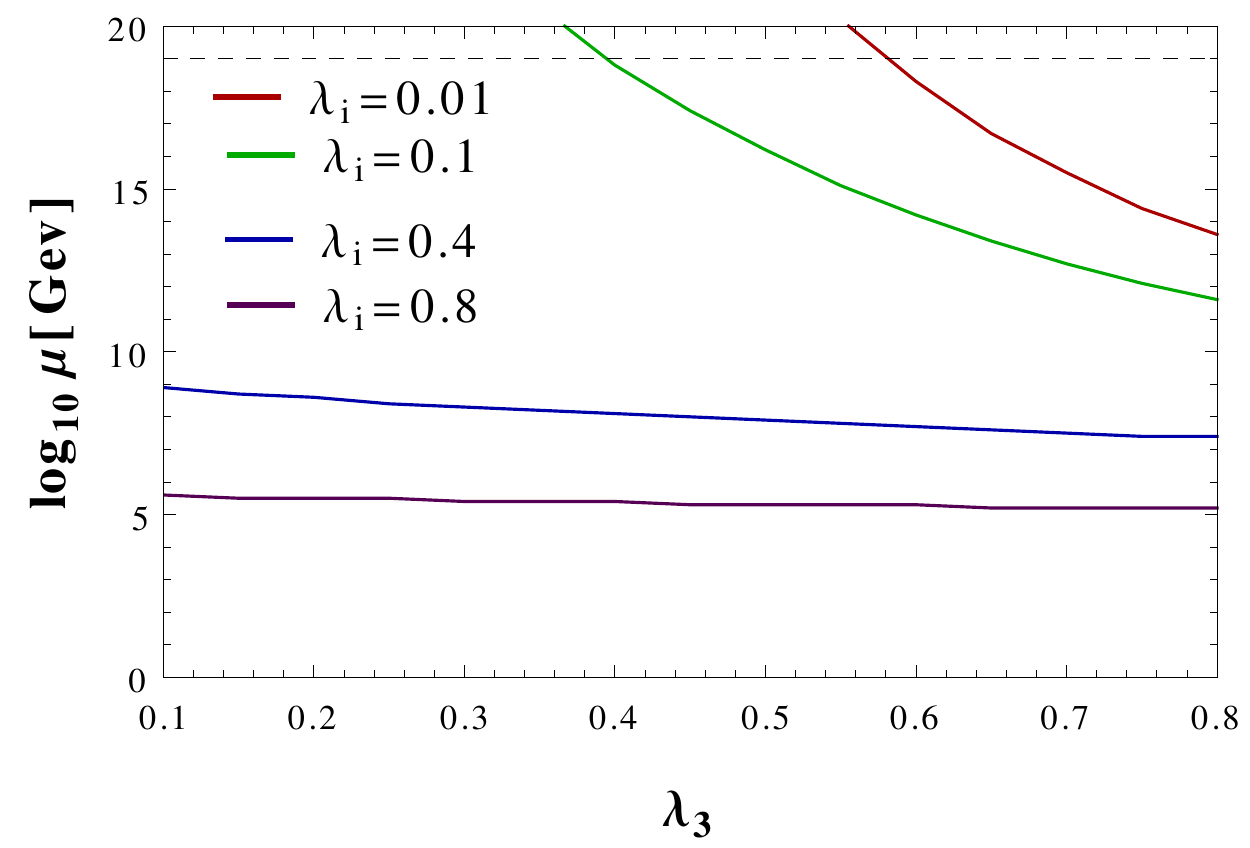}\label{f1}}
				\subfigure[$Y_N=0.4$]{\includegraphics[width=0.5\linewidth,angle=-0]{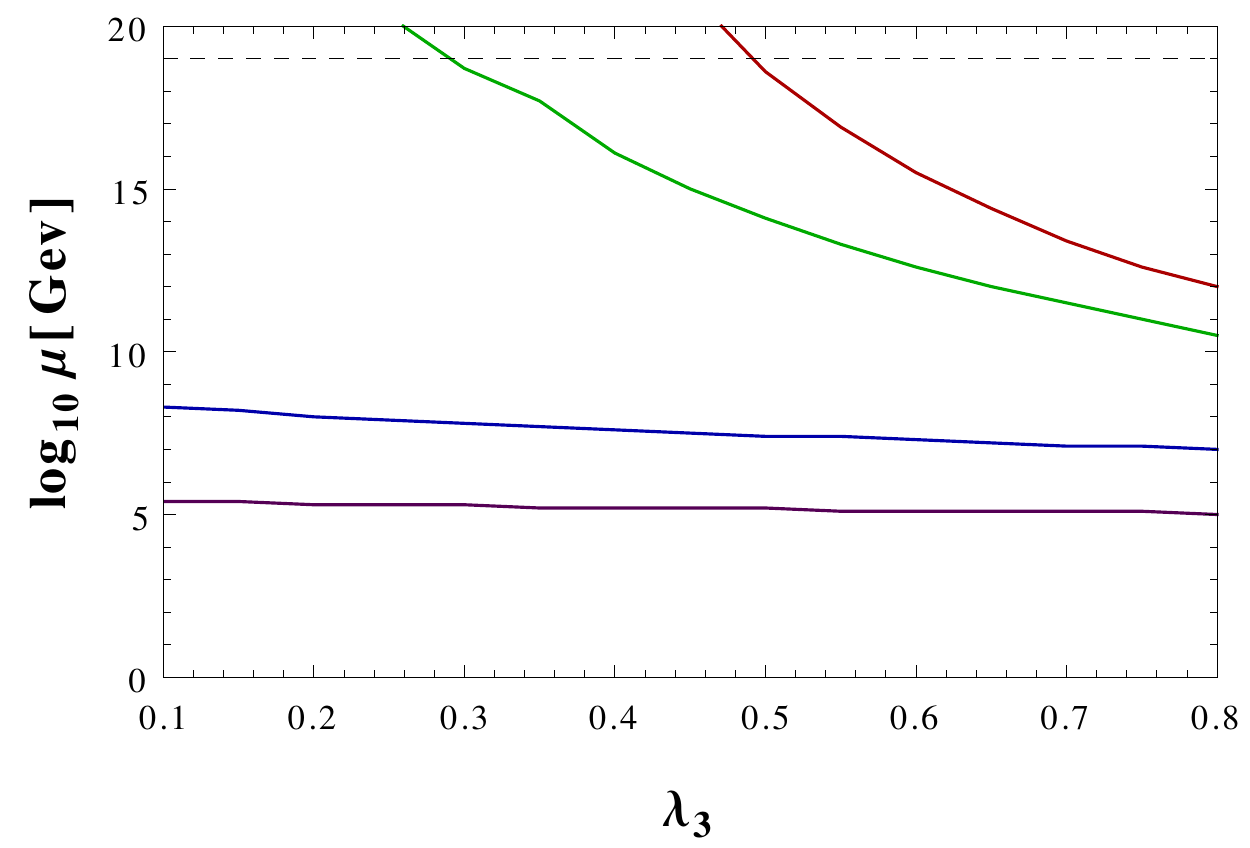}\label{f2}}}
			\mbox{\subfigure[$Y_N=0.01$]{\includegraphics[width=0.5\linewidth,angle=-0]{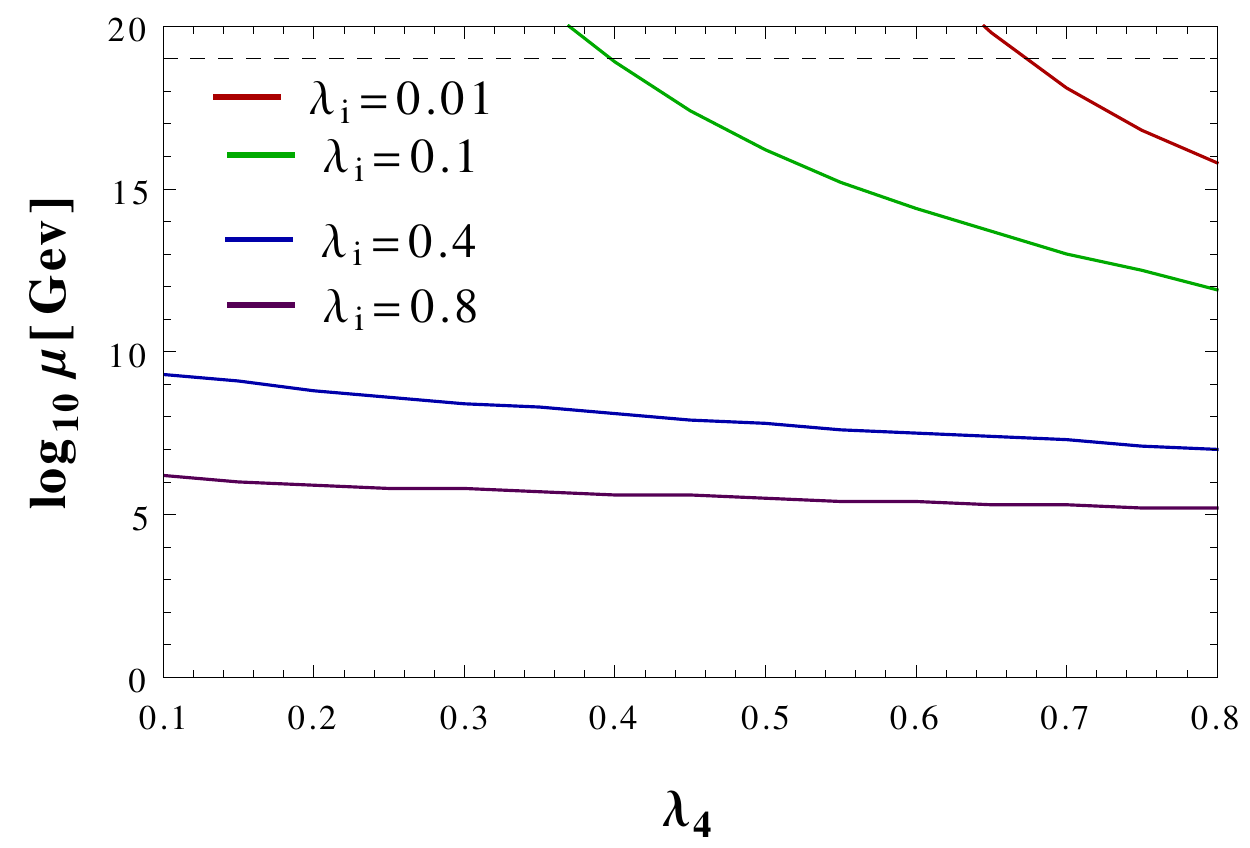}\label{f3}}
				\subfigure[$Y_N=0.4$]{\includegraphics[width=0.5\linewidth,angle=-0]{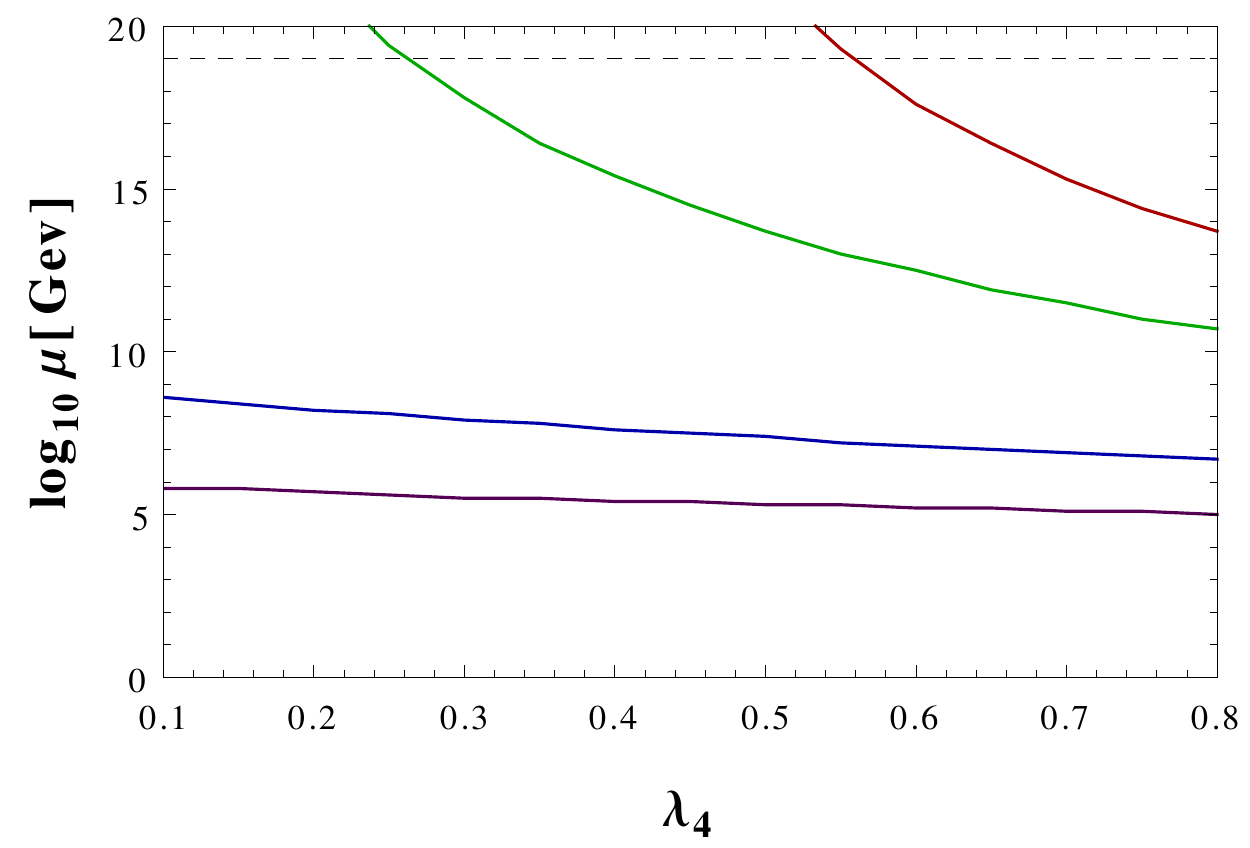}\label{f4}}}
			\mbox{\subfigure[$Y_N=0.01$]{\includegraphics[width=0.5\linewidth,angle=-0]{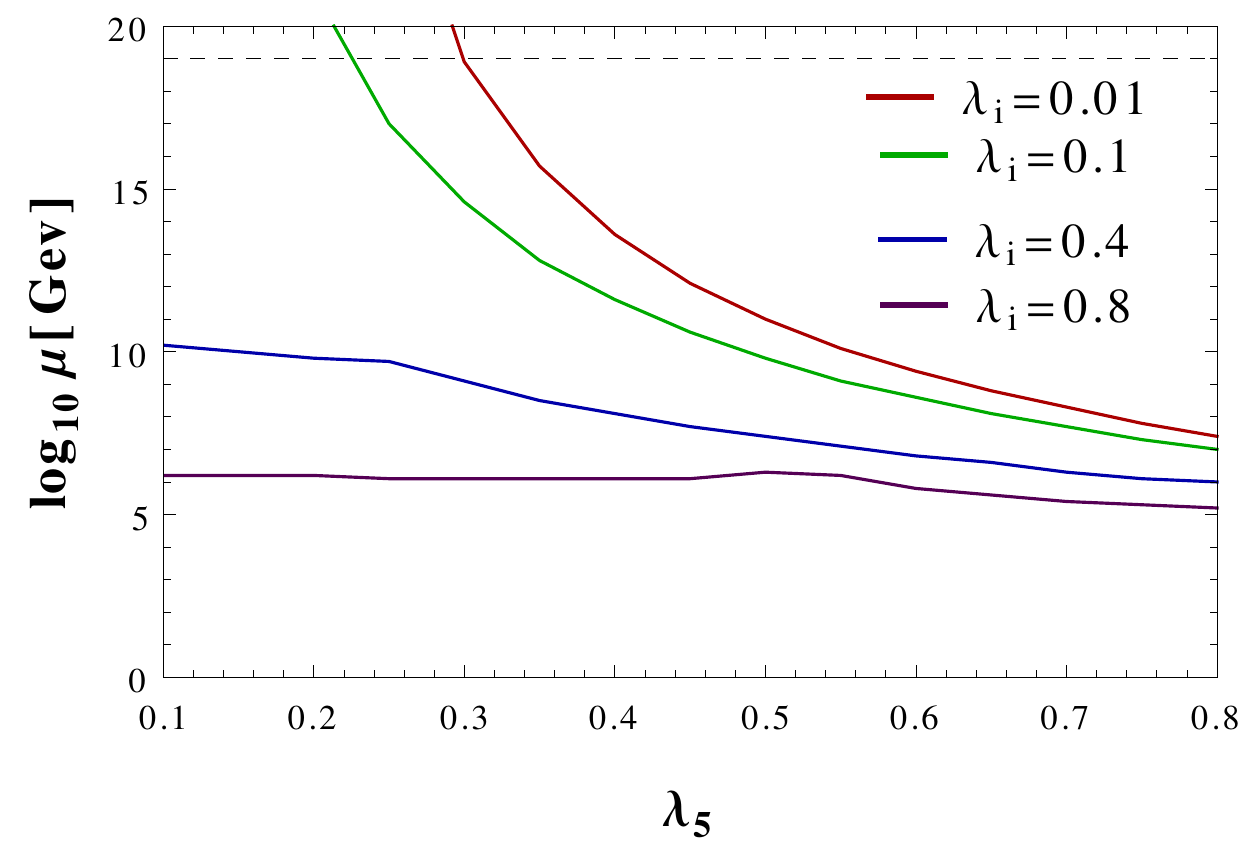}\label{f5}}
				\subfigure[$Y_N=0.4$]{\includegraphics[width=0.5\linewidth,angle=-0]{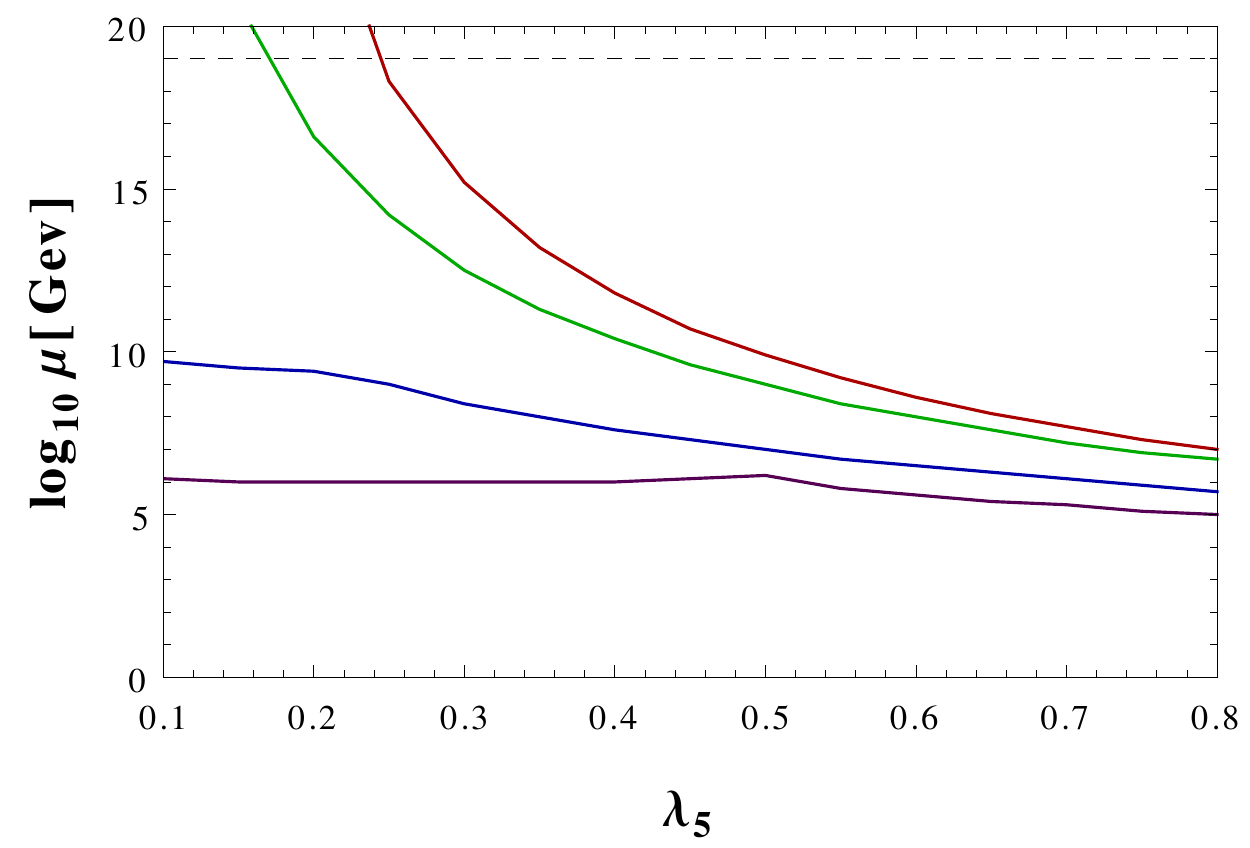}\label{f6}}}
			\caption{Perturbative limits of $\lambda_3, \,  \lambda_4$ and $\lambda_5$ for various  $\lambda_i$s and $Y_N=0.01,\,0.40$ at EW scale.  The colour codes are presented as: very weak ($\lambda_i = 0.01$ in red ), weak ($\lambda_i = 0.10$ in green ), moderate ($\lambda_i = 0.40$ in blue ) and strong ($\lambda_i = 0.80$ in purple) couplings respectively with two generations of fermionic triplet in the SM plus ID plus Type-III ISS scenario.}\label{fig2l}
		\end{center}
	\end{figure}

	In Figure~\ref{fig2l}, we present the perturbative behaviour of the Higgs quartic couplings for SM extension with ID + Type-III + ISS with two generations of fermionic triplets. The perturbativity behaviour of the scalar quartic couplings $\lambda_{3,\,4,\,5}$  are studied in Figure ~\ref{f1} -~\ref{f6} respectively for two different choices of the coupling $Y_N$, i.e. $Y_N=0.01, \, 0.40$. The other   quartic couplings $\lambda_{i}$ are chosen to be $0.01, 0.1, 0.4$ and $0.8$ which are shown by the red, green, blue and purple lines, respectively. Higgs quartic coupling $\lambda_3$ is perturbative till Planck scale for $\lambda_3 \lesssim 0.56, 0.37$ for $\lambda_i(\rm{EW}) =0.01, 0.10$ respectively, as shown in Figure~\ref{f1}. For a larger coupling $\lambda_i (\rm{EW})=0.40, 0.80$ theory becomes non-perturbative at much lower scale $\sim$ $10^{8.9}, \, 10^{5.7}$ GeV  for almost all initial values of $\lambda_3$, and for  a coupling $Y_N=0.01$. 
	Figure~\ref{f2} shows the similar behaviour of $\lambda_3$ for a larger $Y_N$, where we choose  $Y_N=0.40$ and the other quartic couplings $\lambda_i(\rm{EW})=0.01, 0.10$. As is evident from Figure~\ref{f2}, the Higgs quartic coupling $\lambda_3$ is perturbative till Planck scale  for $\lambda_3 \lesssim 0.47, 0.26$ for the above choices of  $\lambda_i(\rm{EW})$. For $\lambda_i (\rm{EW})=0.40, 0.80$, theory again becomes non-perturbative at much lower scale $\sim$ $10^{8.4}, \, 10^{5.4}$ GeV  for almost all initial values of $\lambda_3$. Similarly, the perturbative bounds on Higgs quartic coupling $\lambda_4$ are shown in Figure~\ref{f3}-\ref{f4}. The results are very similar to the case of $\lambda_3$. Here for the choice of $\lambda_i(\rm{EW})=0.01, 0.10$ the perturbative limits remain valid till the Planck scale  for $\lambda_4 \lesssim 0.64, \, 0.36$ and for $Y_N=0.01$. For a larger $Y_N=0.40$, the corresponding perturbative limit turns out to be   $\lambda_4 \lesssim 0.54, \, 0.24$. For higher values of $\lambda_i(\rm{EW})$,  theory becomes non-perturbative at much lower scale, i.e. $\sim$ $10^{9.2}, \, 10^{6.1}$ GeV for $Y_N=0.01$ and $\sim$ $10^{8.8}, \, 10^{5.9}$ GeV  for almost all initial values of $\lambda_4$ and for the choice of Yukawa couplings $Y_N=0.01$ and $Y_N=0.40$,  respectively. As depicted in Figure~\ref{f5}-~\ref{f6}, Higgs quartic coupling $\lambda_5$ is perturbative till Planck scale for $\lambda_5 \lesssim 0.29, \, 0.22$  for $Y_N=0.01$ and $\lambda_5 \lesssim 0.24, \, 0.16$ for $Y_N=0.40$ for the choice of $\lambda_i(\rm{EW})=0.01, 0.10$ respectively. For higher values of $\lambda_i(\rm{EW})$, theory becomes non-perturbative at much lower scale $\sim$ $10^{10.1}, \, 10^{6.2}$ GeV  for $Y_N=0.01$   and $\sim$ $10^{9.8}, \, 10^{6.1}$ GeV  for $Y_N=0.4$ respectively for almost all initial values of $\lambda_5$. The perturbative scale decreases for larger choices of  $Y_N$ as $\lambda_{3,4,5}$ increases with $Y_N$ and even faster than Type-I case \cite{PBBDSJ} with the stringent constraint comes from the perturbativity bound of $\lambda_5$.  Here the theory becomes non-perturbative before  Planck scale for $\lambda_i (\rm{EW})=0.40, 0.80$ respectively.  For   $\lambda_i (\rm{EW})=0.01, 0.10$ the Planck scale validity can be achieved for $\lambda_5 \leq 0.30, 0.22$ and  $\lambda_5 \leq 0.25, 0.17$ for the choices of $Y_N=0.01, 0.40$ respectively. 

\begin{figure}[t] 
	\begin{center}
		\includegraphics[width=0.7\linewidth,angle=-0]{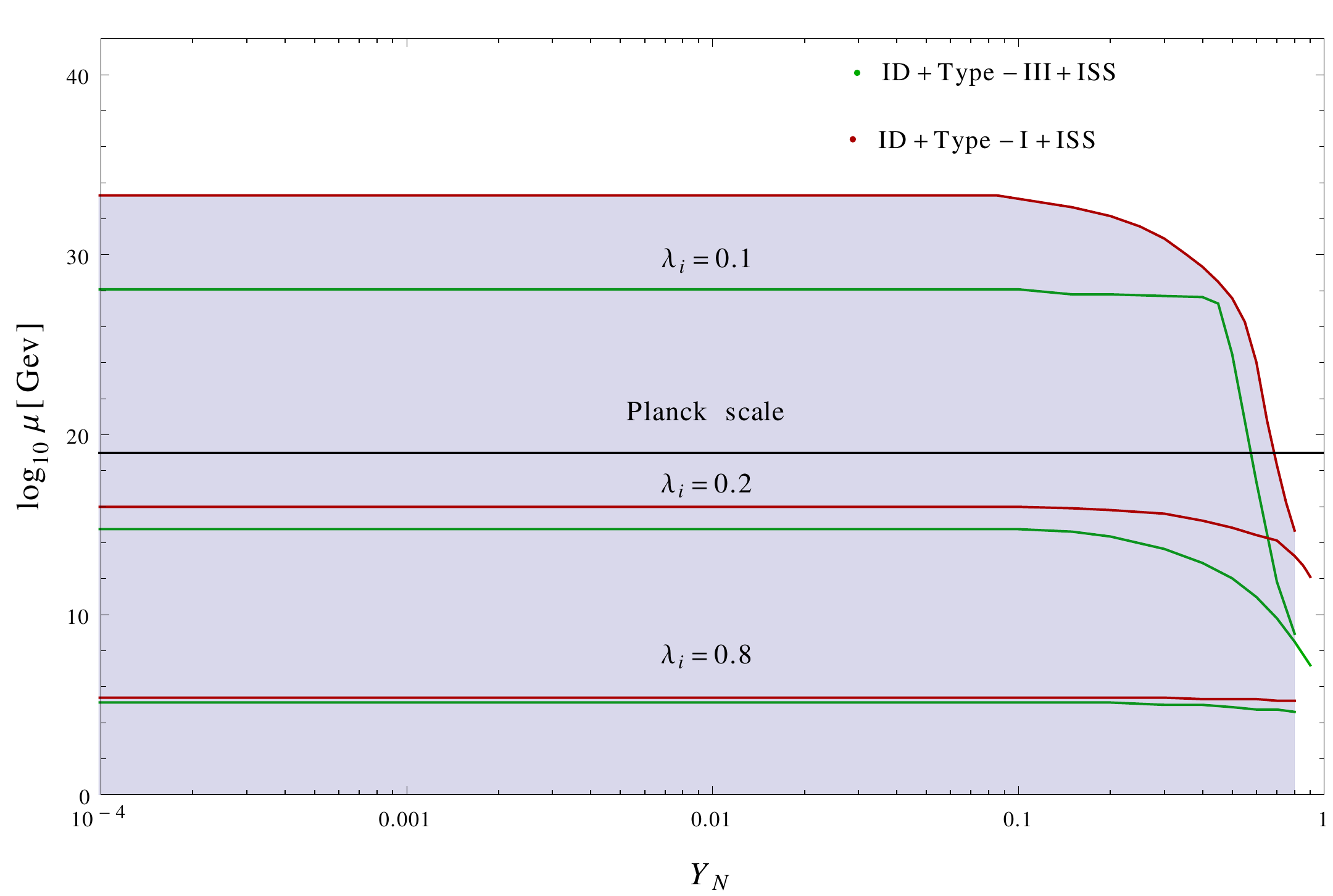}
		\caption{Bounds from perturbativity scale on $Y_N$  as a function of $ \mu$ for different values of $\lambda_i$ with $y_t=0.9369$, $M_N=100$ GeV. The red curve and the green curve corresponds to the ID plus Type-I ISS scenario and ID plus Type-III ISS scenario respectively with two generations of fermionic triplets.}\label{fig3l}
	\end{center}
\end{figure}

 Figure~\ref{fig3l} depicts the bounds on perturbative scale ($log_{10}{\mu}$) with respect to  the Yukawa coupling $Y_N$ for different choices of initial values of  $\lambda_{(i=2,3,4,5)}$.  For $Y_N$ values up to $0.1$, the effect of triplet fermions on the perturbativity of $\lambda_i$ is hardly noticeable as shown in Figure~\ref{fig3l}.  The exact value of $Y_N$ where it starts affecting the perturbativity depends on the initial value of $\lambda_i$ at EW scale. However, due to enhanced $g_2$ in the case of Type-III, $\lambda_i$ attains perturbation limit before than Type-I case and the effect is more prominent for lower values of $\lambda_i(\rm EW )$. For $\lambda_i(\rm EW )=0.1$ in ID+Type-III+ISS, the $\lambda_1$ hits Landau pole before other $\lambda$s as $\lambda_1(\rm EW )=0.1264$ and gets large positive contribution form the enhanced $g_2$ in this case. Effects of other $\lambda_i$s are negligible in this case until a particular value of $Y_N$. Large $g_2$ effect brings the perturbative scale to $10^{28.07}$ GeV with prolonged effect till $Y_N\lesssim 0.23$ after which  $\rm \lambda_1 Tr(Y^\dagger_N Y_N)$ effect takes over bringing perturbative scale further down as can be seen from the green line with the bending effect. Compared to that in
 ID+Type-I+ISS case, for $\lambda_i=0.1$, the perturbativity scale is $\sim$ $10^{33.29}$ GeV and  the effect starts showing up for $Y_N>0.15$ in ID+Type-I+ISS  as can be seen in red line. Here perturbative limits are obtained via the Landau poles of other $\lambda_i$ as $g_2$ decreases with the scale in this case and large $Y_N$ pushes $\lambda_1$ towards negative values of instability \cite{PBBDSJ}. However, other $\lambda_i$ gets positive contributions $\rm \lambda_i Tr(Y^\dagger_N Y_N)$  towards their Landau pole. Like the Type-III case here also being very small $\lambda_i$ effects are negligible.

 For higher values of $\lambda_i(\rm EW )$, the perturbativity limits are obtained when other $\lambda$s hit Landau pole before $\lambda_1$ and mostly controlled by $\lambda_i$ (see Appendix~\ref{betaf1}) and $\lambda_1$ runs towards instability by large $Y_N$ effect. This also results in lesser splitting in the perturbative scale between Type-III and Type-I  until $\rm \lambda_i Tr (Y^\dagger Y_N)$ effects creep in with larger factor for Type-III as compared to Type-I. For $\lambda_i=0.2$, the perturbativity limit comes around the GUT scale  $10^{16-15}$ GeV  and the fermion effect starts for $Y_N \gsim 0.3$ in both scenarios but the effect of $Y_N$ is much stronger in ID+Type-III+ISS.

 For further higher values of $\lambda_i$ =0.8, the perturbativity scales are almost the same  $\sim 10^{5}$ GeV  for both Type-I and Type-III cases mostly governed by the $\lambda_i$ effects. The effect of new fermion comes much later for $Y_N>0.60$.  Higher values of $\lambda_i$s can accommodate higher values of $Y_N$ for vacuum stability in $\lambda_1$ direction. On the other hand, the theory becomes non-perturbative at much lower scale. One can read the upper bounds on  perturbativity from Figure~\ref{fig3l}  for  $\lambda_i$ and $Y_N$ values. We see that for the $\lambda_i\leq 0.15$ and $Y_N\leq 0.25 \,, Y_N\leq 0.3$ the theory  remains perturbative till the Planck scale for Type-III and Type-I respectively.

	 \section{	Stability Bound} \label{stability}
	 In this section we analyse the stability of Higgs potential via two different approaches. Firstly via calculating two-loop scalar quartic couplings and checking if the SM  Higgs like quartic coupling $\lambda_h$ becomes negative at some higher scale. In this case $\lambda_h=\lambda_1$ at tree-level,  but at one-loop and two-loop levels $\lambda_h$ gets contribution from SM fields as well as the ID and   the Type-III fermions. The details are provided  in  subsection~\ref{loops}. For  simplicity,  we only present  the expressions of the corresponding beta functions at one-loop in the next subsection. The expressions for the  two-loop beta functions are given in the Appendix~\ref{betaf1}. Secondly we follow the effective potential approach as described in detail in section~\ref{vstability}. 
	 
	 \subsection{RG Evolution of the Scalar Quartic Couplings }\label{loops}
	 To study the evolutions of dimensionless couplings we have implemented the ID + Type-III Seesaw + ISS scenario in {\tt SARAH 4.13.0}~\cite{Staub:2013tta} for two generations $SU(2)_L$ triplet fermions. The corresponding $\beta$-functions for various gauge, quartic and Yukawa couplings are calculated at one- and two-loop. The full two-loop $\beta$-functions can be found in Appendix~\ref{betaf1}, and used for our numerical analysis of vacuum stability which will be described here.  We first look at the one-loop $\beta$-function of $\lambda_h$  to observe the effect of  the Yukawa and other scalar quartic couplings. $\lambda_h=\lambda_1$ at tree-level and effects of other particles start entering at one-loop level. The $\beta$-function of  the SM-like Higgs quartic coupling $\lambda_h$ in this model receives three different types of contributions: one from the SM gauge, Yukawa, quartic interactions, the second from the Type-III Seesaw Yukawa couplings, and the third from the inert scalar sector (ID):  
	 \begin{align}\label{lfull}
	\beta_{\lambda_h}= \beta_{\lambda_1 } & \ = \ \beta_{\lambda _1}^{\rm SM} +\beta_{\lambda _{1}}^{\rm Type-III + ISS} + \beta_{\lambda _1}^{\rm ID} \, ,
	 \end{align}
	 with  
	 \begin{eqnarray}\label{b1}
	 \beta_{\lambda_1}^{\rm SM} & \ = \ & \frac{1}{16\pi^2}\Bigg[
	 \frac{27}{200} g_{1}^{4} +\frac{9}{20} g_{1}^{2} g_{2}^{2} +\frac{9}{8} g_{2}^{4} -\frac{9}{5} g_{1}^{2} \lambda_1 -9 g_{2}^{2} \lambda_1 +24 \lambda_1^{2}\nonumber \\
	 && \qquad +12 \lambda_1 \mbox{Tr}\Big({Y_u  Y_{u}^{\dagger}}\Big) +12 \lambda_1 {\rm Tr}\Big({Y_d  Y_{d}^{\dagger}}\Big) +4 \lambda_1 \mbox{Tr}\Big({Y_e  Y_{e}^{\dagger}}\Big)  \nonumber\\
	 && \qquad 
	 -6 \mbox{Tr}\Big({Y_u  Y_{u}^{\dagger}  Y_u  Y_{u}^{\dagger}}\Big)-6 \mbox{Tr}\Big({Y_d  Y_{d}^{\dagger}  Y_d  Y_{d}^{\dagger}}\Big)\nonumber\\
	 && \qquad  -2 \mbox{Tr}\Big({Y_e  Y_{e}^{\dagger}  Y_e  Y_{e}^{\dagger}}\Big) \Bigg], \label{eq:4.2} \\
	 \beta_{\lambda_1}^{\rm Type-III + ISS} & \ = \ & \frac{1}{16 \pi^2}\Big[12 \lambda_1 \mbox{Tr}\Big({Y_{N}  Y_{N}^{\dagger}}\Big) - 10 \mbox{Tr}\Big({Y_{N}  Y_{N}^{\dagger}  Y_{N}  Y_{N}^{\dagger}}\Big)\Big], \label{eq:4.3} \\
	 \beta_{\lambda_1}^{\rm ID} & \ = \ & \frac{1}{16 \pi^2}\Big[ 2 \lambda_{3}^{2} +2 \lambda_3 \lambda_4 +\lambda_{4}^{2}+4 \lambda_{5}^{2}\Big]. \label{eq:4.4}
	 \end{eqnarray}	
	 Here $Y_u, Y_d, Y_e$ represent the up, down and electron-type Yukawa couplings and other Yukawa couplings are neglected~\cite{Buttazzo:2013uya} and $g_1,g_2$ are the $U(1)_Y$, $SU(2)_L$ gauge couplings respectively their values at the EW scale~\cite{Tanabashi:2018oca} are: $\lambda_1=0.1264$, $g_1=0.3583$, $g_2=0.6478$, $y_t=0.9369$. In this analysis the ISS contribution to the RG evolution of $\lambda_1$ starts only  at the scale above $M_{N}$.

\begin{figure}[t]
		\hspace{-0.2cm}
		\mbox{
			\subfigure[]{\includegraphics[width=0.5\linewidth,angle=-0]{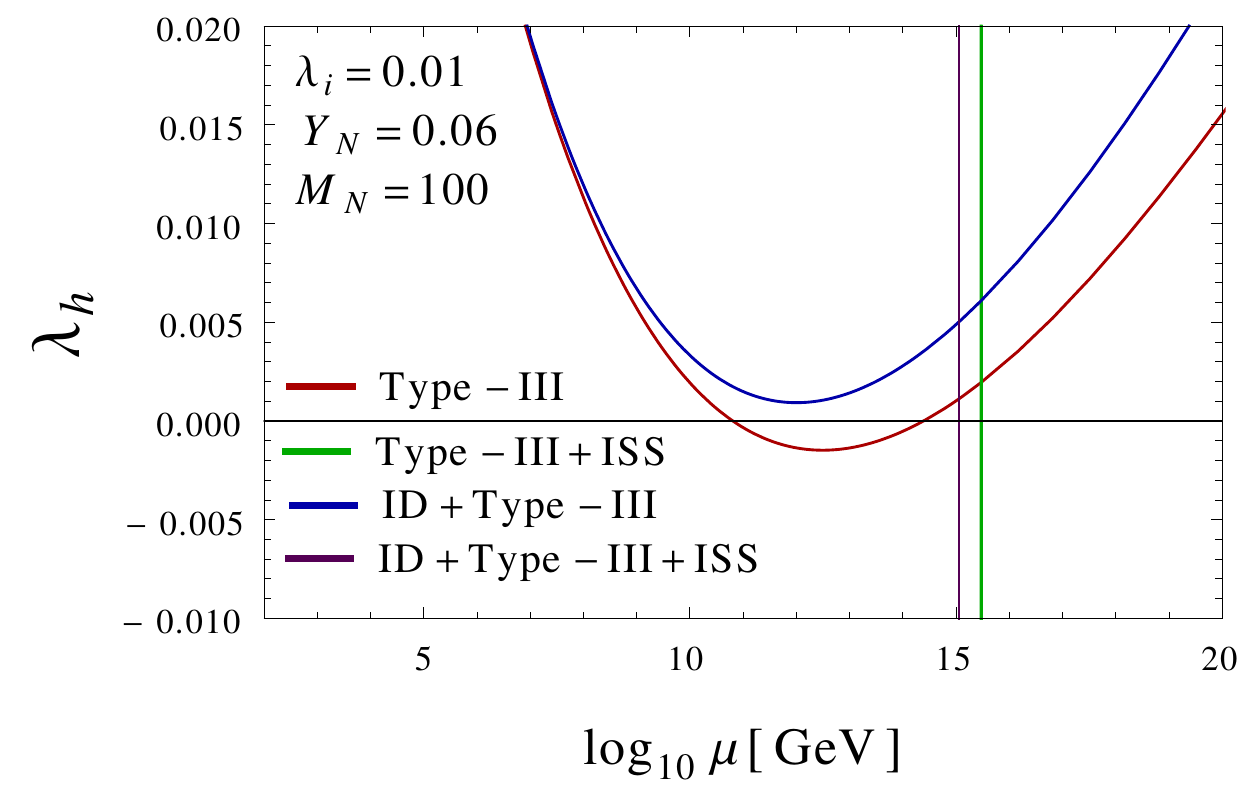}}
			\subfigure[]{\includegraphics[width=0.5\linewidth,angle=-0]{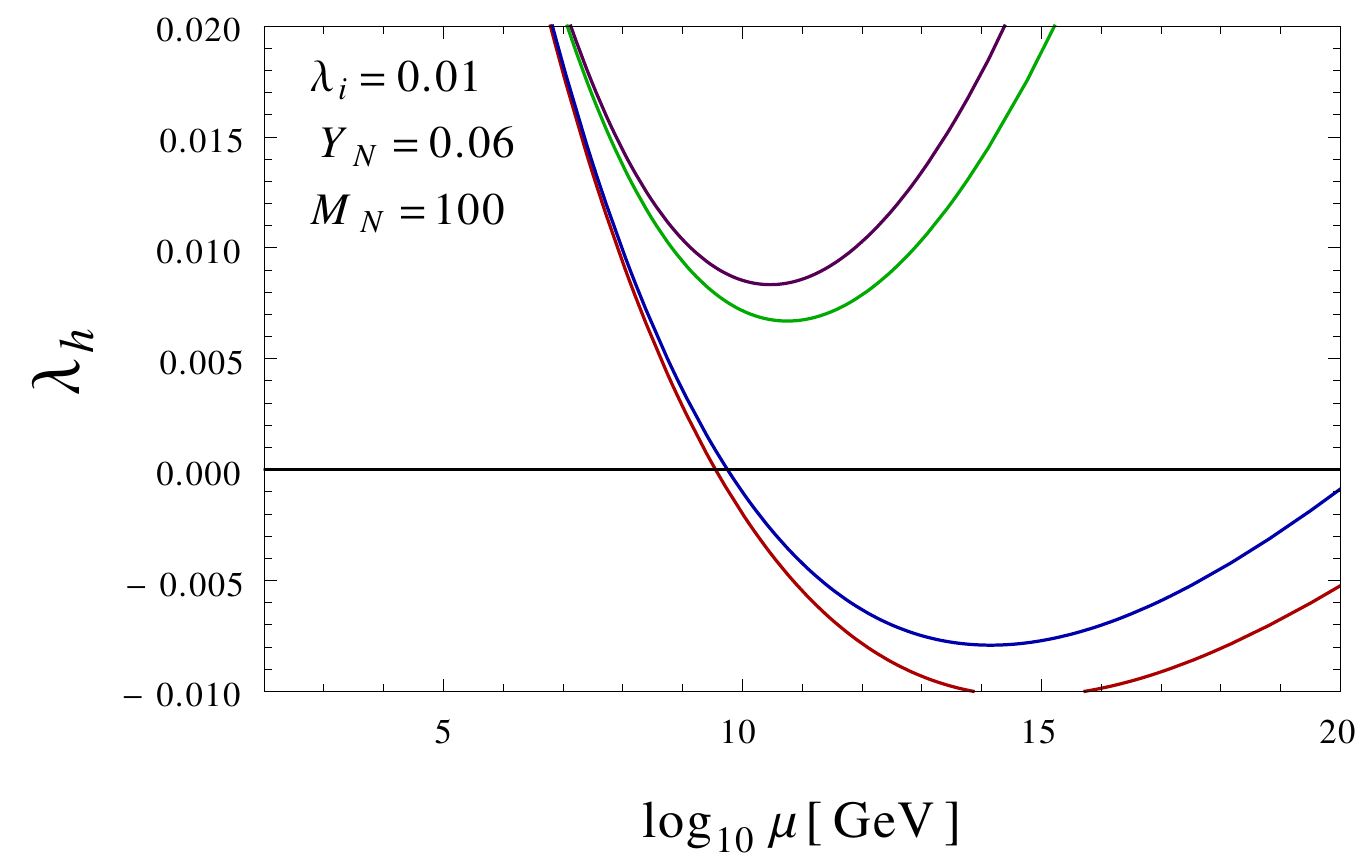}}}
		\caption{Two loop running of the SM like Higgs quartic coupling as a function of energy. Here the red, green, blue and the purple curve correspond to the Type-III Seesaw,  Type-III Seesaw+ISS, ID+Type-III Seesaw and ID+Type-III Seesaw+ISS scenarios. The left and right plot represent   three generations and two generations of fermionic triplets,  respectively. }\label{fig4l}
	\end{figure}

	 In Figure~\ref{fig4l}(a)-\ref{fig4l}(b), we show the respective $\lambda_h$ running with the scale $\mu$.  The  plot in the left panel  represents the three generation scenario, and the plot in the right panel   represents two generations scenario. The two plots describe the behaviour of $\lambda_h$ at two-loop level for four different cases as before with three and two generations of triplet fermions, where 
$M_{N}=100$ GeV, $\lambda_1=0.1264$, $\lambda_{i}=0.01$ (with $i=2,3,4,5$)  at the EW scale. The RG evolution of $\lambda_1\equiv \lambda_h$  is described by the red curve as a function of the scale $\mu$  where $\lambda_h$ has contributions of $\beta_{\lambda_1}^{\rm SM}+\beta_{\lambda_1}^{\rm Type-III}$ only. The green curve shows the evolution using $\beta_{\lambda_1}^{\rm SM}+\beta_{\lambda_1}^{\rm Type-III}+\beta_{\lambda_1}^{\rm ISS}$, blue curve describes the evolution using  $\beta_{\lambda_1}^{\rm SM}+\beta_{\lambda_1}^{\rm Type-III}+\beta_{\lambda_1}^{\rm ID}$, and finally the purple curve shows the full evolution using $\beta_{\lambda_1}\equiv \beta_{\lambda_1}^{\rm SM}+\beta_{\lambda_1}^{\rm Type-III}+\beta_{\lambda_1}^{\rm ISS}+\beta_{\lambda_1}^{\rm ID}$ [cf.~Eq.~\eqref{lfull}].  The added effects of the new contributions to $\lambda_1\equiv \lambda_h$ at one-loop are given in Eq.~\eqref{lfull} and the detailed two-loop expressions are written in Appendix~\ref{betaf1}. 
	
	 In Figure~\ref{fig4l}(a),  the three generations of fermionic triplet make the $g_2$ contribution too large (see Eq.~\ref{g2t3}) for both Type-III+ISS (green) and ID+ Type-III +ISS (purple),  such that $\beta_{\lambda_1}$  hits the Landau pole at $\sim 10^{15.4}$ GeV and $\sim 10^{15.1}$ GeV respectively before hitting the instability scale (at which $\lambda_h \leq 0$ ). This makes the theory non-perturbative below Planck scale. Without ISS the $\beta_{g_2}$ is relatively smaller (see subsection~\ref{gaugecoup}) which restrains $\lambda_h$ from hitting the Landau pole. Thus for Type-III Seesaw scenario  $\beta_{\lambda_h}$ becomes unstable $\sim 10^{10.7}$ GeV but bounce back to stability (where $\lambda_h \geq 0$) at $\sim 10^{14}$ GeV. For  ID+ Type-III scenario,  $\beta_{\lambda_h}$ remains stable till Planck scale for $\lambda_i=0.01$ and $Y_N=0.06$ respectively. 
	 
	 Figure~\ref{fig4l} (b) describes the behaviour of  two generations of fermionic triplet with reduced positive $g_2$ effect, which prohibits the Landau pole of  $\lambda_1$. We can see the ID+Type-III+ ISS (purple curve) is more stable than Type-III+ISS (green curve). Again without ISS, the $g_2$ contribution is less so Type-III  (red) and ID+Type-III (blue) have gone more negative; especially Type-III (red) is even more negative due to the lack of positive effects of ID scalars.

	 \begin{figure}[t]
	 	\begin{center}
	 		\mbox{\subfigure[]{\includegraphics[width=0.5\linewidth,angle=-0]{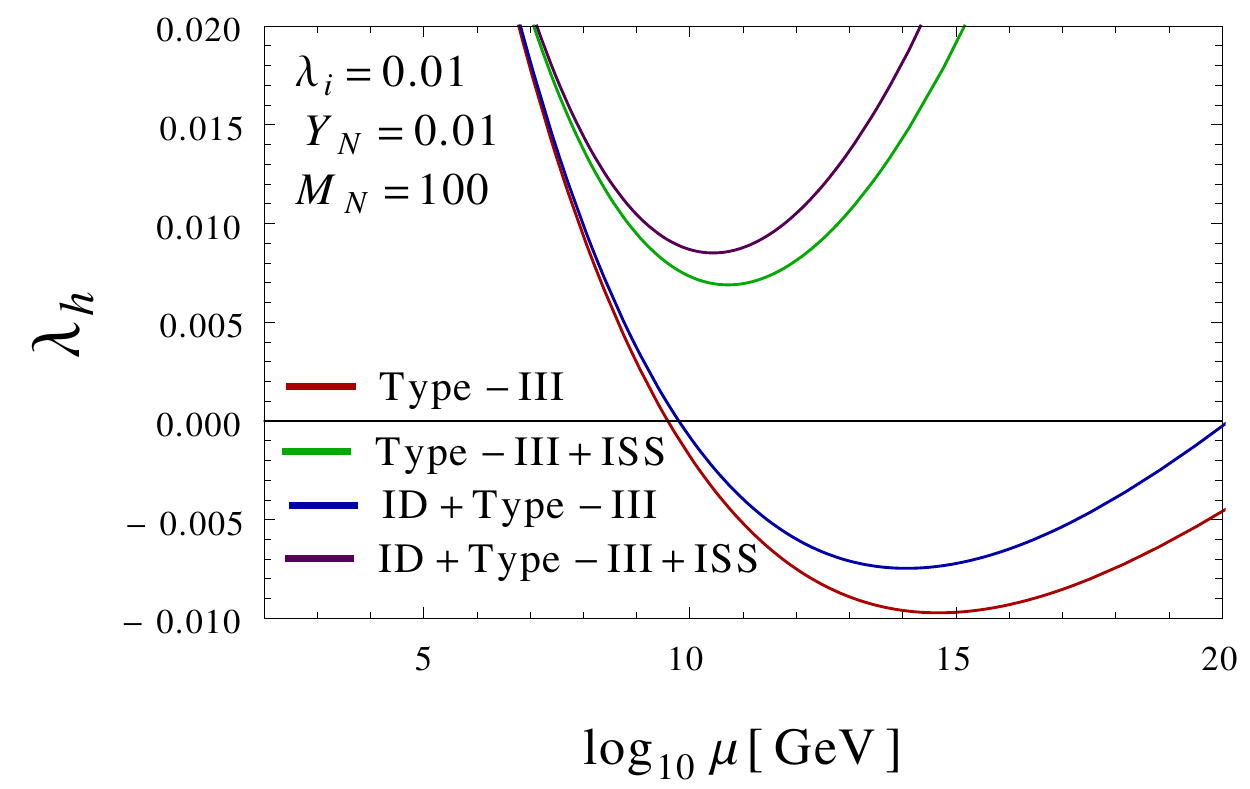}\label{f7}}
	 			\subfigure[]{\includegraphics[width=0.5\linewidth,angle=-0]{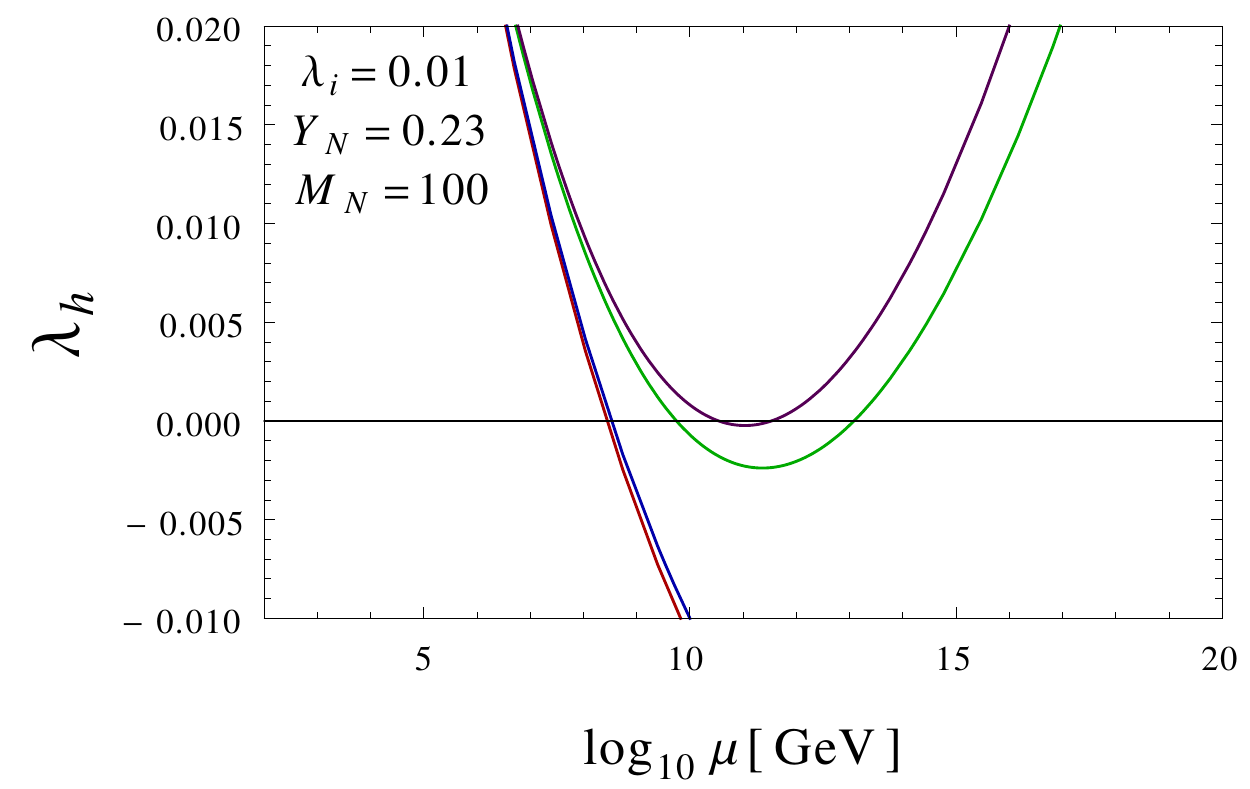}\label{f8}}}
	 		\mbox{\subfigure[3 gen]{\includegraphics[width=0.5\linewidth,angle=-0]{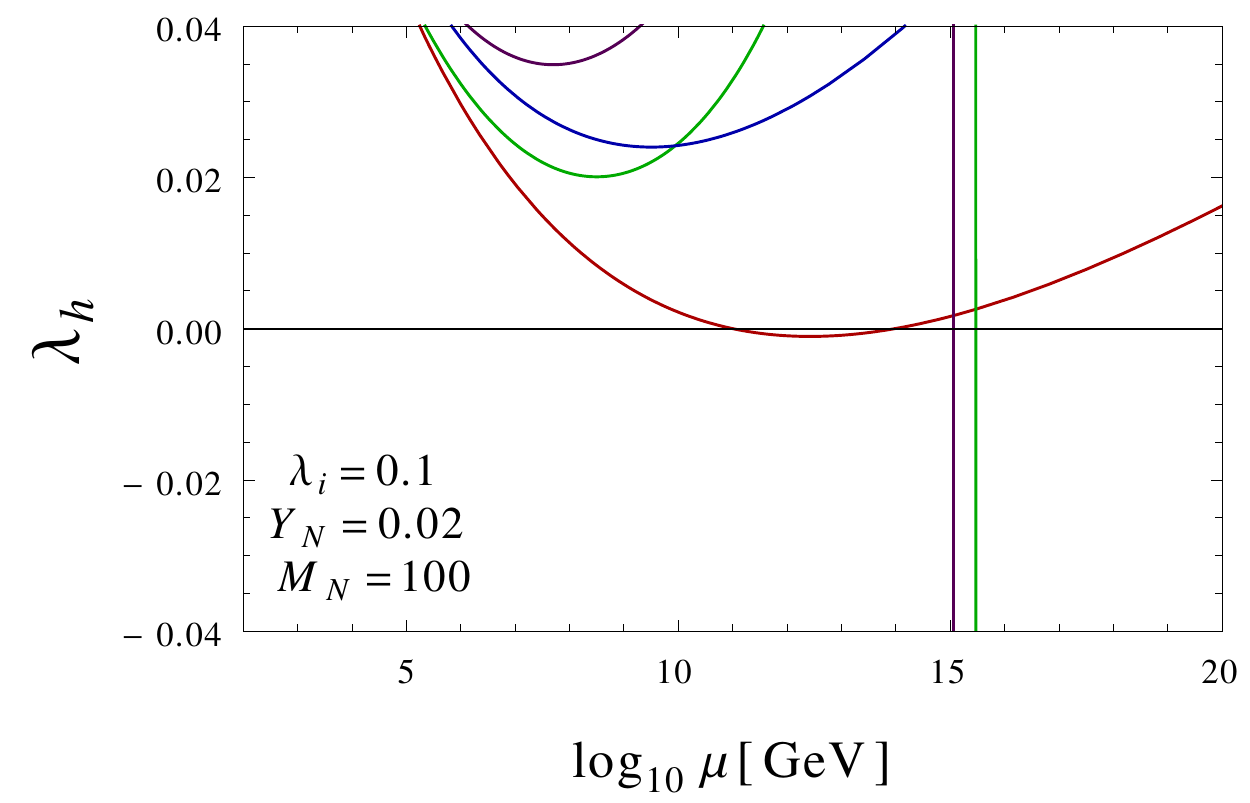}\label{f8a}}
	 			\subfigure[]{\includegraphics[width=0.5\linewidth,angle=-0]{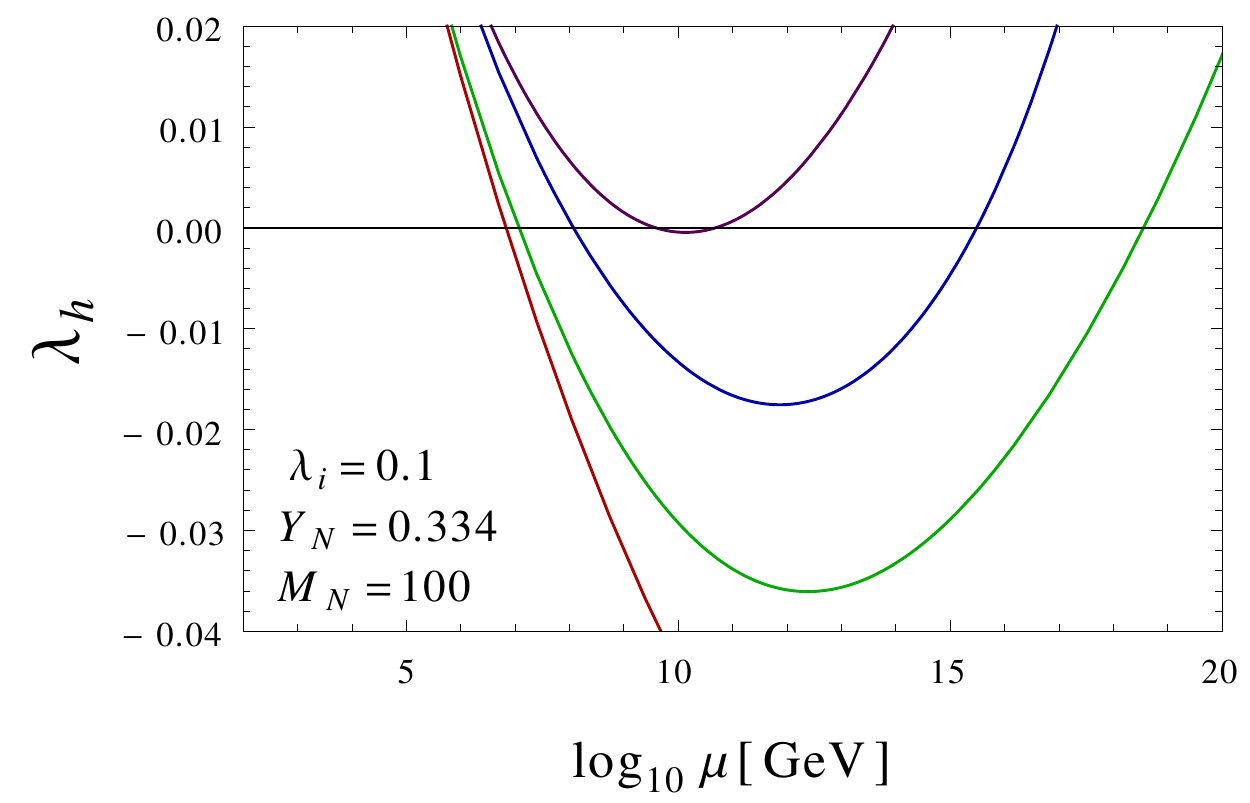}\label{f10}}}	\mbox{\subfigure[]{\includegraphics[width=0.5\linewidth,angle=-0]{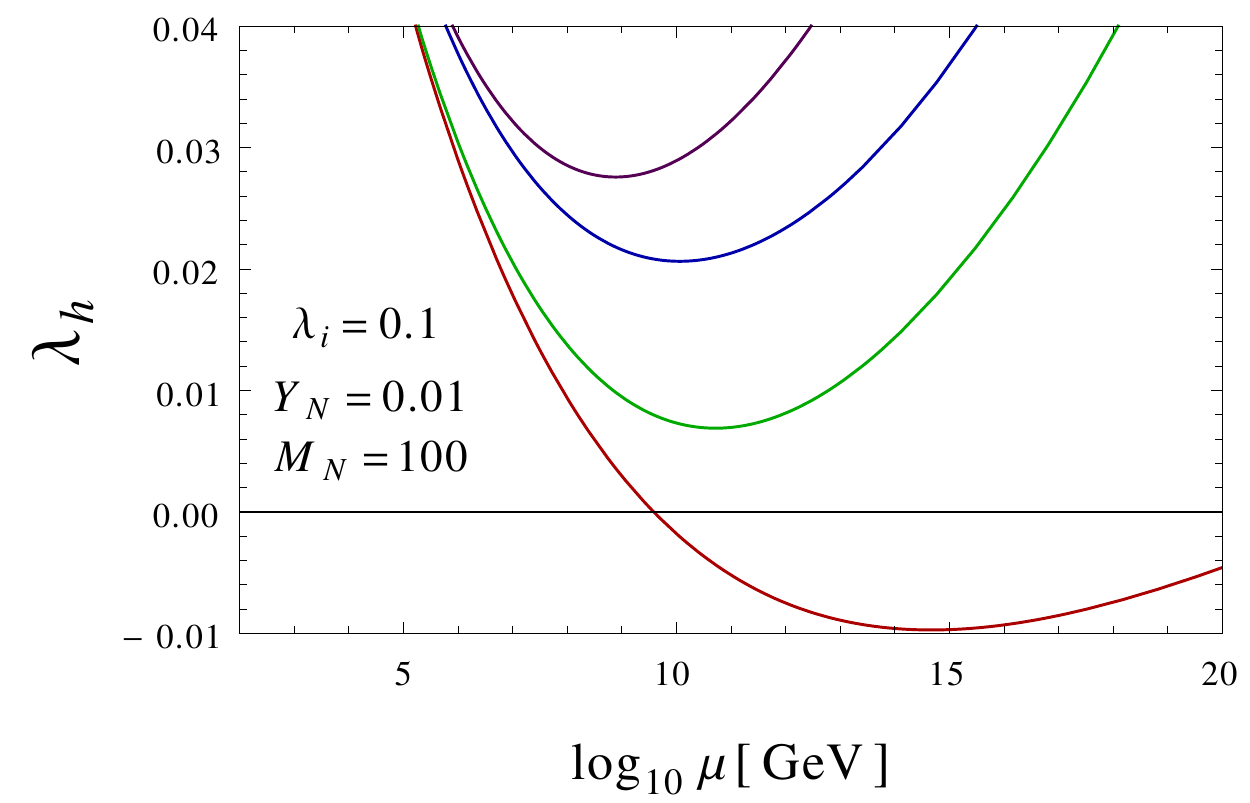}\label{f9}}
	 			\subfigure[3 gen]{\includegraphics[width=0.5\linewidth,angle=-0]{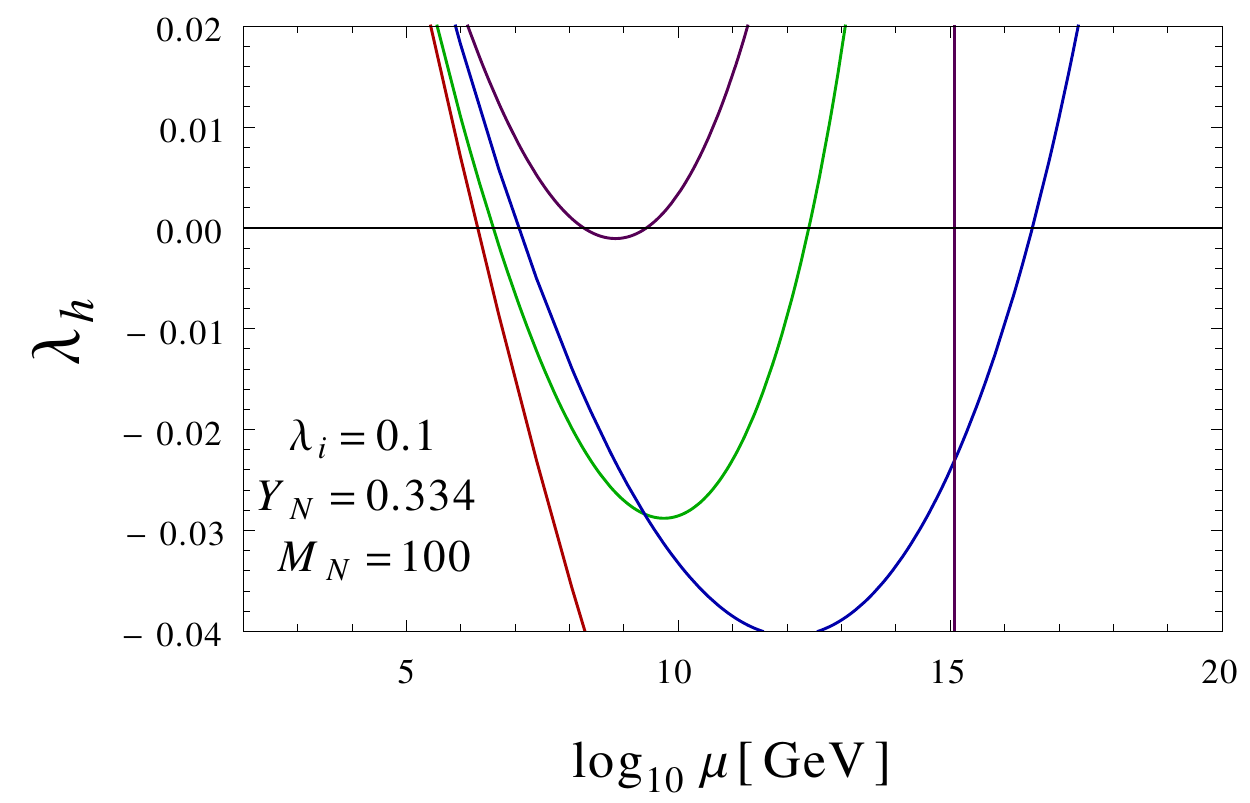}\label{f9a}}}
	 		\caption{Two-loop running of the SM-like Higgs quartic coupling $\lambda_h$ as a function of scale ($\mu$). Here the red, green, blue and the purple curve corresponds to the Type-III,  Type-III+ISS, ID+Type-III and ID+Type-III +ISS scenarios respectively for four different benchmark values of $Y_N$ and Higgs quartic couplings $(\lambda_{1=2,3,4,5})$ with two generations of fermionic triplet. (c) and (f) are with three-generations of triplet fermions.}\label{fig5l}
	 	\end{center}
	 \end{figure}
	 
	 In Figure~\ref{fig5l} we describe the behaviour of two-loop running of the SM-like Higgs quartic coupling $\lambda_h$ as a function of scale ($\mu$) for six benchmark points. We follow the same colour code as Figure~\ref{fig4l} and $\lambda_1(EW)=0.1264$ chosen for all the graphs. Figure~\ref{f7} describes for the benchmark points of $\lambda_i=0.01, \, Y_N=0.01$ and $M_N=100$ GeV  where $i=2,3,4,5$ with two generations of  Type-III fermions. We can clearly see that the stability scales for ID+Type-III+ISS (purple curve) and Type-III+ISS (green) are enhanced to Planck scale. Type-III (red) and ID+ Type-III (blue) hit instability around $10^{9.5}$ GeV. Now if we increase $Y_N=0.23$ as shown in Figure~\ref{f8}, we can see that red and blue curve hit instability earlier around $10^{8.5}$ GeV due to larger negative effects of $Y_N$. Even green and purple curves also hit $\lambda_h =0$ due to this negative effect. For a comparative study with Figure~\ref{f7} we plot the running of the $\lambda_h$ for three generations of Type-III fermions in Figure~\ref{f8a}.  We see that the purple and the green curve hits Landau pole again due to larger positive $g_2$ contributions as explained earlier. Blue and red curve also move towards stability in this case. Figure~\ref{f9} we restrict the fermion generation only to two and with the reduced positive effect the Landau poles are gone with overall shift towards the left.

Figure~\ref{f10} shows the comparison with Figure~\ref{f8} with two Type-III fermion generation 
with  $Y_N$ enhanced to $0.334$. We notice the overall negative effect that shifts all the curves towards the left reducing the instability scale with red curve hitting instability at first at $10^{7}$ GeV. If we increase the number of fermion generation to three in Figure~\ref{f9a} the effect will be enhanced as red curve crosses zero at around $10^{6}$ GeV before acquiring  the Landau pole at $10^{15}$ GeV. The shapes of all the curves becomes more steeper as compared to Figure~\ref{f10}. 
Compared to Figure~\ref{f8a} where $Y_N=0.02$, if we increase $Y_N=0.334$ in Figure~\ref{f9a}, we see that the negative effect creeps in and the Type-III + ISS i.e. green the green curve does not hit the Landau pole. However, in ID+Type-III +ISS case due to the positive contributions from $\lambda_i$ the purple curve hits the Landau pole at $10^{15}$ GeV.

\subsection{Variation perturbativity and stability with $Y_N$}
	 \begin{figure}[t]
	 	\begin{center}
	 		\includegraphics[width=0.7\linewidth,angle=-0]{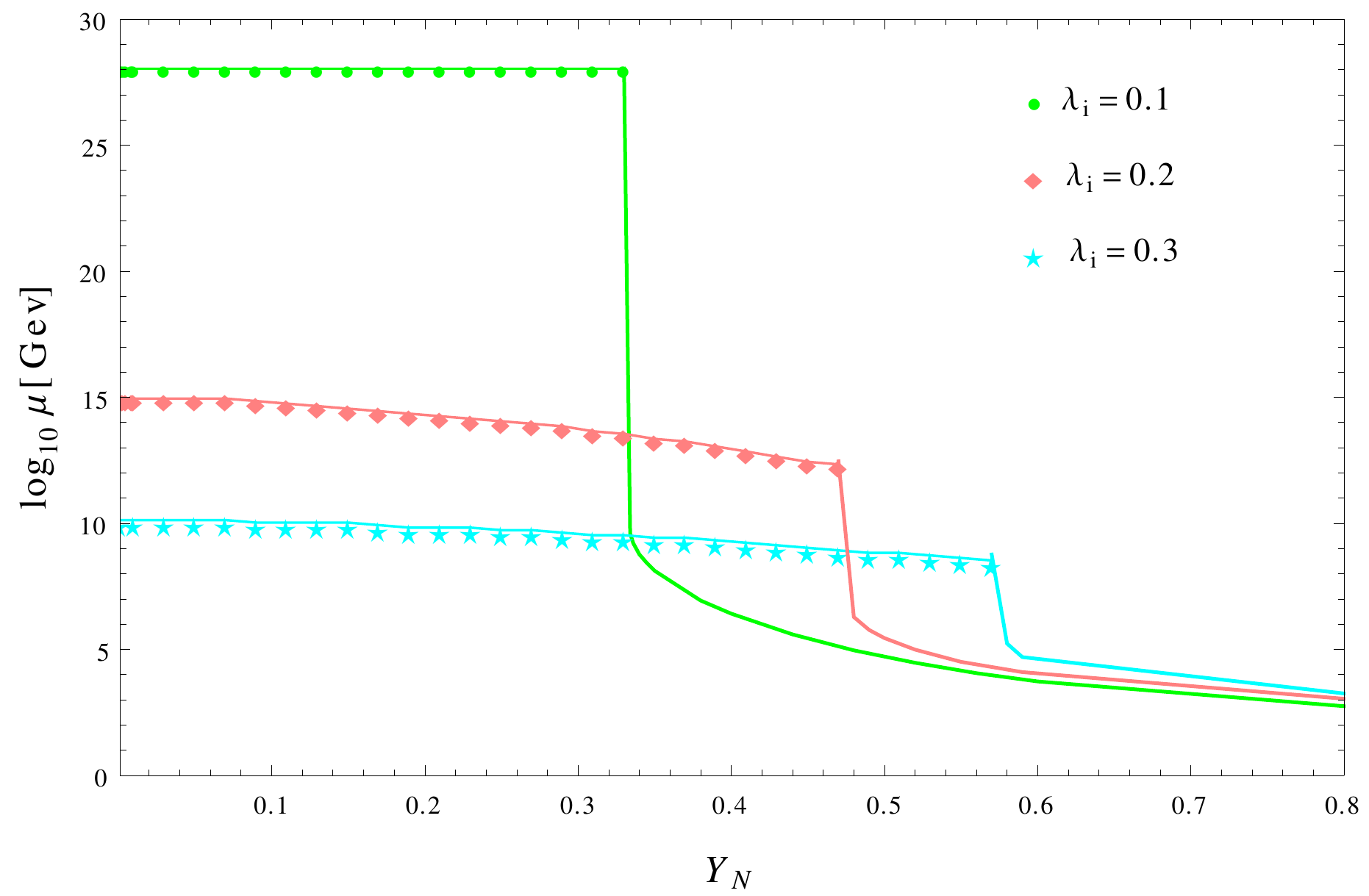}
	 		\caption{Effect of Yukawa coupling on the perturbativity and stability bound for different  values of $\lambda_i$. $\lambda_i=0.10$ (green curve) ensures Planck scale perturbativity and stability for $Y_N \leq 0.32$. The pink curve corresponds to $\lambda_i$=0.2 which gives Landau pole  for $Y_N\leq 0.47$ (shown by diamond). After $Y_N=0.47$, the negative fermionic contribution starts dominating which makes the  $\lambda_{h}\leq 0$ for $10^{6.24}$ GeV (shown by pink line) before hitting the Landau pole.  For $\lambda_i=0.3$, the running of $\beta_{\lambda_1}$ hits the Landau pole till $Y_N\leq 0.57$ (described by cyan color  star) and later the effect of fermions makes the $\lambda_{h}\leq 0$ for $10^{5.23}$ GeV (described by cyan line). }\label{fig6l}
	 	\end{center}
	 \end{figure}
	 Figure~\ref{fig6l} describes the variation of the stability scale with $Y_N$ and $\lambda_i$. We notice that for the  smaller values of $\lambda_i \sim 0.1$ (green  circles), the Planck scale stability can be achieved  till $Y_N\leq 0.32$. In this case $\lambda_1$ strikes the perturbative bounds before other quartic couplings due to the strong $g_2$ effect and as $\lambda_1(EW)=0.1264$. The other quartic coupling effects are negligible in this region as explained earlier. After $Y_N\geq 0.32$ (green line) the negative contribution from the new fermions become very strong and push $\lambda_h$ towards negative values at around $10^{9}$ GeV. 

For example as we enhance the $\lambda_i$ at EW scale the effect of quartic couplings in their beta functions increase as $\rm \lambda_i Tr(Y^\dagger_N Y_N)$ along with the enhanced $g_2$ effect.
These inflate $\beta_{\lambda_1}$ towards the higher scale stability compared to the SM  but leads to non-perturbative limit for the other  $\lambda_i$s at lower scale. The point to be noted here is that for the choices of $\lambda_i > \lambda_1$ at the EW scale,  other $\lambda_i$( $i=2,3,4,5$) are most likely to hit the Landau pole before $\lambda_1$.  For example as we enhance the $\lambda_i=0.2$ with $\lambda_1=0.1264$ at the EW scale, one of the $\lambda_i$s afflicts the landau pole (shown by pink color  diamond)  around $10^{15}-10^{12}$ GeV, even before $\lambda_1$  enters into instability at $Y_N=0.47$ for $10^{6.24}$ GeV (shown by pink line). The bending of the curves for higher $Y_N$ are due to the positive $\rm \lambda_i Tr(Y^\dagger_N Y_N)$ effect for perturbativity and negative $\rm Tr(Y^\dagger_N Y_NY^\dagger_N Y_N)$ effect for the instability respectively.
	  
 For $\lambda_i=0.3$, one of the $\lambda_i$s hits the Landau pole at much lower scale around $10^{10}$ GeV (cyan colour  star) due to large quartic coupling contributions along with $\rm \lambda_i Tr(Y^\dagger_N Y_N)$. In this case the negative effect of $Y_N$ in $\lambda_1$ starts much later due to larger value of $\lambda_i$ and at $Y_N\geq 0.57$ the effect of fermions make the $\lambda_{h}\leq 0$ for $10^{5.23}$ GeV (cyan line).

	\subsection{Vacuum Stability: RG-improved potential Approach }\label{vstability}
We study the stability of the EW vacuum via the Coleman and Weinberg~\cite{Coleman:1973jx} prescription of RG-improved effective potential at one-loop.  The effective potential at one-loop  has been calcualted for SM + Type-III-ISS + ID with two generations of fermions. The potential then analysed for the  stability, metastability and instability  by observing the behaviour of the effective Higgs quartic coupling. For this purpose we scan the parameter space of the model and then segregate them as above mentioned three regions. 
	
 Eq.~\eqref{eq:2.3} describes the tree-level Higgs potential for the model. The tree-level stability conditions of the potential are given by~\cite{Branco:2011iw}
\begin{align} \label{stabTHDM1}
\lambda_1 \ \geq \ 0 \, , \quad \lambda_2 \ \geq \ 0 \, , \quad 
\lambda_3 \ \geq \ -\sqrt{\lambda_1 \lambda_2} \, ,\quad \lambda_3+\lambda_4- |\lambda_5| \ \geq \ -\sqrt{\lambda_1 \lambda_2}  \, .
\end{align}

In SM $\lambda_h$ receives negative effects from the top quark loop which makes the potential unstable at  around $10^{10-11}$ GeV. There is also a possibility of second minima in the $h$ direction due to quantum fluctuation. In this article we will see how different beyond SM fields can contribute to that possibility of second minima. However, such possibility can only occur at higher field values which justifies the choice of effective potential as given below:

\begin{align}
V_{\rm eff}(h,\mu) \ \simeq \ \lambda_{\rm eff}(h,\mu)\frac{h^4}{4},\quad {\rm with}~h\gg v \, ,
\label{eq:4.6}
\end{align}
where $\lambda_{\rm eff}(h,\mu)$ is the effective Higgs quartic coupling and the calculation of which is described below . Then the stability of the vacuum then be guaranteed  for the scale $\mu$ by assuring that $\lambda_{\rm eff}(h,\mu)\geq 0$. 
\subsection{Effective Potential}
In our model the one-loop RG-improved effective potential can be written as 
\begin{align}
V_{\rm eff} \ = \ V_0+V_1^{\rm SM}+V_{1}^{\rm ID}+V_{1}^{\rm ISS+Type-III} \, ,
\label{eq:4.7}
\end{align}
where $V_0$ is the tree-level potential given by Eq.~\eqref{eq:2.3}, $V_1^{\rm SM}$ is the effective Coleman-Weinberg potential in the SM that contains all the one-loop corrections involving the SM particles at zero temperature with vanishing momenta. The other two terms  $V_{1}^{\rm ID}$ and $V_{1}^{\rm ISS+Type-III\, seesaw}$ represent the corresponding one-loop effective potential terms from the inert scalar doublet, and fermionic triplet,  respectively. $V_1$ describes one-loop Coleman and Weinberg effective potential which can be written as
\begin{align}\label{qc}
V_1(h, \mu) \ = \ \frac{1}{64\pi^2}\sum_{i} (-1)^F n_iM_i^4(h) \Bigg[\log\frac{M_i^2(h)}{\mu^2}-c_i\Bigg],
\end{align}
where the summation is over all the fields interacting with the $h$-field. $F=0,1$ for bosons and fermions  respectively, $n_i$ is the total of degrees of freedom of the particle and $M_i^2$ are the field-dependent mass terms
\begin{align}
M_i^2(h) \ = \ \kappa_i h^2-\kappa'_i \, ,
\label{eq:4.9}
\end{align} 
with the corresponing coefficients are shown in Table~\ref{table:1} in the last column. Note that the   massless fileds do not contribute to Eq.~\eqref{eq:4.9}, and to Eq.~\eqref{qc}. Hence, we only include  top quark from SM, and the other  contributions are neglected. For Type-III+ISS, their contributions comes after the mass threshold  $M_{N_i}$. 
\begin{table}[h!]
	\begin{center}
		\begin{tabular}{||c|c|c|c|c|c|c||}\hline\hline
			Particles & $i$ & $F$ & $n_i$ & $c_i$ & $\kappa_i$ & $\kappa'_i$ \\ \hline\hline
			& $W^\pm$ & 0 & 6 & 5/6 & $g_2^2/4$ & 0\\
			& $Z$ & 0 & 3 & 5/6 & $(g_1^2+g_2^2)/4$ & 0\\
			SM & $t$ & 1 & 12 & 3/2 & $Y_t^2$ & 0\\
			& $h$ & 0 & 1 & 3/2 & $\lambda_h$ & $m^2_1$\\
			& $G^\pm$ & 0 & 2 & 3/2 & $\lambda_h$ & $m^2_1$\\
			& $G^0$ & 0 & 1 & 3/2 & $\lambda_h$ & $m^2_1$\\ \hline
			& $H^\pm$ & 0 & 2 & 3/2 & $\lambda_3/2$ & $m^2_2$\\
			Inert & $H$ & 0 & 1 & 3/2 & $(\lambda_3+\lambda_4+2\lambda_5)/2$ & $m^2_2$\\
			& $A$ & 0 & 1 & 3/2 & $(\lambda_3+\lambda_4-2\lambda_5)/2$ & $m^2_2$\\ \hline
				Type3seesaw +ISS & $\Sigma_{1i}$ & 1 & 2 & 3/2 & $Y_{N}^2/2$ & 0\\ 
				 \hline\hline
		\end{tabular}
	\end{center}
	\caption{Coefficients in the Coleman-Weinberg effective potential needed for the study, cf.~Eq.~\eqref{qc}.}	\label{table:1}
\end{table}
The full effective potential can be read from Eq.~\eqref{eq:4.7}, where each of these extra one-loop terms can be explained from Eq.~\eqref{qc}. The effective poential with only SM like Higgs field can be explained  as  Eq.~\eqref{eq:4.6} with the effective coupling given as follows: 
\begin{align} \label{totalL}
\lambda_{\rm eff}\left(h,\mu\right) & \ \simeq \  \underbrace{\lambda_h\left(\mu\right)}_{\text{tree-level}}+\frac{1}{16\pi^2}\Bigg\{\underbrace{\sum_{\substack{i=W^\pm, Z, t, \\ h, G^\pm, G^0}} n_i\kappa_i^2 \Big[\log\frac{\kappa_i h^2}{\mu^2}-c_i\Big]}_{\text{Contribution from SM}}   +\underbrace{\sum_{i = H,A,H^\pm}n_i\kappa_i^2 \Big[\log\frac{\kappa_i h^2}{\mu^2}-c_i\Big]}_{\text{ Contribution from ID}}
\nonumber \\ 
&
+ \underbrace{2\sum_{i = 1,2}  n_i  \kappa_i^2 \Big[\log\frac{\kappa_i h^2}{\mu^2}-c_i\Big]}_{\text{ Contribution from Type-III+ISS}}\Bigg\}.
\end{align}
In the case of inverse seesaw in the limit $\mu_{\Sigma}\to 0$, the triplet fermion masses are double-degenerate per generation. Thus we  take an extra factor of two for each new Type-III fermion contribution in Eq.~\eqref{totalL}. We then study the $\lambda_{\rm eff}(h,\mu)$ in our model to distinguish the stability, metastability
and instability regions. $h=\mu$ is taken for the numerical analysis as at this scale  we have scale-invariant potential \cite{Casas:1994us}.
\subsection{Phase diagrams: Stable, Metastable and Unstable Regions}
For $\lambda_{\rm eff }\geq 0$ the potential is bounded from below and the region is termed as the {\it stable} region. In this region we expect to have a global EW minimum. On the other hand, for $\lambda_{\rm eff}<0$ the potential develops a second minima and there can be tunnelling to the global second minima sufficiently faster. But there exists  regions when such tunnelling is not fast enough and EW minimum still can survive due to the tunnelling rate which is greater than the age of the Today's universe. Such possibilities are known as  {\it metastable}  vacuua.  The tunnelling rate at zero temperature has been calculated as 
\begin{equation}
P \ = \ T_0^4 {\mu}^4 \exp\left[\frac{-8 {\pi}^2}{3 \lambda_{\rm eff}(\mu)}\right] \, ,
\label{eq:P}
\end{equation}
where $T_0$ is time corresponds to the age of the universe,  and $\mu$ is  the scale with maximum  probability, i.e. $\frac{\partial P}{\partial \mu}=0$. The $\lambda$ values at different scales can be obtained as: 
\begin{align}
\lambda_{\rm eff}(\mu) \ = \ \frac{\lambda_{\rm eff}(v)}{1-\frac{3}{2\pi^2}\log\left(\frac{v}{\mu}\right)\lambda_{\rm eff}(v)} \, ,
\label{eq:lamb}
\end{align}
where $v\simeq 246$ GeV.  If we substitute  $P=1$, $T= 10^{10}$ years and $\mu=v$ in Eq.~\eqref{eq:P}, one finds $\lambda_{\rm eff}(v)$ =0.0623. Demanding $P< 1$, for $T= 10^{10}$ years old universe would estimate the condition that such tunnelling rates corresponds to the {\it \ metastability }as given by  \cite{Isidori:2001bm}: 
\begin{align}\label{meta}
0 \ > \ \lambda_{\rm eff}(\mu) \ \gtrsim \ \frac{-0.065}{1-0.01 \log\left(\frac{v}{\mu}\right)} \, .
\end{align}
The condition of $\lambda_{\rm eff}<0$ but outside region Eq.~\eqref{meta}, corresponds to when potential develops a second minima as mentioned before and if the tunneling rate from the EW vacuum to this second vacuua is less than the age of the universe then it depicts the {\it unstable} region. It is evident from Eq.~\eqref{totalL}, these different solutions are sensitive to the choices of the scale $\mu$ and other model parameters such as bare masses and couplings. 

\begin{figure}[t] 
\hspace*{-2.0cm}
\mbox{\subfigure[ID+Type-III+ISS]{\includegraphics[width=0.5\linewidth,angle=-0]{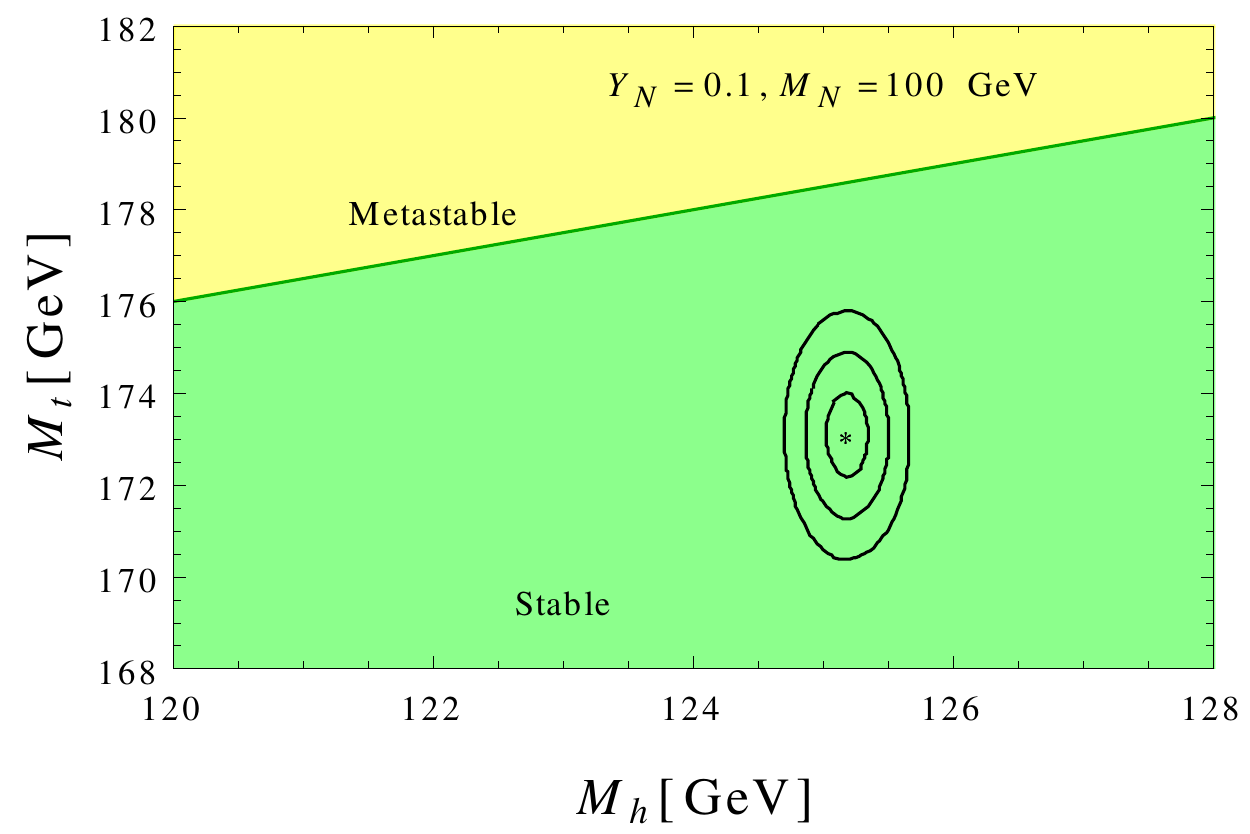}\label{f12}}	\subfigure[ID+Type-III+ISS]{
\includegraphics[width=0.5\linewidth,angle=-0]{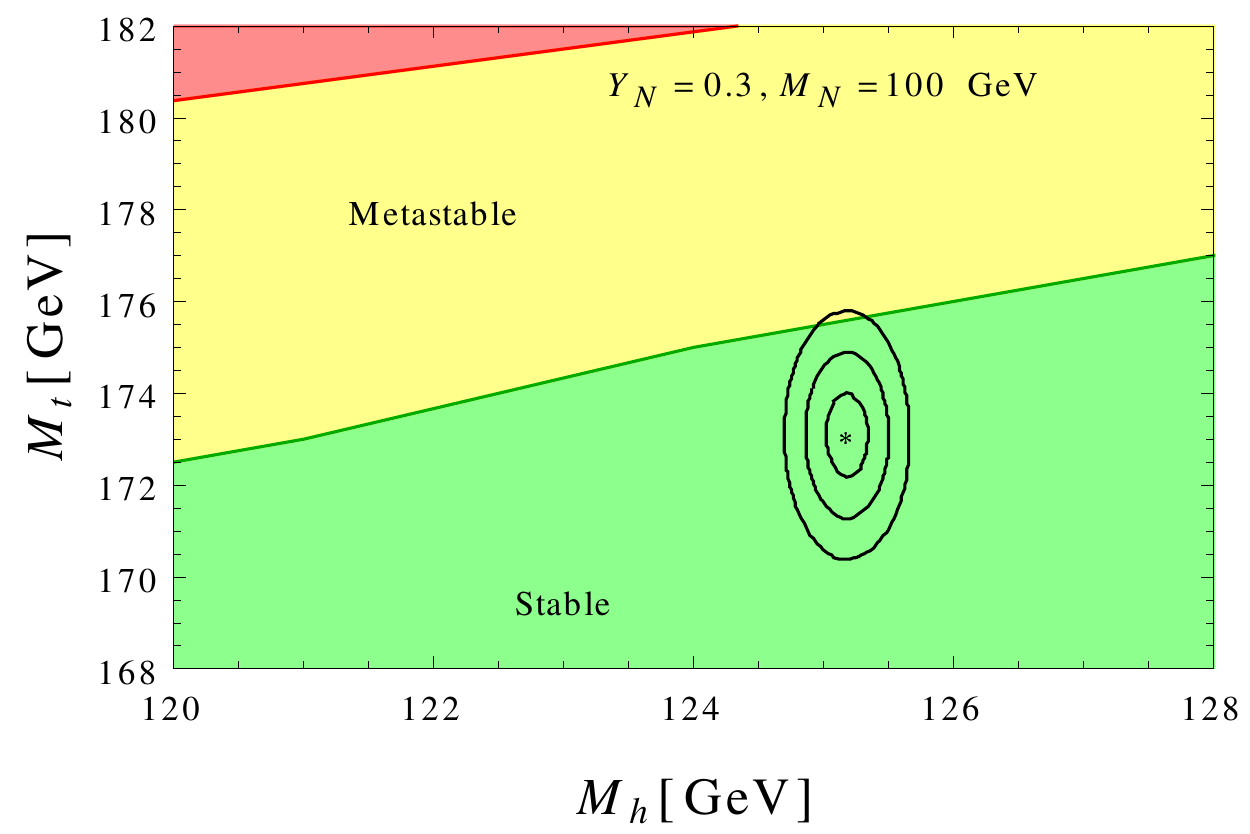}\label{f13}}}
\mbox{\subfigure[ID+Type-III+ISS]{\includegraphics[width=0.5\linewidth,angle=-0]{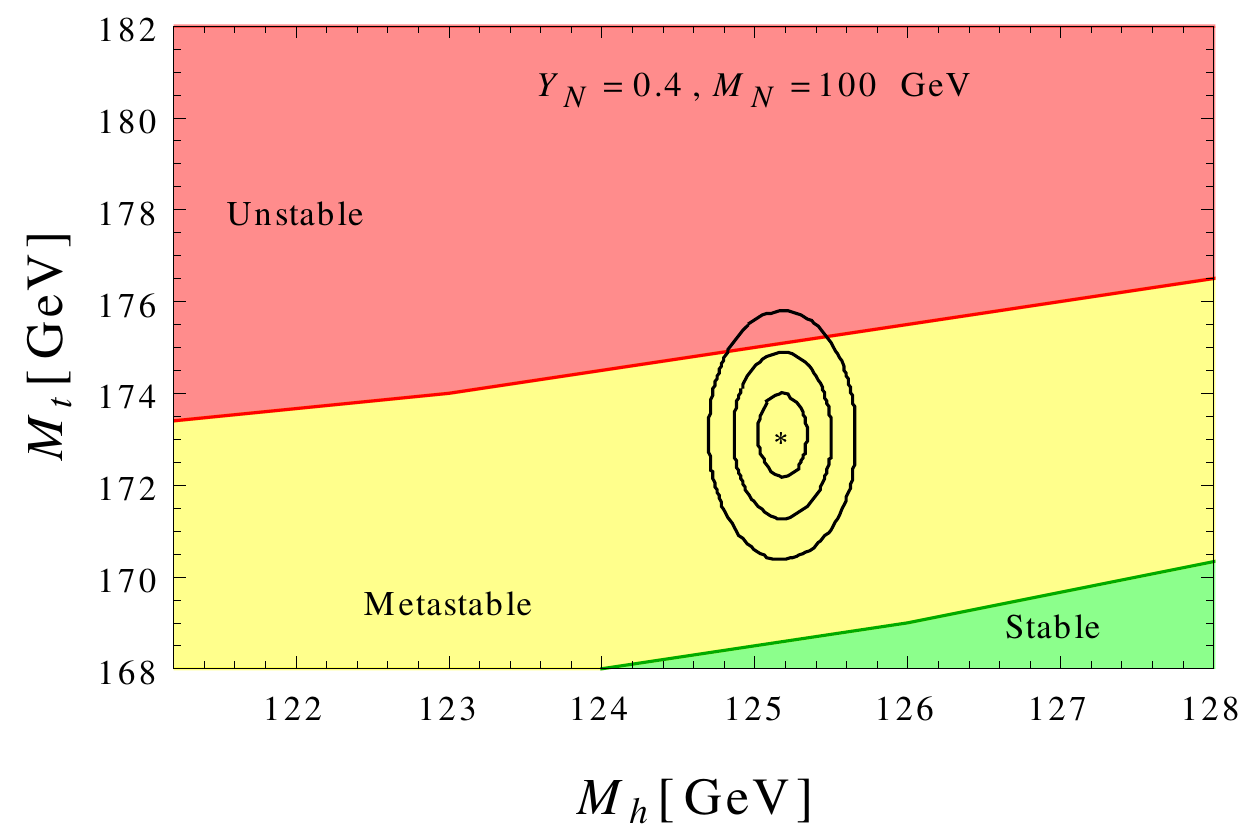}\label{f14}}	\subfigure[Type-III]
{\includegraphics[width=0.5\linewidth,angle=-0]{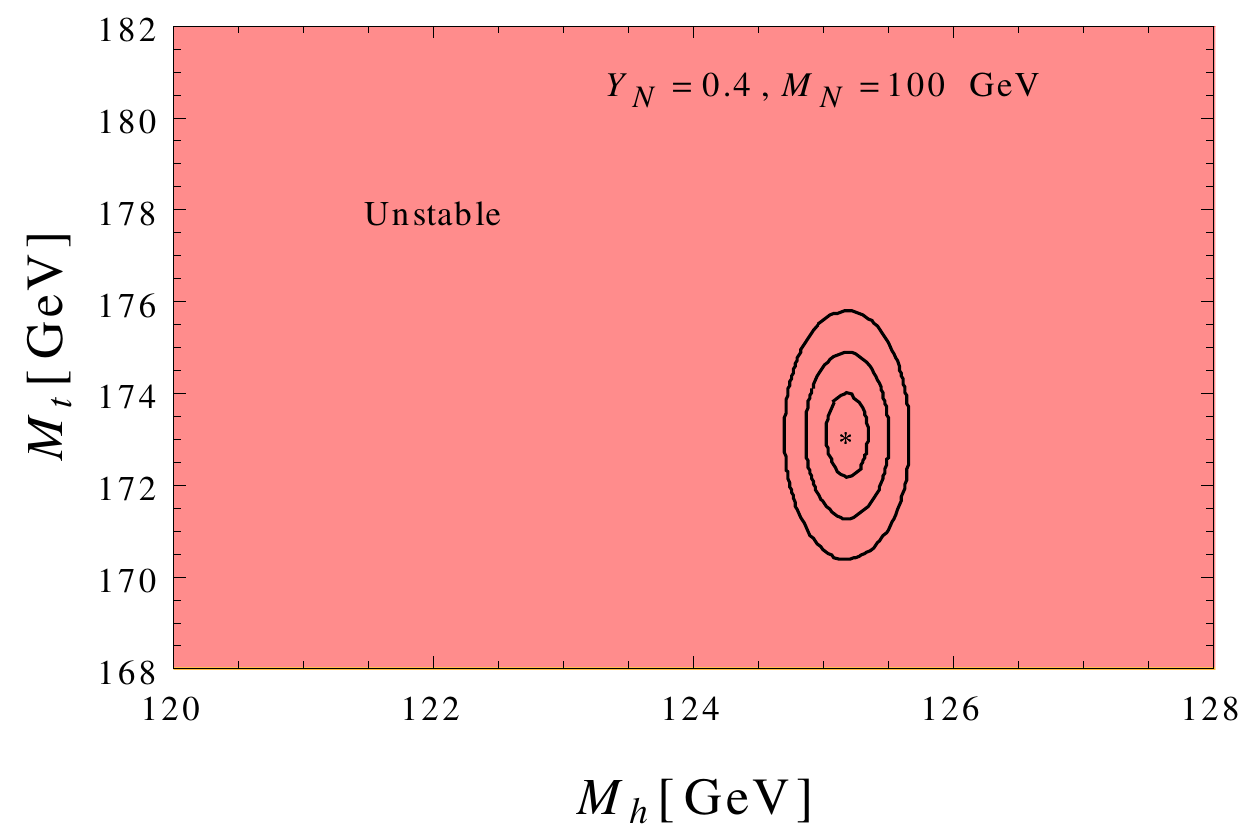}\label{f15}}}
\mbox{\subfigure[Type-III+ISS]{\includegraphics[width=0.5\linewidth,angle=-0]{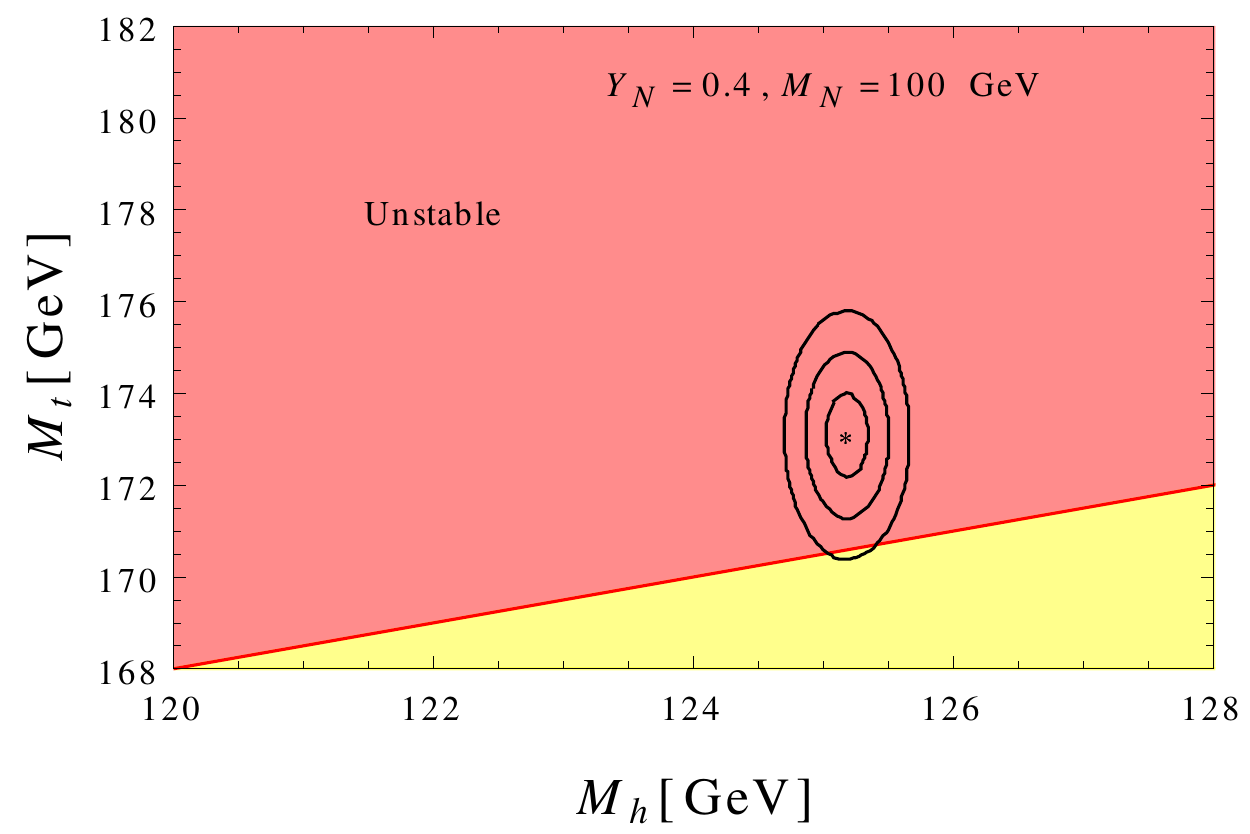}\label{f16}}	\subfigure[ID+Type-III]{
\includegraphics[width=0.5\linewidth,angle=-0]{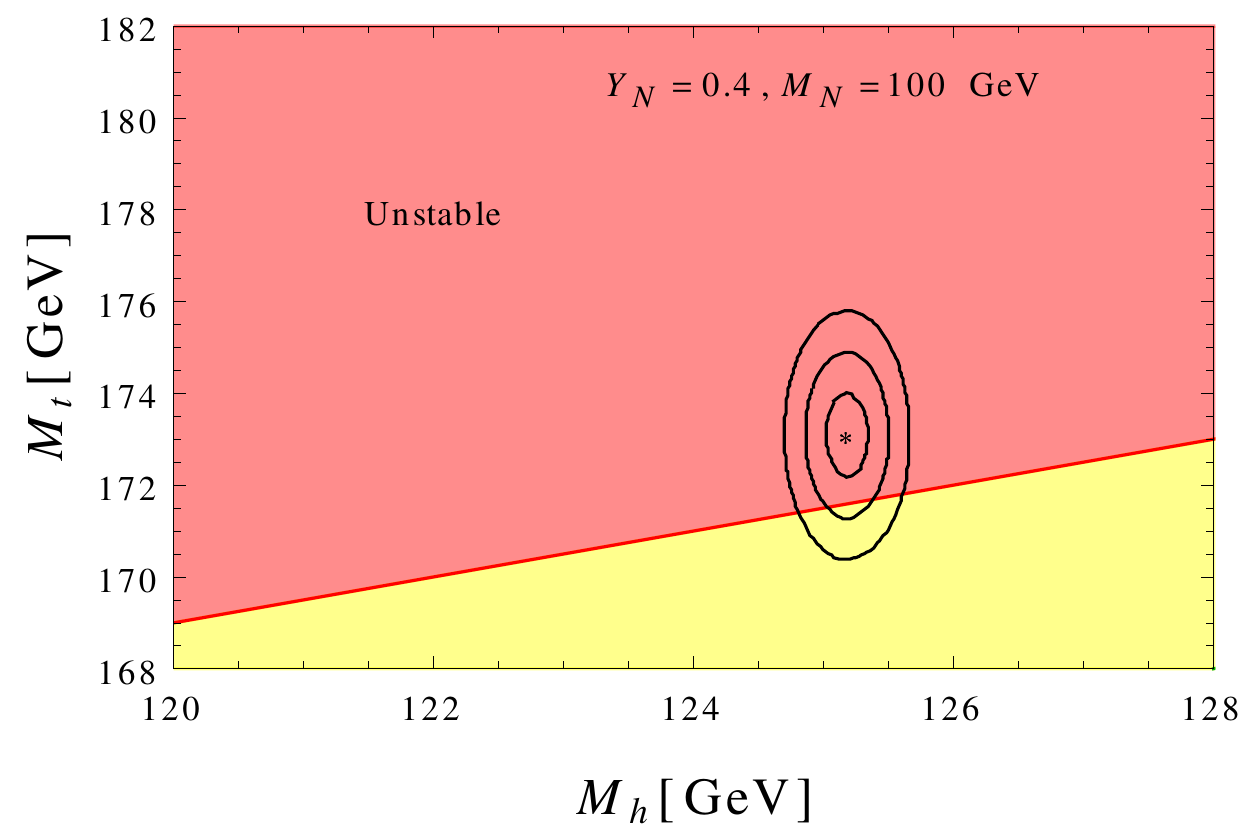}\label{f17}}}
\caption{Phase diagram in terms of Higgs and top pole masses in GeV  with $M_N=100$ GeV in Figure~\ref{f12}, \ref{f13}, \ref{f14}  for ID+Type-III+ISS scenarios with  $Y_N=0.1, 0.3, 0.4$ respectively. Figure~\ref{f15} is for Type-III only and Figure~\ref{f16}, \ref{f17} are for Type-III+ISS with $Y_N=0.4$. The unstable, metastable and stable regions are depicted by the colours red, yellow and green respectively .  The contours describe the current experimental $1\sigma,2\sigma,3\sigma$ values with the dot specifying the central value in the $(M_h,M_t)$ plane.}\label{fig7l}
\end{figure}

Figure~\ref{fig7l} represents the phase diagrams in terms of Higgs boson mass and top quark pole mass in GeV. Different regions of the solution spaces corresponding to the unstable, metastable and stable regions are depicted in red, yellow and green colours respectively.  The contours describe the current experimental $1\sigma,2\sigma,3\sigma$ values with the dot specifying the central value in the $(M_h,M_t)$ plane\cite{Buttazzo:2013uya,Masina:2012tz}. To obtain the regions we vary all the $\lambda_i =0.1-0.8$  while  the $\lambda_1$ and $y_t$ are varied to attain the Higgs boson mass within  $120-128$ GeV and  top quark mass within $168-182$ GeV respectively. In Figure~\ref{fig7l}  we fix $M_N=100$ GeV and vary $Y_N=0.1-0.4$ for two generations of $SU(2)$ triplet fermions. Figure~\ref{f12} and Figure~\ref{f13} present the scenarios with ID+Type-III+ISS  for relatively lower values of $Y_N=0.1,\, 0.3$. It is realized that the scenarios are  stable till Planck scale.

As shown in Figure~\ref{f12},  in this scenario, $\lambda_{\rm{eff}}$ becomes more positive and the region is fully in the stable region till Planck scale. This occurs, as there is more  positive contribution from $g_2$  compared to negative effect from fermions for lower values of  $Y_N$. Additionally,  the inert doublet also adds more scalars to the effective potential, leading to the enhanced stability. In Figure~\ref{f13} we depict the scenario for $Y_N=0.3$, where negative fermionic effect starts showing up which is compensated by the scalar effect of IDM. As is evident,   the stability is still more than SM, and hence, the $3\sigma$ contour in $m_h-m_t$ plane just touches the  region of metastability. Further enhancement in the value of $Y_N$ counters the positive scalar effect of IDM, and for $Y_N=0.4$ the $2\sigma$ region enters in the unstable region as described in Figure~\ref{f14}.  

Figure~\ref{f15} describes the scenario for  Type-III seesaw with two generations of triplet fermions, assuming  $Y_N=0.4$.  It can be seen that the whole region is unstable. Further addition of   ISS $SU(2)_L$ triplet fermions which directly do not give negative contributions but enhance $g_2$ to more positive value as discussed before leads to  Type-III seesaw+ISS scenario marginally  extending into the metastable region for $Y_N=0.4$,  as depicted in Figure~\ref{f16}. Instead of ISS fermions addition of inert doublet also have the similar effect and pushes the potential into the  metastable region as shown in  Figure~\ref{f17}. This further motivates the extension of Type-III seesaw scenario with ISS and IDM to achieve the stability at larger values of $Y_N$ as described in Figure~\ref{f15}.

\begin{figure}[H] 
	\begin{center}
		\mbox
		{\subfigure[ID+Type-III+ISS]{\includegraphics[width=0.5\linewidth,angle=-0]{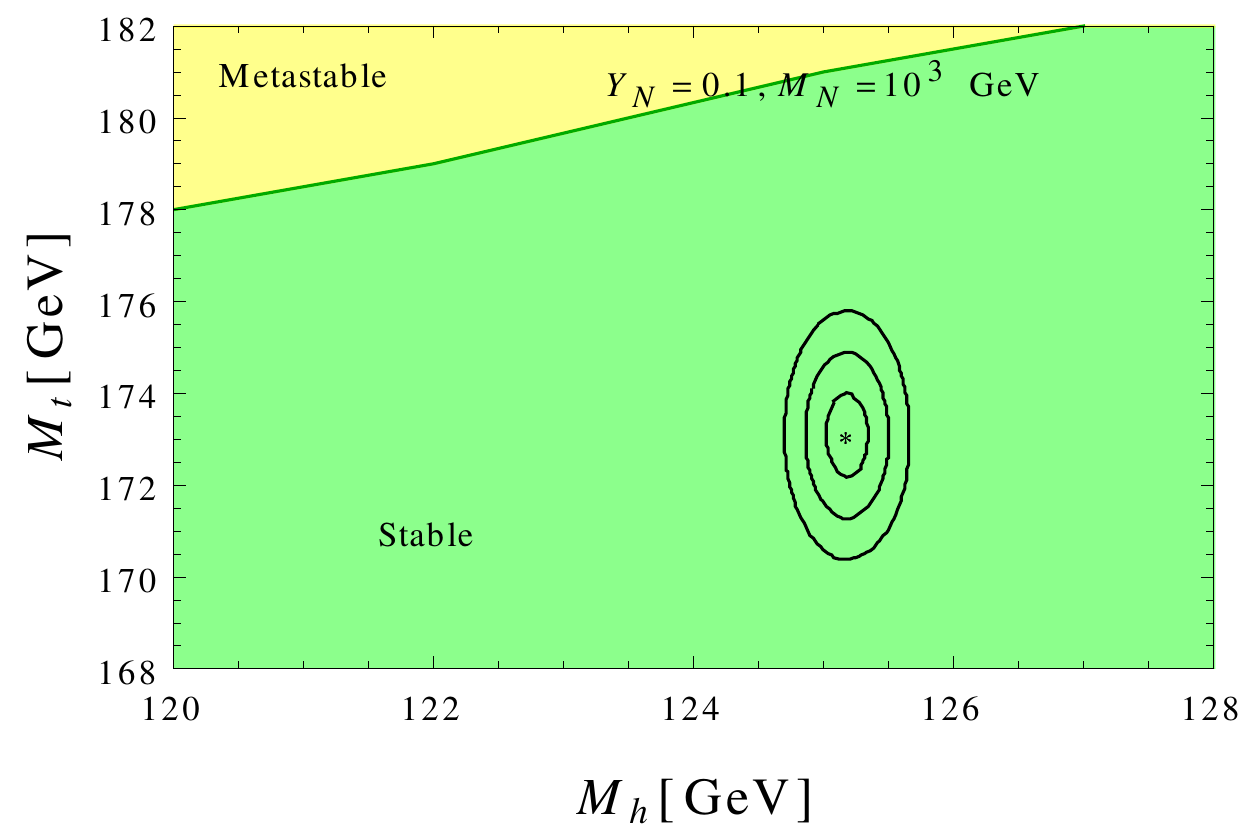}\label{f18}}	\subfigure[ID+Type-III+ISS]{
				\includegraphics[width=0.5\linewidth,angle=-0]{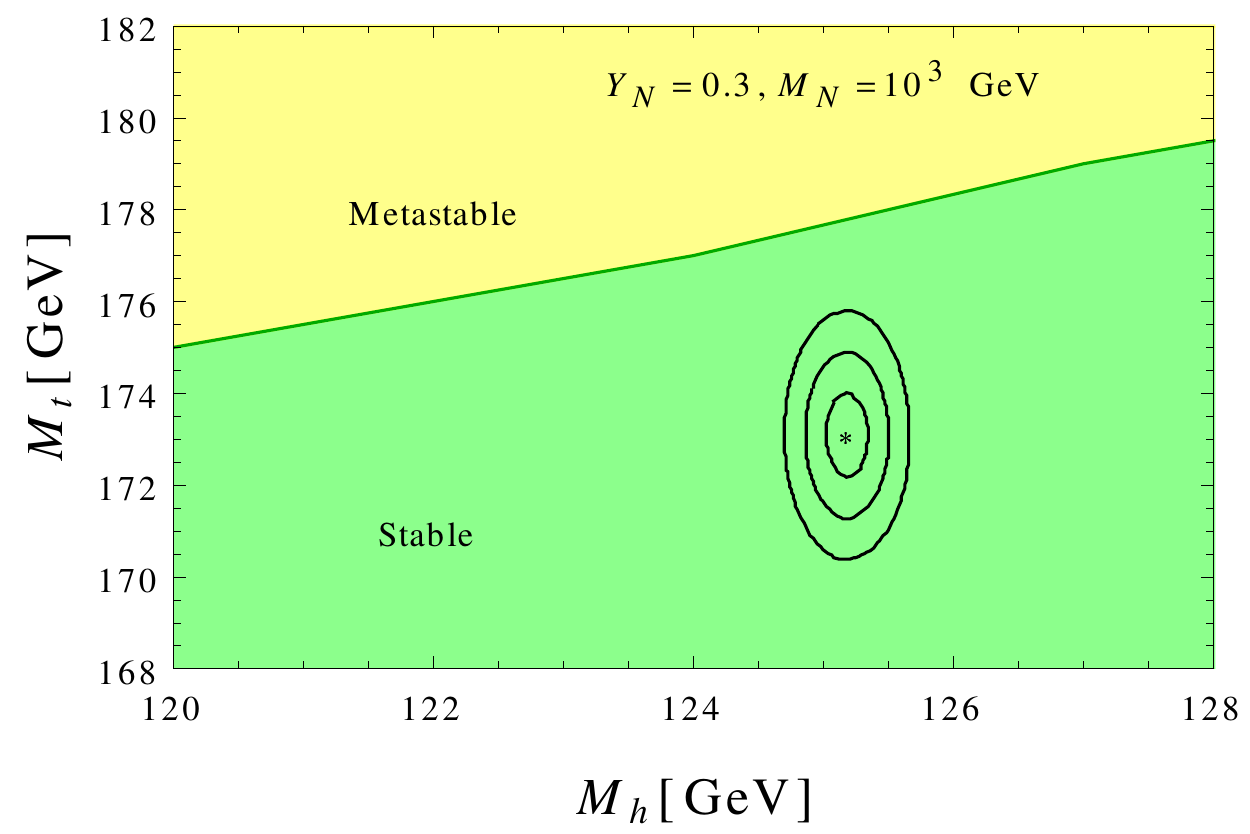}\label{f19}
			}}
				\mbox
				{\subfigure[ID+Type-III+ISS]{\includegraphics[width=0.5\linewidth,angle=-0]{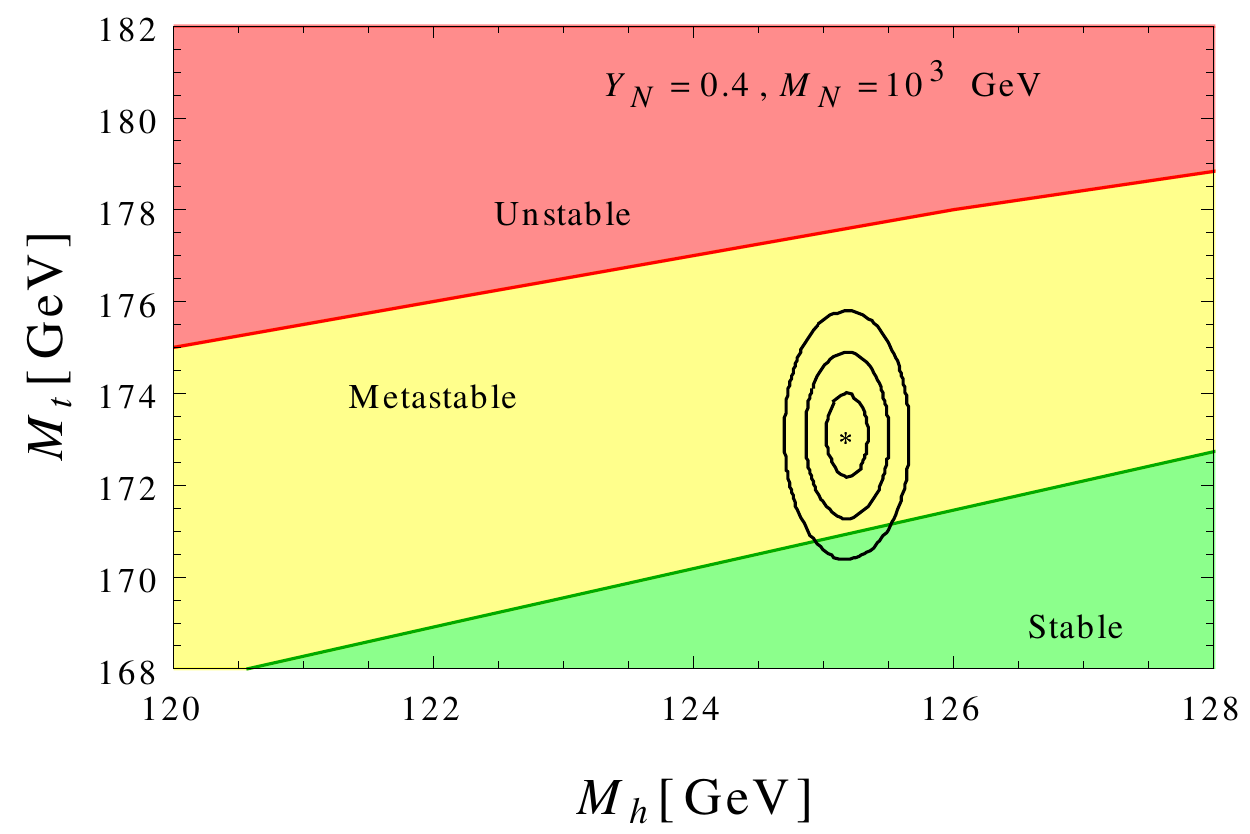}\label{f20}}	\subfigure[Type-III]{
						\includegraphics[width=0.5\linewidth,angle=-0]{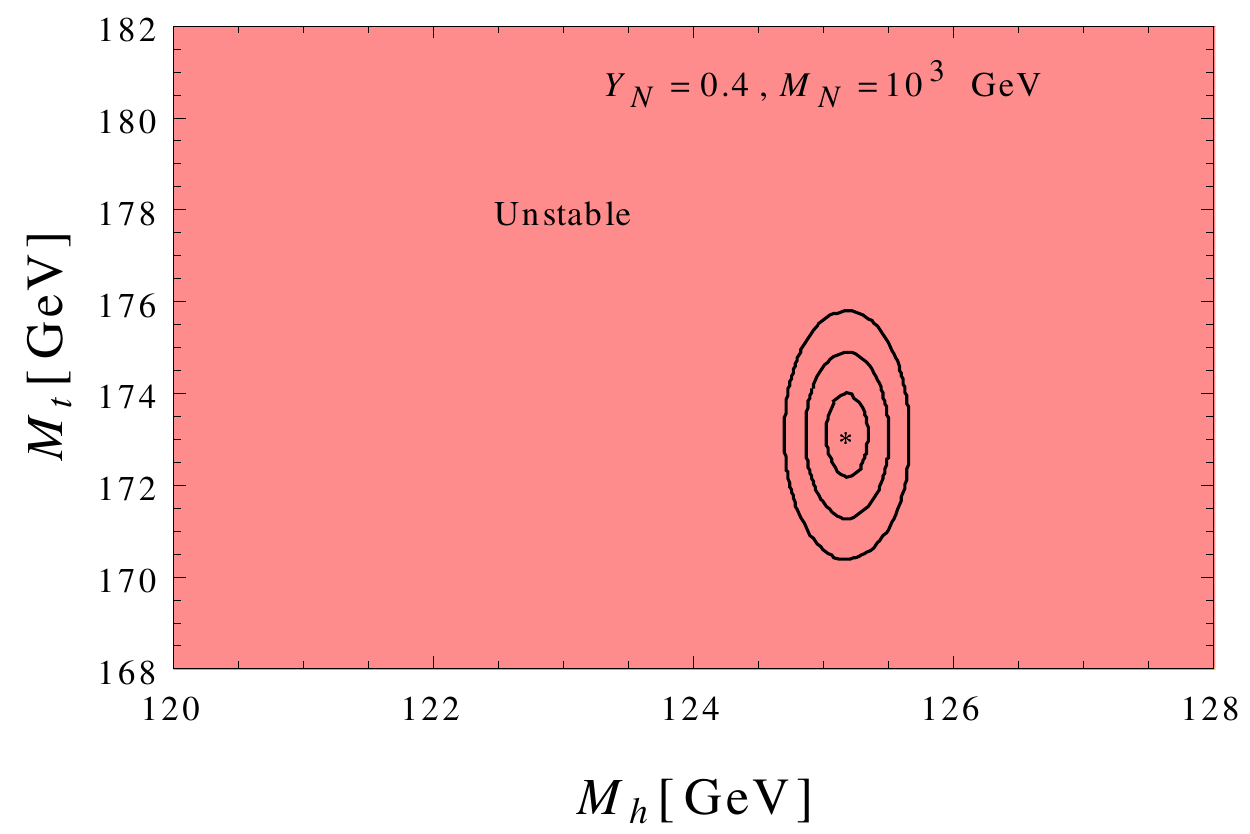}\label{f21}
					}}
					\mbox
						{\subfigure[Type-III+ISS]{\includegraphics[width=0.5\linewidth,angle=-0]{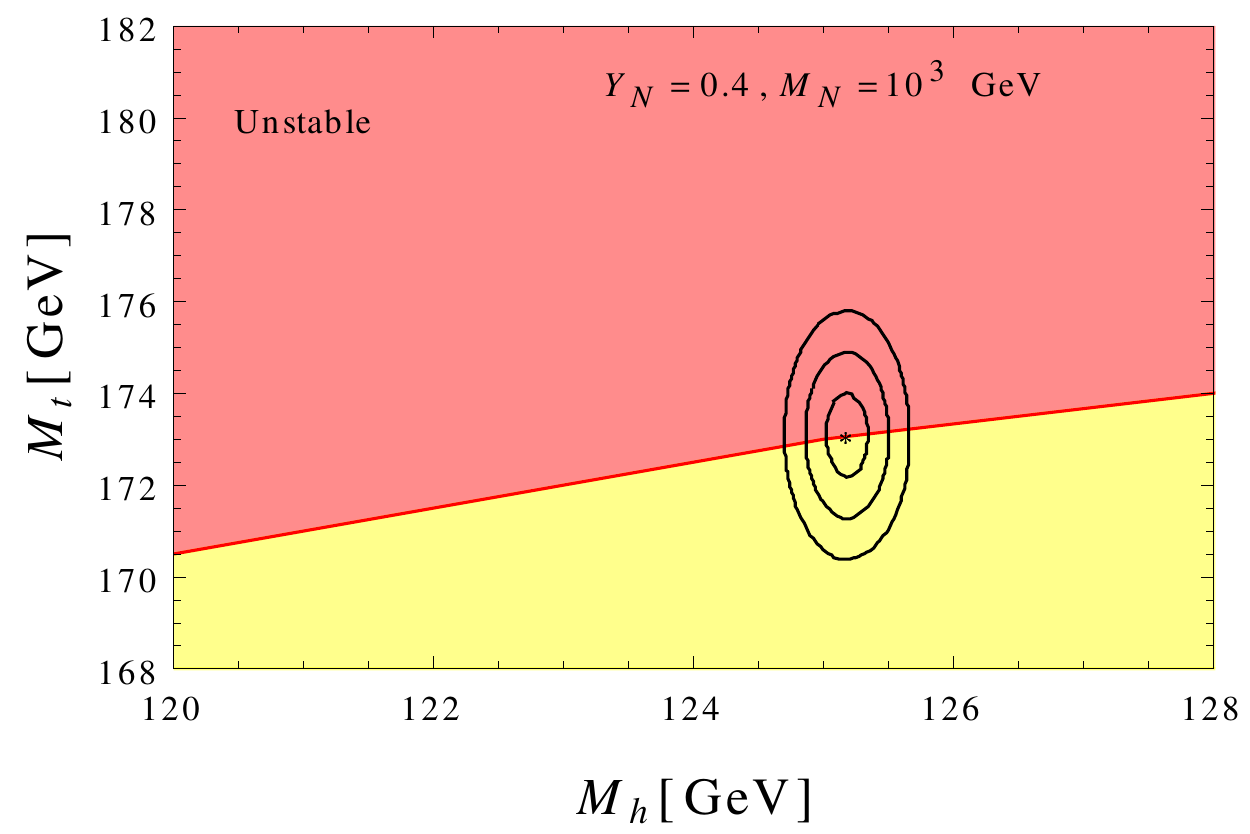}\label{f22}}	\subfigure[ID+Type-III]{
								\includegraphics[width=0.5\linewidth,angle=-0]{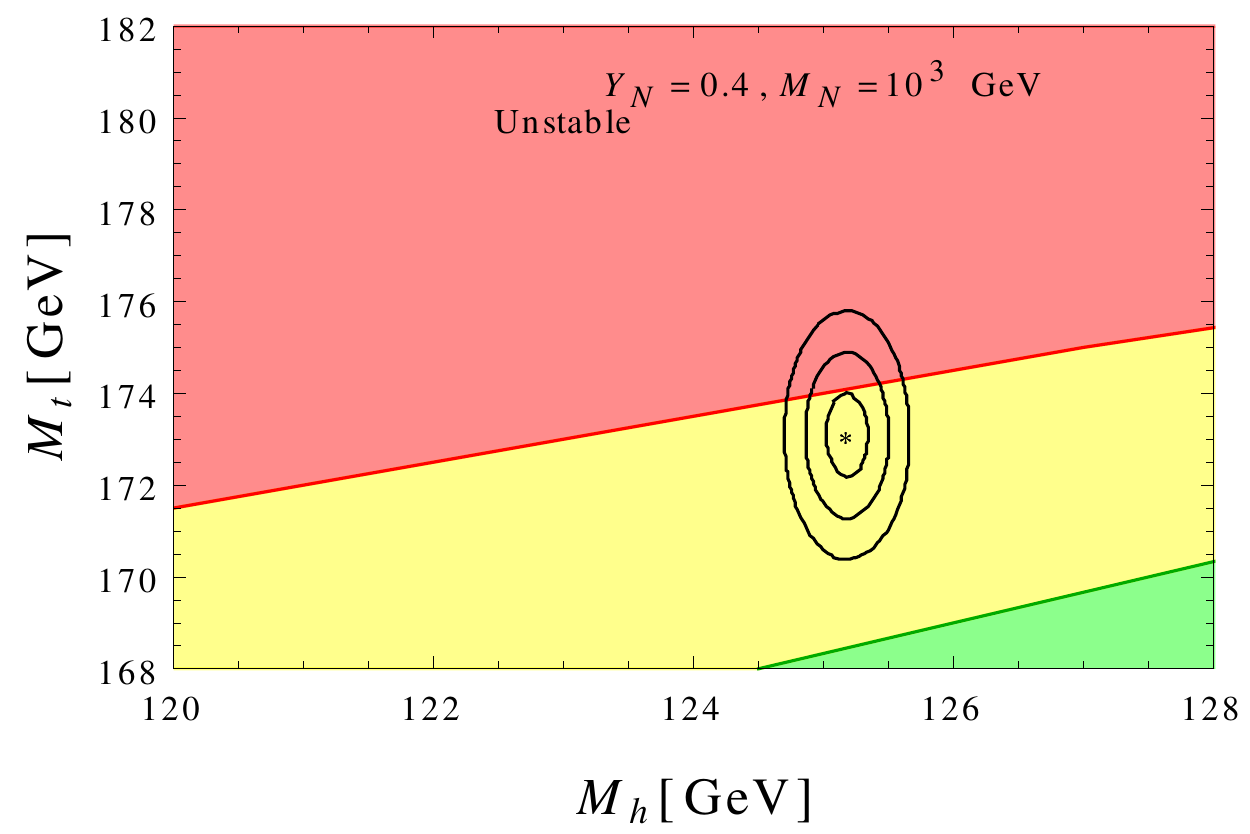}\label{f23}
							}}
			\caption{Phase diagram in terms of Higgs and top pole masses in GeV with $M_N=10^3$ GeV in Figure~\ref{f18}, \ref{f19}, \ref{f20}  for ID+Type-III+ISS scenarios with  $Y_N=0.1, 0.3, 0.4$ respectively. Figure~\ref{f21} is for Type-III only and Figure~\ref{f22}, \ref{f23} are for Type-III+ISS with $Y_N=0.4$. The unstable, metastable and stable regions are depicted by the colours red, yellow and green respectively .  The contours describe the current experimental $1\sigma,2\sigma,3\sigma$ values with the dot specifying the central value in the $(M_h,M_t)$ plane.}\label{fig8l}
		\end{center}
	\end{figure}

Figure~\ref{fig8l} we increase $M_N=10^3$ GeV and analyse the phase diagrams as before. Due to this enhancement in $SU(2)_L$ triplet fermion mass their negative loop effects will now be reduced. This can be realised from Figure~\ref{f18}-\ref{f19}, where $\lambda_{\rm{eff}}$ is highly positive and the regions are fully stable  as compared to  Figure~\ref{f12}-\ref{f13}.  As we increase $Y_N=0.4$,  the negative fermionic effect starts showing up in Figure~\ref{f20}. The region now lies in the metastable region, however  touching the stable region, contrary to Figure~\ref{f14} where some part is in unstable region. 

Figure~\ref{f21} describes the only Type-III scenario for $Y_N=0.4$ in which the central value in the $(M_h,M_t)$ plane  lies  in the unstable region similar to Figure~\ref{f15}. An extension of Type-III seesaw with ISS  in  Figure-\ref{f22} and an extension of Type-III seesaw with inert doublet in Figure~\ref{f23} moves the potential into the metastable region similar to $M_N=100$ GeV case in Figure~\ref{f16}-\ref{f17} respectively. 

The inert Higgs doublet model provides a dark matter in terms of either $A$ or $H_0$ whichever is lighter \cite{PBSJ,PBBDSJ,Khan:2012zw}. Being $SU(2)$ doublet the dominant mode of annihilation is  $\rm {DM\, DM} \to W^\pm W^\mp$ with sub-dominant annihilation and co-annihilation modes of $ZZ$ and $ZW^\pm$ depends on the gauge couplings for $M_A > M_h$.  It was shown that correct DM relic puts a lower bounds on the DM mass of $700$ GeV  and  Eq.~\eqref{mass} implies $m_{22}\gsim 665$ GeV. This is very crucial in terms of Figure~\ref{fig7l} and Figure~\ref{fig8l} as we vary $\rm\lambda_i (EW)$=0.01-0.8 while the $\lambda_1$ and $y_t$ are varied to attain the Higgs boson mass within 120-128 GeV and top quark mass within 168-182 GeV. However, the RGE evolution of the quartic couplings that we use in Figure 8 and 9 is independent of the value of $m_{22}$. Since, the mass spectra of $H_0$ and $A$ depend on the $m_{22}$ and $\lambda_{345}(=\lambda_3 +\lambda_4 -2\lambda_5)$ the quartic coupling combination (see Eq.~\eqref{mass}) and we can always choose $m_{22}$ in such a way that the DM mass will satisfy the correct DM relic density. Thus above conclusions are consistent with the DM relic bounds.

Similarly, the EW value of  $\mu_{\Sigma}$ will not effect the RGE evolution of the quartic couplings. Figure 8 and 9 fix $M_N$  as 100 GeV and 1000 GeV for three diffrent fixed values of $Y_N$. Still, we can choose $\mu_{\Sigma}$ in such a way to satisfy the eV order light neutrino mass (see Eq.~\eqref{neueigen}). Thus the chosen benchmark points can satisfy both DM mass compatible with correct relic density and eV scale neutrino mass.  
\section{Discussions and conclusion}\label{concl}
In this article we studied the vacuum stability of the electroweak vacuum in the presence of Type-III fermions along with inverse seesaw fermions. Unlike Type-I seesaw case, Type-III fermions are in the triplet representation of $SU(2)$ which contribute to the beta-function of the gauge coupling $g_2$ even at one-loop level. This is a positive effect and  increases from Type-III to Type-III+ISS+ID case 
 step wise and makes the $g_2$ grow to higher values as scale increases. $g_2$ certainly  becomes not perturbative below Planck scale or GUT scale for three generations of Type-III fermions. It is only with the two generation that we are able to acquire the Planck or GUT scale stability. 
 
 The enhancement of $g_2$ also has impact on the scalar quartic couplings which makes them non-perturbative much before compared to SM or SM+Type-I+ISS \cite{PBBDSJ}. For lower values of quartic couplings the $g_2$ effect is the dominant. At larger values of $\lambda_i$ (except the $\lambda_1$ which is fixed by the Higgs mass at EW scale) $\rm \lambda_i Tr(Y^\dagger_N Y_N)$ effect creeps in making the quartic couplings further divergent.  However, a further increment of  $Y_N$ will bring down the stability bound by pushing $\lambda_1$ to negative direction which is proportional to $\rm Tr(Y^\dagger_N Y_N Y^\dagger_N Y_N)$. For Planck scale perturbativity we can go up to $\lambda_5 \sim 0.17$ with  $\lambda_1=0.126$ and other  $\lambda_i =0.10$ at the EW scale for $Y_N =0.40$. The effective potential approach calculations show that even for $Y_N=0.30$ for $m_N=100,\,1000$ GeV the model ID+Type-III+ISS  with two generations of new fermions lies in the stable region for lower values of $Y_N$ and draws to metastable region for higher values of $Y_N$. However, only  Type-III scenario belongs to unstable regions in both cases, whereas ID+Type-III and Type-III+ISS  scenarios can be in between metastable and unstable regions.

IDM is generally motivated to provide the much needed DM to explain the DM relic and other experimental observations. Nevertheless, it is also supported to enhance the stability of electroweak vacuum. Being in $Z_2$-odd multiplet it does not couple to  the $SU(2)$ triplet fermions which makes their phenomenology more illusive. No two-body decays are allowed for the Type-III fermions into any of the inert Higgs bosons. In \cite{PBBDSJ,PBSJ} authors have shown that due to compressed spectrum only some three- and four-body decays are allowed maintaining the $Z_2$ symmetry of the Lagrangian. In \cite{Belyaev:2016lok,PBSJ,Eiteneuer:2017hoh,Diaz:2015pyv,Garcia-Cely:2015ysa} a detailed relic calculations has been carried out including the direct, indirect DM searches and collider phenomenology. It has been found out that the lightest $Z_2$-odd particles should be heavier than $700$ GeV in ID and $1176$ GeV in IT to satisfy the DM relic constraints which can be in the desired range to explain the AMS-02 positron excess observation \cite{PBSJ,AMS}. The decays of charged Higgs boson can give rise to mono-lepton plus missing energy signatures \cite{PBSJ} which can be isolated from  displaced mono-leptonic signatures in real scalar and complex triplet scenarios \cite{PBSJ, PBAC} and other charged Higgs  signatures \cite{ISS2, chNSSM,chTNSSM,chTESSM}. Similar displaced charged leptonic signatures can be observed in the models with Type-I seesaw \cite{ISS2,RHNU1,RHNLFV,RHNLFV2,disTypeI,Bandyopadhyay:2017bgh,Chiang} and Type-III seesaw\cite{TypeIII2,disTypeIII,disTypeIII2}. 

The triplet fermions are searched a the LHC at 13 TeV centre of mass energy with democratic branching fractions \cite{CMSTypeIII} and a lower bound of $620-840$ GeV has been put at $2\sigma$ level. However, due the presence and mixing with other set of $SU(2)$
triplet fermions involving in inverse seesaw and non-democratic branching can substantially reduce the mass limit allowing even smaller triplet fermion mass. Type-III fermions can also be looked via their angular distributions at the LHC \cite{PBSD}.

\acknowledgments
 SJ thanks DST/INSPIRES/03/2018/001207 for the financial support towards the PhD program. SJ also wants to thank Dr. Anirban Karan for useful discussions. PB wants to thank  SERB CORE Grant CRG/2018/004971, MTR/2020/000668, Anomalies 2019-IUSSTF and Anomalies 2020 for the support. PB thanks Prof. Debajyoti Chaudhury for useful discussion. PB and SJ also want to thank Mr. Saiyad Ashanujjaman for the help in SARAH. PB and SJ also thank IOPB for the visit and local hospitality in the early part of the project. 

\appendix
\section{Two-loop $\beta$-functions-With two generations} \label{betaf1}
\subsection{Scalar Quartic Couplings}
\footnotesize{
\begingroup
\allowdisplaybreaks
\begin{align*}
	\beta_{\lambda_1} \ =  \ &
\frac{1}{16\pi^2} \Bigg[\frac{27}{200} g_{1}^{4} +\frac{9}{20} g_{1}^{2} g_{2}^{2} +\frac{9}{8} g_{2}^{4} -\frac{9}{5} g_{1}^{2} \lambda_1 -9 g_{2}^{2} \lambda_1 +24 \lambda_{1}^{2} +2 \lambda_{3}^{2} +2 \lambda_3 \lambda_4 +\lambda_{4}^{2}+4 \lambda_{5}^{2} \nonumber \\ 
&+12 \lambda_1 \mbox{Tr}\Big({Y_d  Y_{d}^{\dagger}}\Big) +4 \lambda_1 \mbox{Tr}\Big({Y_e  Y_{e}^{\dagger}}\Big) +12 \lambda_1 \mbox{Tr}\Big({Y_{N}  Y_{N}^{\dagger}}\Big) +12 \lambda_1 \mbox{Tr}\Big({Y_u  Y_{u}^{\dagger}}\Big) -6 \mbox{Tr}\Big({Y_d  Y_{d}^{\dagger}  Y_d  Y_{d}^{\dagger}}\Big) \nonumber \\ 
&-2 \mbox{Tr}\Big({Y_e  Y_{e}^{\dagger}  Y_e  Y_{e}^{\dagger}}\Big) -8 \mbox{Tr}\Big({Y_e  Y_{N}^{\dagger}  Y_{N}  Y_{e}^{\dagger}}\Big) -10 \mbox{Tr}\Big({Y_{N}  Y_{N}^{\dagger}  Y_{N}  Y_{N}^{\dagger}}\Big) -6 \mbox{Tr}\Big({Y_u  Y_{u}^{\dagger}  Y_u  Y_{u}^{\dagger}}\Big)\Bigg] \nonumber \\
&+\frac{1}{(16\pi^2)^2}\Bigg[-\frac{3537}{2000} g_{1}^{6} -\frac{1719}{400} g_{1}^{4} g_{2}^{2} -\frac{559}{80} g_{1}^{2} g_{2}^{4} +\frac{35}{16} g_{2}^{6} +\frac{1953}{200} g_{1}^{4} \lambda_1 +\frac{117}{20} g_{1}^{2} g_{2}^{2} \lambda_1 +\frac{269}{8} g_{2}^{4} \lambda_1  \nonumber \\ 
&+108 g_{2}^{2} \lambda_{1}^{2} -312 \lambda_{1}^{3} +\frac{9}{10} g_{1}^{4} \lambda_3 +\frac{15}{2} g_{2}^{4} \lambda_3 +\frac{12}{5} g_{1}^{2} \lambda_{3}^{2} +12 g_{2}^{2} \lambda_{3}^{2} -20 \lambda_1 \lambda_{3}^{2} -8 \lambda_{3}^{3} +\frac{9}{20} g_{1}^{4} \lambda_4 \nonumber \\ 
&+\frac{3}{2} g_{1}^{2} g_{2}^{2} \lambda_4 +\frac{15}{4} g_{2}^{4} \lambda_4 +\frac{12}{5} g_{1}^{2} \lambda_3 \lambda_4 +12 g_{2}^{2} \lambda_3 \lambda_4 -20 \lambda_1 \lambda_3 \lambda_4 -12 \lambda_{3}^{2} \lambda_4 +\frac{6}{5} g_{1}^{2} \lambda_{4}^{2} \nonumber \\ 
&+3 g_{2}^{2} \lambda_{4}^{2} -12 \lambda_1 \lambda_{4}^{2} -16 \lambda_3 \lambda_{4}^{2} -6 \lambda_{4}^{3} -\frac{12}{5} g_{1}^{2} \lambda_{5}^{2} -56 \lambda_1 \lambda_{5}^{2} -80 \lambda_3 \lambda_{5}^{2} -88 \lambda_4 \lambda_{5}^{2}+\frac{108}{5} g_{1}^{2} \lambda_{1}^{2} \nonumber \\ 
&+\frac{1}{20} \Big(-5 \Big(64 \lambda_1 \Big(-5 g_{3}^{2}  + 9 \lambda_1 \Big) -90 g_{2}^{2} \lambda_1  + 9 g_{2}^{4} \Big) + 9 g_{1}^{4}  + g_{1}^{2} \Big(50 \lambda_1  + 54 g_{2}^{2} \Big)\Big)\mbox{Tr}\Big({Y_d  Y_{d}^{\dagger}}\Big) \nonumber \\ 
&-\frac{3}{20} \Big(15 g_{1}^{4}  -2 g_{1}^{2} \Big(11 g_{2}^{2}  + 25 \lambda_1 \Big) + 5 \Big(-10 g_{2}^{2} \lambda_1  + 64 \lambda_{1}^{2}  + g_{2}^{4}\Big)\Big)\mbox{Tr}\Big({Y_e  Y_{e}^{\dagger}}\Big) -\frac{27}{100} g_{1}^{4} \mbox{Tr}\Big({Y_{N}  Y_{N}^{\dagger}}\Big) \nonumber \\ 
&-\frac{57}{10} g_{1}^{2} g_{2}^{2} \mbox{Tr}\Big({Y_{N}  Y_{N}^{\dagger}}\Big) +\frac{7}{4} g_{2}^{4} \mbox{Tr}\Big({Y_{N}  Y_{N}^{\dagger}}\Big) +\frac{9}{2} g_{1}^{2} \lambda_1 \mbox{Tr}\Big({Y_{N}  Y_{N}^{\dagger}}\Big) +\frac{165}{2} g_{2}^{2} \lambda_1 \mbox{Tr}\Big({Y_{N}  Y_{N}^{\dagger}}\Big) \nonumber \\ 
&-144 \lambda_{1}^{2} \mbox{Tr}\Big({Y_{N}  Y_{N}^{\dagger}}\Big) -\frac{171}{100} g_{1}^{4} \mbox{Tr}\Big({Y_u  Y_{u}^{\dagger}}\Big) +\frac{63}{10} g_{1}^{2} g_{2}^{2} \mbox{Tr}\Big({Y_u  Y_{u}^{\dagger}}\Big) -\frac{9}{4} g_{2}^{4} \mbox{Tr}\Big({Y_u  Y_{u}^{\dagger}}\Big) \nonumber \\ 
&+\frac{17}{2} g_{1}^{2} \lambda_1 \mbox{Tr}\Big({Y_u  Y_{u}^{\dagger}}\Big) +\frac{45}{2} g_{2}^{2} \lambda_1 \mbox{Tr}\Big({Y_u  Y_{u}^{\dagger}}\Big) +80 g_{3}^{2} \lambda_1 \mbox{Tr}\Big({Y_u  Y_{u}^{\dagger}}\Big) -144 \lambda_{1}^{2} \mbox{Tr}\Big({Y_u  Y_{u}^{\dagger}}\Big) \nonumber \\ 
&+\frac{4}{5} g_{1}^{2} \mbox{Tr}\Big({Y_d  Y_{d}^{\dagger}  Y_d  Y_{d}^{\dagger}}\Big) -32 g_{3}^{2} \mbox{Tr}\Big({Y_d  Y_{d}^{\dagger}  Y_d  Y_{d}^{\dagger}}\Big) -3 \lambda_1 \mbox{Tr}\Big({Y_d  Y_{d}^{\dagger}  Y_d  Y_{d}^{\dagger}}\Big) -42 \lambda_1 \mbox{Tr}\Big({Y_d  Y_{u}^{\dagger}  Y_u  Y_{d}^{\dagger}}\Big) \nonumber \\ 
&-\frac{12}{5} g_{1}^{2} \mbox{Tr}\Big({Y_e  Y_{e}^{\dagger}  Y_e  Y_{e}^{\dagger}}\Big) - \lambda_1 \mbox{Tr}\Big({Y_e  Y_{e}^{\dagger}  Y_e  Y_{e}^{\dagger}}\Big) -\frac{24}{5} g_{1}^{2} \mbox{Tr}\Big({Y_e  Y_{N}^{\dagger}  Y_{N}  Y_{e}^{\dagger}}\Big) +30 \mbox{Tr}\Big({Y_u  Y_{u}^{\dagger}  Y_u  Y_{u}^{\dagger}  Y_u  Y_{u}^{\dagger}}\Big) \nonumber \\ 
&-16 g_{2}^{2} \mbox{Tr}\Big({Y_e  Y_{N}^{\dagger}  Y_{N}  Y_{e}^{\dagger}}\Big) +38 \lambda_1 \mbox{Tr}\Big({Y_e  Y_{N}^{\dagger}  Y_{N}  Y_{e}^{\dagger}}\Big) -40 g_{2}^{2} \mbox{Tr}\Big({Y_{N}  Y_{N}^{\dagger}  Y_{N}  Y_{N}^{\dagger}}\Big) \nonumber \\ 
&-5 \lambda_1 \mbox{Tr}\Big({Y_{N}  Y_{N}^{\dagger}  Y_{N}  Y_{N}^{\dagger}}\Big) -\frac{8}{5} g_{1}^{2} \mbox{Tr}\Big({Y_u  Y_{u}^{\dagger}  Y_u  Y_{u}^{\dagger}}\Big) -32 g_{3}^{2} \mbox{Tr}\Big({Y_u  Y_{u}^{\dagger}  Y_u  Y_{u}^{\dagger}}\Big) -3 \lambda_1 \mbox{Tr}\Big({Y_u  Y_{u}^{\dagger}  Y_u  Y_{u}^{\dagger}}\Big) \nonumber \\ 
&+30 \mbox{Tr}\Big({Y_d  Y_{d}^{\dagger}  Y_d  Y_{d}^{\dagger}  Y_d  Y_{d}^{\dagger}}\Big) -12 \mbox{Tr}\Big({Y_d  Y_{d}^{\dagger}  Y_d  Y_{u}^{\dagger}  Y_u  Y_{d}^{\dagger}}\Big) +6 \mbox{Tr}\Big({Y_d  Y_{u}^{\dagger}  Y_u  Y_{d}^{\dagger}  Y_d  Y_{d}^{\dagger}}\Big) \nonumber \\ 
&-6 \mbox{Tr}\Big({Y_d  Y_{u}^{\dagger}  Y_u  Y_{u}^{\dagger}  Y_u  Y_{d}^{\dagger}}\Big) +10 \mbox{Tr}\Big({Y_e  Y_{e}^{\dagger}  Y_e  Y_{e}^{\dagger}  Y_e  Y_{e}^{\dagger}}\Big) +36 \mbox{Tr}\Big({Y_e  Y_{e}^{\dagger}  Y_e  Y_{N}^{\dagger}  Y_{N}  Y_{e}^{\dagger}}\Big) \nonumber \\ 
&+38 \mbox{Tr}\Big({Y_e  Y_{N}^{\dagger}  Y_{N}  Y_{e}^{\dagger}  Y_e  Y_{e}^{\dagger}}\Big) +150 \mbox{Tr}\Big({Y_e  Y_{N}^{\dagger}  Y_{N}  Y_{N}^{\dagger}  Y_{N}  Y_{e}^{\dagger}}\Big) +94 \mbox{Tr}\Big({Y_{N}  Y_{N}^{\dagger}  Y_{N}  Y_{N}^{\dagger}  Y_{N}  Y_{N}^{\dagger}}\Big) \Bigg] \, . \\
\beta_{\lambda_2} \  = \ &
\frac{1}{16\pi^2}\Bigg[24 \lambda_{2}^{2}  + 2 \lambda_{3}^{2}  + 2 \lambda_3 \lambda_4  + 4 \lambda_{5}^{2}  -9 g_{2}^{2} \lambda_2  + \frac{27}{200} g_{1}^{4}  + \frac{9}{20} g_{1}^{2} \Big(-4 \lambda_2  + g_{2}^{2}\Big) + \frac{9}{8} g_{2}^{4}  + \lambda_{4}^{2}\Bigg] \nonumber \\
&+\frac{1}{(16\pi^2)^2}\Bigg[-\frac{3537}{2000} g_{1}^{6} -\frac{1719}{400} g_{1}^{4} g_{2}^{2} -\frac{559}{80} g_{1}^{2} g_{2}^{4} +\frac{35}{16} g_{2}^{6} +\frac{1953}{200} g_{1}^{4} \lambda_2 +\frac{117}{20} g_{1}^{2} g_{2}^{2} \lambda_2 +\frac{269}{8} g_{2}^{4} \lambda_2  \nonumber \\ 
&+108 g_{2}^{2} \lambda_{2}^{2} -312 \lambda_{2}^{3} +\frac{9}{10} g_{1}^{4} \lambda_3 +\frac{15}{2} g_{2}^{4} \lambda_3 +\frac{12}{5} g_{1}^{2} \lambda_{3}^{2} +12 g_{2}^{2} \lambda_{3}^{2} -20 \lambda_2 \lambda_{3}^{2} -8 \lambda_{3}^{3} +\frac{9}{20} g_{1}^{4} \lambda_4 \nonumber \\ 
&+\frac{3}{2} g_{1}^{2} g_{2}^{2} \lambda_4 +\frac{15}{4} g_{2}^{4} \lambda_4 +\frac{12}{5} g_{1}^{2} \lambda_3 \lambda_4 +12 g_{2}^{2} \lambda_3 \lambda_4 -20 \lambda_2 \lambda_3 \lambda_4 -12 \lambda_{3}^{2} \lambda_4 +\frac{6}{5} g_{1}^{2} \lambda_{4}^{2} +\frac{108}{5} g_{1}^{2} \lambda_{2}^{2}\nonumber \\ 
&+3 g_{2}^{2} \lambda_{4}^{2} -12 \lambda_2 \lambda_{4}^{2} -16 \lambda_3 \lambda_{4}^{2} -6 \lambda_{4}^{3} -\frac{12}{5} g_{1}^{2} \lambda_{5}^{2} -56 \lambda_2 \lambda_{5}^{2} -80 \lambda_3 \lambda_{5}^{2} -88 \lambda_4 \lambda_{5}^{2} \nonumber \\ 
&-6 \Big(2 \lambda_{3}^{2}  + 2 \lambda_3 \lambda_4  + 4 \lambda_{5}^{2}  + \lambda_{4}^{2}\Big)\mbox{Tr}\Big({Y_d  Y_{d}^{\dagger}}\Big) -2 \Big(2 \lambda_{3}^{2}  + 2 \lambda_3 \lambda_4  + 4 \lambda_{5}^{2}  + \lambda_{4}^{2}\Big)\mbox{Tr}\Big({Y_e  Y_{e}^{\dagger}}\Big) \nonumber \\ 
&-12 \lambda_{3}^{2} \mbox{Tr}\Big({Y_{N}  Y_{N}^{\dagger}}\Big) -12 \lambda_3 \lambda_4 \mbox{Tr}\Big({Y_{N}  Y_{N}^{\dagger}}\Big) -6 \lambda_{4}^{2} \mbox{Tr}\Big({Y_{N}  Y_{N}^{\dagger}}\Big) -24 \lambda_{5}^{2} \mbox{Tr}\Big({Y_{N}  Y_{N}^{\dagger}}\Big) \nonumber \\ 
&-12 \lambda_{3}^{2} \mbox{Tr}\Big({Y_u  Y_{u}^{\dagger}}\Big) -12 \lambda_3 \lambda_4 \mbox{Tr}\Big({Y_u  Y_{u}^{\dagger}}\Big) -6 \lambda_{4}^{2} \mbox{Tr}\Big({Y_u  Y_{u}^{\dagger}}\Big) -24 \lambda_{5}^{2} \mbox{Tr}\Big({Y_u  Y_{u}^{\dagger}}\Big)\Bigg] \, .  \\
	\beta_{\lambda_3} \ =  \ &
	\frac{1}{16\pi^2}\Bigg[\frac{27}{100} g_{1}^{4} -\frac{9}{10} g_{1}^{2} g_{2}^{2} +\frac{9}{4} g_{2}^{4} -\frac{9}{5} g_{1}^{2} \lambda_3 -9 g_{2}^{2} \lambda_3 +12 \lambda_1 \lambda_3 +12 \lambda_2 \lambda_3 +4 \lambda_{3}^{2} +4 \lambda_1 \lambda_4 +4 \lambda_2 \lambda_4  \nonumber \\ 
	&+2 \lambda_{4}^{2}+8 \lambda_{5}^{2} +6 \lambda_3 \mbox{Tr}\Big({Y_d  Y_{d}^{\dagger}}\Big) +2 \lambda_3 \mbox{Tr}\Big({Y_e  Y_{e}^{\dagger}}\Big) +6 \lambda_3 \mbox{Tr}\Big({Y_{N}  Y_{N}^{\dagger}}\Big) +6 \lambda_3 \mbox{Tr}\Big({Y_u  Y_{u}^{\dagger}}\Big)\Bigg]\nonumber \\
&+\frac{1}{(16\pi^2)^2}\Bigg[-\frac{3537}{1000} g_{1}^{6} +\frac{909}{200} g_{1}^{4} g_{2}^{2} +\frac{289}{40} g_{1}^{2} g_{2}^{4} +\frac{35}{8} g_{2}^{6} +\frac{27}{10} g_{1}^{4} \lambda_1 -3 g_{1}^{2} g_{2}^{2} \lambda_1 +\frac{45}{2} g_{2}^{4} \lambda_1 +\frac{27}{10} g_{1}^{4} \lambda_2 \nonumber \\ 
&-3 g_{1}^{2} g_{2}^{2} \lambda_2 +\frac{45}{2} g_{2}^{4} \lambda_2 +\frac{1773}{200} g_{1}^{4} \lambda_3 +\frac{33}{20} g_{1}^{2} g_{2}^{2} \lambda_3 +\frac{209}{8} g_{2}^{4} \lambda_3 +\frac{72}{5} g_{1}^{2} \lambda_1 \lambda_3 +72 g_{2}^{2} \lambda_1 \lambda_3 \nonumber \\ 
&-60 \lambda_{1}^{2} \lambda_3 +\frac{72}{5} g_{1}^{2} \lambda_2 \lambda_3 +72 g_{2}^{2} \lambda_2 \lambda_3 -60 \lambda_{2}^{2} \lambda_3 +\frac{6}{5} g_{1}^{2} \lambda_{3}^{2} +6 g_{2}^{2} \lambda_{3}^{2} -72 \lambda_1 \lambda_{3}^{2} -72 \lambda_2 \lambda_{3}^{2} \nonumber \\ 
&-12 \lambda_{3}^{3} +\frac{9}{10} g_{1}^{4} \lambda_4 -\frac{9}{5} g_{1}^{2} g_{2}^{2} \lambda_4 +\frac{15}{2} g_{2}^{4} \lambda_4 +\frac{24}{5} g_{1}^{2} \lambda_1 \lambda_4 +36 g_{2}^{2} \lambda_1 \lambda_4 -16 \lambda_{1}^{2} \lambda_4 +\frac{24}{5} g_{1}^{2} \lambda_2 \lambda_4 \nonumber \\ 
&+36 g_{2}^{2} \lambda_2 \lambda_4 -16 \lambda_{2}^{2} \lambda_4 -12 g_{2}^{2} \lambda_3 \lambda_4 -32 \lambda_1 \lambda_3 \lambda_4 -32 \lambda_2 \lambda_3 \lambda_4 -4 \lambda_{3}^{2} \lambda_4 -\frac{6}{5} g_{1}^{2} \lambda_{4}^{2} \nonumber \\ 
&+6 g_{2}^{2} \lambda_{4}^{2} -28 \lambda_1 \lambda_{4}^{2} -28 \lambda_2 \lambda_{4}^{2} -16 \lambda_3 \lambda_{4}^{2} -12 \lambda_{4}^{3} +\frac{48}{5} g_{1}^{2} \lambda_{5}^{2} -144 \lambda_1 \lambda_{5}^{2} -144 \lambda_2 \lambda_{5}^{2} \nonumber \\ 
&-72 \lambda_3 \lambda_{5}^{2} -176 \lambda_4 \lambda_{5}^{2}+\frac{1}{20} \Big(-5 \Big(-45 g_{2}^{2} \lambda_3  + 8 \Big(-20 g_{3}^{2} \lambda_3  + 3 \Big(2 \lambda_{3}^{2}  + 4 \lambda_1 \Big(3 \lambda_3  + \lambda_4\Big) \nonumber \\ 
& + 4 \lambda_{5}^{2}  + \lambda_{4}^{2}\Big)\Big) + 9 g_{2}^{4} \Big) + 9 g_{1}^{4}  + g_{1}^{2} \Big(25 \lambda_3  -54 g_{2}^{2} \Big)\Big)\mbox{Tr}\Big({Y_d  Y_{d}^{\dagger}}\Big)-\frac{1}{20} \Big(45 g_{1}^{4}  \nonumber \\ 
& + 5 \Big(-15 g_{2}^{2} \lambda_3  + 3 g_{2}^{4}  + 8 \Big(2 \lambda_{3}^{2}  + 4 \lambda_1 \Big(3 \lambda_3  + \lambda_4\Big) + 4 \lambda_{5}^{2}  + \lambda_{4}^{2}\Big)\Big) + g_{1}^{2} \Big(66 g_{2}^{2}  -75 \lambda_3 \Big)\Big)\mbox{Tr}\Big({Y_e  Y_{e}^{\dagger}}\Big) \nonumber \\ 
&-\frac{27}{100} g_{1}^{4} \mbox{Tr}\Big({Y_{N}  Y_{N}^{\dagger}}\Big) +\frac{57}{10} g_{1}^{2} g_{2}^{2} \mbox{Tr}\Big({Y_{N}  Y_{N}^{\dagger}}\Big) +\frac{7}{4} g_{2}^{4} \mbox{Tr}\Big({Y_{N}  Y_{N}^{\dagger}}\Big) +\frac{9}{4} g_{1}^{2} \lambda_3 \mbox{Tr}\Big({Y_{N}  Y_{N}^{\dagger}}\Big) \nonumber \\ 
&+\frac{165}{4} g_{2}^{2} \lambda_3 \mbox{Tr}\Big({Y_{N}  Y_{N}^{\dagger}}\Big) -72 \lambda_1 \lambda_3 \mbox{Tr}\Big({Y_{N}  Y_{N}^{\dagger}}\Big) -12 \lambda_{3}^{2} \mbox{Tr}\Big({Y_{N}  Y_{N}^{\dagger}}\Big) -24 \lambda_1 \lambda_4 \mbox{Tr}\Big({Y_{N}  Y_{N}^{\dagger}}\Big) \nonumber \\ 
&-6 \lambda_{4}^{2} \mbox{Tr}\Big({Y_{N}  Y_{N}^{\dagger}}\Big) -24 \lambda_{5}^{2} \mbox{Tr}\Big({Y_{N}  Y_{N}^{\dagger}}\Big) -\frac{171}{100} g_{1}^{4} \mbox{Tr}\Big({Y_u  Y_{u}^{\dagger}}\Big) -\frac{63}{10} g_{1}^{2} g_{2}^{2} \mbox{Tr}\Big({Y_u  Y_{u}^{\dagger}}\Big) \nonumber \\ 
&-\frac{9}{4} g_{2}^{4} \mbox{Tr}\Big({Y_u  Y_{u}^{\dagger}}\Big) +\frac{17}{4} g_{1}^{2} \lambda_3 \mbox{Tr}\Big({Y_u  Y_{u}^{\dagger}}\Big) +\frac{45}{4} g_{2}^{2} \lambda_3 \mbox{Tr}\Big({Y_u  Y_{u}^{\dagger}}\Big) +40 g_{3}^{2} \lambda_3 \mbox{Tr}\Big({Y_u  Y_{u}^{\dagger}}\Big) \nonumber \\ 
&-72 \lambda_1 \lambda_3 \mbox{Tr}\Big({Y_u  Y_{u}^{\dagger}}\Big) -12 \lambda_{3}^{2} \mbox{Tr}\Big({Y_u  Y_{u}^{\dagger}}\Big) -24 \lambda_1 \lambda_4 \mbox{Tr}\Big({Y_u  Y_{u}^{\dagger}}\Big) -6 \lambda_{4}^{2} \mbox{Tr}\Big({Y_u  Y_{u}^{\dagger}}\Big) \nonumber \\ 
&-24 \lambda_{5}^{2} \mbox{Tr}\Big({Y_u  Y_{u}^{\dagger}}\Big) -\frac{27}{2} \lambda_3 \mbox{Tr}\Big({Y_d  Y_{d}^{\dagger}  Y_d  Y_{d}^{\dagger}}\Big) -21 \lambda_3 \mbox{Tr}\Big({Y_d  Y_{u}^{\dagger}  Y_u  Y_{d}^{\dagger}}\Big) -24 \lambda_4 \mbox{Tr}\Big({Y_d  Y_{u}^{\dagger}  Y_u  Y_{d}^{\dagger}}\Big) \nonumber \\ 
&-\frac{9}{2} \lambda_3 \mbox{Tr}\Big({Y_e  Y_{e}^{\dagger}  Y_e  Y_{e}^{\dagger}}\Big) +3 \lambda_3 \mbox{Tr}\Big({Y_e  Y_{N}^{\dagger}  Y_{N}  Y_{e}^{\dagger}}\Big) +8 \lambda_4 \mbox{Tr}\Big({Y_e  Y_{N}^{\dagger}  Y_{N}  Y_{e}^{\dagger}}\Big) -\frac{45}{2} \lambda_3 \mbox{Tr}\Big({Y_{N}  Y_{N}^{\dagger}  Y_{N}  Y_{N}^{\dagger}}\Big) \nonumber \\ 
&-\frac{27}{2} \lambda_3 \mbox{Tr}\Big({Y_u  Y_{u}^{\dagger}  Y_u  Y_{u}^{\dagger}}\Big)\Bigg] \, . \\
\beta_{\lambda_4} \ = \  &
\frac{1}{16\pi^2}\Bigg[\frac{9}{5} g_{1}^{2} g_{2}^{2} -\frac{9}{5} g_{1}^{2} \lambda_4 -9 g_{2}^{2} \lambda_4 +4 \lambda_1 \lambda_4 +4 \lambda_2 \lambda_4 +8 \lambda_3 \lambda_4 +4 \lambda_{4}^{2} +32 \lambda_{5}^{2} +6 \lambda_4 \mbox{Tr}\Big({Y_d  Y_{d}^{\dagger}}\Big) \nonumber \\ 
&+2 \lambda_4 \mbox{Tr}\Big({Y_e  Y_{e}^{\dagger}}\Big) +6 \lambda_4 \mbox{Tr}\Big({Y_{N}  Y_{N}^{\dagger}}\Big) +6 \lambda_4 \mbox{Tr}\Big({Y_u  Y_{u}^{\dagger}}\Big) \Bigg]\nonumber \\
&+\frac{1}{(16\pi^2)^2} \Bigg[-\frac{657}{50} g_{1}^{4} g_{2}^{2} -\frac{106}{5} g_{1}^{2} g_{2}^{4} +6 g_{1}^{2} g_{2}^{2} \lambda_1 +6 g_{1}^{2} g_{2}^{2} \lambda_2 +\frac{6}{5} g_{1}^{2} g_{2}^{2} \lambda_3 +\frac{1413}{200} g_{1}^{4} \lambda_4 +\frac{153}{20} g_{1}^{2} g_{2}^{2} \lambda_4 \nonumber \\ 
&+\frac{89}{8} g_{2}^{4} \lambda_4 +\frac{24}{5} g_{1}^{2} \lambda_1 \lambda_4 -28 \lambda_{1}^{2} \lambda_4 +\frac{24}{5} g_{1}^{2} \lambda_2 \lambda_4 -28 \lambda_{2}^{2} \lambda_4 +\frac{12}{5} g_{1}^{2} \lambda_3 \lambda_4 +36 g_{2}^{2} \lambda_3 \lambda_4 \nonumber \\ 
&-80 \lambda_1 \lambda_3 \lambda_4 -80 \lambda_2 \lambda_3 \lambda_4 -28 \lambda_{3}^{2} \lambda_4 +\frac{24}{5} g_{1}^{2} \lambda_{4}^{2} +18 g_{2}^{2} \lambda_{4}^{2} -40 \lambda_1 \lambda_{4}^{2} -40 \lambda_2 \lambda_{4}^{2} -28 \lambda_3 \lambda_{4}^{2} \nonumber \\ 
&+\frac{192}{5} g_{1}^{2} \lambda_{5}^{2} +216 g_{2}^{2} \lambda_{5}^{2} -192 \lambda_1 \lambda_{5}^{2} -192 \lambda_2 \lambda_{5}^{2} -192 \lambda_3 \lambda_{5}^{2} -104 \lambda_4 \lambda_{5}^{2} \nonumber \\ 
&+\Big(4 \Big(10 g_{3}^{2} \lambda_4  -3 \Big(2 \lambda_1 \lambda_4  + 2 \lambda_3 \lambda_4  + 8 \lambda_{5}^{2}  + \lambda_{4}^{2}\Big)\Big) + \frac{45}{4} g_{2}^{2} \lambda_4  + g_{1}^{2} \Big(\frac{27}{5} g_{2}^{2}  + \frac{5}{4} \lambda_4 \Big)\Big)\mbox{Tr}\Big({Y_d  Y_{d}^{\dagger}}\Big) \nonumber \\ 
&+\Big(-4 \Big(2 \lambda_1 \lambda_4  + 2 \lambda_3 \lambda_4  + 8 \lambda_{5}^{2}  + \lambda_{4}^{2}\Big) + \frac{15}{4} g_{2}^{2} \lambda_4  + \frac{3}{20} g_{1}^{2} \Big(25 \lambda_4  + 44 g_{2}^{2} \Big)\Big)\mbox{Tr}\Big({Y_e  Y_{e}^{\dagger}}\Big) \nonumber \\ 
&-\frac{57}{5} g_{1}^{2} g_{2}^{2} \mbox{Tr}\Big({Y_{N}  Y_{N}^{\dagger}}\Big) +\frac{9}{4} g_{1}^{2} \lambda_4 \mbox{Tr}\Big({Y_{N}  Y_{N}^{\dagger}}\Big) +\frac{165}{4} g_{2}^{2} \lambda_4 \mbox{Tr}\Big({Y_{N}  Y_{N}^{\dagger}}\Big) -24 \lambda_1 \lambda_4 \mbox{Tr}\Big({Y_{N}  Y_{N}^{\dagger}}\Big) \nonumber \\ 
&-24 \lambda_3 \lambda_4 \mbox{Tr}\Big({Y_{N}  Y_{N}^{\dagger}}\Big) -12 \lambda_{4}^{2} \mbox{Tr}\Big({Y_{N}  Y_{N}^{\dagger}}\Big) -96 \lambda_{5}^{2} \mbox{Tr}\Big({Y_{N}  Y_{N}^{\dagger}}\Big) +\frac{63}{5} g_{1}^{2} g_{2}^{2} \mbox{Tr}\Big({Y_u  Y_{u}^{\dagger}}\Big) \nonumber \\ 
&+\frac{17}{4} g_{1}^{2} \lambda_4 \mbox{Tr}\Big({Y_u  Y_{u}^{\dagger}}\Big) +\frac{45}{4} g_{2}^{2} \lambda_4 \mbox{Tr}\Big({Y_u  Y_{u}^{\dagger}}\Big) +40 g_{3}^{2} \lambda_4 \mbox{Tr}\Big({Y_u  Y_{u}^{\dagger}}\Big) -24 \lambda_1 \lambda_4 \mbox{Tr}\Big({Y_u  Y_{u}^{\dagger}}\Big) \nonumber \\ 
&-24 \lambda_3 \lambda_4 \mbox{Tr}\Big({Y_u  Y_{u}^{\dagger}}\Big) -12 \lambda_{4}^{2} \mbox{Tr}\Big({Y_u  Y_{u}^{\dagger}}\Big) -96 \lambda_{5}^{2} \mbox{Tr}\Big({Y_u  Y_{u}^{\dagger}}\Big) -\frac{27}{2} \lambda_4 \mbox{Tr}\Big({Y_d  Y_{d}^{\dagger}  Y_d  Y_{d}^{\dagger}}\Big) \nonumber \\ 
&+27 \lambda_4 \mbox{Tr}\Big({Y_d  Y_{u}^{\dagger}  Y_u  Y_{d}^{\dagger}}\Big) -\frac{9}{2} \lambda_4 \mbox{Tr}\Big({Y_e  Y_{e}^{\dagger}  Y_e  Y_{e}^{\dagger}}\Big) -13 \lambda_4 \mbox{Tr}\Big({Y_e  Y_{N}^{\dagger}  Y_{N}  Y_{e}^{\dagger}}\Big) -\frac{45}{2} \lambda_4 \mbox{Tr}\Big({Y_{N}  Y_{N}^{\dagger}  Y_{N}  Y_{N}^{\dagger}}\Big) \nonumber \\ 
&-\frac{27}{2} \lambda_4 \mbox{Tr}\Big({Y_u  Y_{u}^{\dagger}  Y_u  Y_{u}^{\dagger}}\Big)\Bigg] \, . \\
\beta_{\lambda_5} \ = \ &
\frac{1}{16\pi^2}\Bigg[-\frac{9}{5} g_{1}^{2} \lambda_5 -9 g_{2}^{2} \lambda_5 +4 \lambda_1 \lambda_5 +4 \lambda_2 \lambda_5 +8 \lambda_3 \lambda_5 +12 \lambda_4 \lambda_5 +6 \lambda_5 \mbox{Tr}\Big({Y_d  Y_{d}^{\dagger}}\Big) +2 \lambda_5 \mbox{Tr}\Big({Y_e  Y_{e}^{\dagger}}\Big) \nonumber \\ 
&+6 \lambda_5 \mbox{Tr}\Big({Y_{N}  Y_{N}^{\dagger}}\Big) +6 \lambda_5 \mbox{Tr}\Big({Y_u  Y_{u}^{\dagger}}\Big)\Bigg] 
\nonumber \\
 &+\frac{1}{(16\pi^2)^2}\Bigg[\frac{1413}{200} g_{1}^{4} \lambda_5 +\frac{57}{20} g_{1}^{2} g_{2}^{2} \lambda_5 +\frac{89}{8} g_{2}^{4} \lambda_5 -\frac{12}{5} g_{1}^{2} \lambda_1 \lambda_5 -28 \lambda_{1}^{2} \lambda_5 -\frac{12}{5} g_{1}^{2} \lambda_2 \lambda_5 -28 \lambda_{2}^{2} \lambda_5 \nonumber \\ 
 &+\frac{48}{5} g_{1}^{2} \lambda_3 \lambda_5 +36 g_{2}^{2} \lambda_3 \lambda_5 -80 \lambda_1 \lambda_3 \lambda_5 -80 \lambda_2 \lambda_3 \lambda_5 -28 \lambda_{3}^{2} \lambda_5 +\frac{72}{5} g_{1}^{2} \lambda_4 \lambda_5 +72 g_{2}^{2} \lambda_4 \lambda_5 \nonumber \\ 
 &-88 \lambda_1 \lambda_4 \lambda_5 -88 \lambda_2 \lambda_4 \lambda_5 -76 \lambda_3 \lambda_4 \lambda_5 -32 \lambda_{4}^{2} \lambda_5 +24 \lambda_{5}^{3} \nonumber \\ 
 &+\frac{1}{4} \Big(16 \Big(10 g_{3}^{2}  -6 \lambda_1  -6 \lambda_3  -9 \lambda_4 \Big) + 45 g_{2}^{2}  + 5 g_{1}^{2} \Big)\lambda_5 \mbox{Tr}\Big({Y_d  Y_{d}^{\dagger}}\Big) \nonumber \\ 
 &+\frac{1}{4} \Big(15 g_{1}^{2}  + 15 g_{2}^{2}  -16 \Big(2 \lambda_1  + 2 \lambda_3  + 3 \lambda_4 \Big)\Big)\lambda_5 \mbox{Tr}\Big({Y_e  Y_{e}^{\dagger}}\Big) +\frac{9}{4} g_{1}^{2} \lambda_5 \mbox{Tr}\Big({Y_{N}  Y_{N}^{\dagger}}\Big) +\frac{165}{4} g_{2}^{2} \lambda_5 \mbox{Tr}\Big({Y_{N}  Y_{N}^{\dagger}}\Big) \nonumber \\ 
 &-24 \lambda_1 \lambda_5 \mbox{Tr}\Big({Y_{N}  Y_{N}^{\dagger}}\Big) -24 \lambda_3 \lambda_5 \mbox{Tr}\Big({Y_{N}  Y_{N}^{\dagger}}\Big) -36 \lambda_4 \lambda_5 \mbox{Tr}\Big({Y_{N}  Y_{N}^{\dagger}}\Big) +\frac{17}{4} g_{1}^{2} \lambda_5 \mbox{Tr}\Big({Y_u  Y_{u}^{\dagger}}\Big) \nonumber \\ 
 &+\frac{45}{4} g_{2}^{2} \lambda_5 \mbox{Tr}\Big({Y_u  Y_{u}^{\dagger}}\Big) +40 g_{3}^{2} \lambda_5 \mbox{Tr}\Big({Y_u  Y_{u}^{\dagger}}\Big) -24 \lambda_1 \lambda_5 \mbox{Tr}\Big({Y_u  Y_{u}^{\dagger}}\Big) -24 \lambda_3 \lambda_5 \mbox{Tr}\Big({Y_u  Y_{u}^{\dagger}}\Big) \nonumber \\ 
 &-36 \lambda_4 \lambda_5 \mbox{Tr}\Big({Y_u  Y_{u}^{\dagger}}\Big) -\frac{3}{2} \lambda_5 \mbox{Tr}\Big({Y_d  Y_{d}^{\dagger}  Y_d  Y_{d}^{\dagger}}\Big) +3 \lambda_5 \mbox{Tr}\Big({Y_d  Y_{u}^{\dagger}  Y_u  Y_{d}^{\dagger}}\Big) -\frac{1}{2} \lambda_5 \mbox{Tr}\Big({Y_e  Y_{e}^{\dagger}  Y_e  Y_{e}^{\dagger}}\Big) \nonumber \\ 
 &-21 \lambda_5 \mbox{Tr}\Big({Y_e  Y_{N}^{\dagger}  Y_{N}  Y_{e}^{\dagger}}\Big) -\frac{5}{2} \lambda_5 \mbox{Tr}\Big({Y_{N}  Y_{N}^{\dagger}  Y_{N}  Y_{N}^{\dagger}}\Big) -\frac{3}{2} \lambda_5 \mbox{Tr}\Big({Y_u  Y_{u}^{\dagger}  Y_u  Y_{u}^{\dagger}}\Big) \Bigg] \, .
\end{align*}
\endgroup

\subsection{Yukawa Coupling}
\begingroup
\allowdisplaybreaks
\begin{align*}
\beta_{Y_u} \  = \ & 
\frac{1}{16\pi^2}\Bigg[-\frac{3}{2} \Big(- {Y_u  Y_{u}^{\dagger}  Y_u}  + {Y_u  Y_{d}^{\dagger}  Y_d}\Big)\nonumber \\ 
&+Y_u \Big(3 \mbox{Tr}\Big({Y_d  Y_{d}^{\dagger}}\Big)  + 3 \mbox{Tr}\Big({Y_{N}  Y_{N}^{\dagger}}\Big)  + 3 \mbox{Tr}\Big({Y_u  Y_{u}^{\dagger}}\Big)  -8 g_{3}^{2}  -\frac{17}{20} g_{1}^{2}  -\frac{9}{4} g_{2}^{2}  + \mbox{Tr}\Big({Y_e  Y_{e}^{\dagger}}\Big)\Big)\Bigg]\nonumber \\
&+
\frac{1}{(16\pi^2)^2}\Bigg[\frac{1}{80} \Big(20 \Big(11 {Y_u  Y_{d}^{\dagger}  Y_d  Y_{d}^{\dagger}  Y_d}  -4 {Y_u  Y_{u}^{\dagger}  Y_u  Y_{d}^{\dagger}  Y_d}  + 6 {Y_u  Y_{u}^{\dagger}  Y_u  Y_{u}^{\dagger}  Y_u}  - {Y_u  Y_{d}^{\dagger}  Y_d  Y_{u}^{\dagger}  Y_u} \Big)\nonumber \\ 
&+{Y_u  Y_{u}^{\dagger}  Y_u} \Big(1280 g_{3}^{2}  -180 \mbox{Tr}\Big({Y_e  Y_{e}^{\dagger}}\Big)  + 223 g_{1}^{2}  -540 \mbox{Tr}\Big({Y_d  Y_{d}^{\dagger}}\Big)  -540 \mbox{Tr}\Big({Y_{N}  Y_{N}^{\dagger}}\Big)  -540 \mbox{Tr}\Big({Y_u  Y_{u}^{\dagger}}\Big)  \nonumber \\ 
&+{Y_u  Y_{d}^{\dagger}  Y_d} \Big(100 \mbox{Tr}\Big({Y_e  Y_{e}^{\dagger}}\Big)  -1280 g_{3}^{2}  + 300 \mbox{Tr}\Big({Y_d  Y_{d}^{\dagger}}\Big)  + 300 \mbox{Tr}\Big({Y_{N}  Y_{N}^{\dagger}}\Big)  + 300 \mbox{Tr}\Big({Y_u  Y_{u}^{\dagger}}\Big)  -43 g_{1}^{2}  + 45 g_{2}^{2} \Big)\Big)\nonumber \\ 
&+Y_u \Big(\frac{1267}{600} g_{1}^{4} -\frac{9}{20} g_{1}^{2} g_{2}^{2} -\frac{5}{4} g_{2}^{4} +\frac{19}{15} g_{1}^{2} g_{3}^{2} +9 g_{2}^{2} g_{3}^{2} -108 g_{3}^{4} +6 \lambda_{1}^{2} +\lambda_{3}^{2}+\lambda_3 \lambda_4 +\lambda_{4}^{2}+6 \lambda_{5}^{2} \nonumber \\ 
&+\frac{5}{8} \Big(32 g_{3}^{2}  + 9 g_{2}^{2}  + g_{1}^{2}\Big)\mbox{Tr}\Big({Y_d  Y_{d}^{\dagger}}\Big) +\frac{15}{8} \Big(g_{1}^{2} + g_{2}^{2}\Big)\mbox{Tr}\Big({Y_e  Y_{e}^{\dagger}}\Big) +\frac{9}{8} g_{1}^{2} \mbox{Tr}\Big({Y_{N}  Y_{N}^{\dagger}}\Big) +\frac{165}{8} g_{2}^{2} \mbox{Tr}\Big({Y_{N}  Y_{N}^{\dagger}}\Big) \nonumber \\ 
&+\frac{17}{8} g_{1}^{2} \mbox{Tr}\Big({Y_u  Y_{u}^{\dagger}}\Big) +\frac{45}{8} g_{2}^{2} \mbox{Tr}\Big({Y_u  Y_{u}^{\dagger}}\Big) +20 g_{3}^{2} \mbox{Tr}\Big({Y_u  Y_{u}^{\dagger}}\Big) -\frac{27}{4} \mbox{Tr}\Big({Y_d  Y_{d}^{\dagger}  Y_d  Y_{d}^{\dagger}}\Big) \nonumber \\ 
&+\frac{3}{2} \mbox{Tr}\Big({Y_d  Y_{u}^{\dagger}  Y_u  Y_{d}^{\dagger}}\Big) -\frac{9}{4} \mbox{Tr}\Big({Y_e  Y_{e}^{\dagger}  Y_e  Y_{e}^{\dagger}}\Big) -\frac{21}{2} \mbox{Tr}\Big({Y_e  Y_{N}^{\dagger}  Y_{N}  Y_{e}^{\dagger}}\Big) -\frac{45}{4} \mbox{Tr}\Big({Y_{N}  Y_{N}^{\dagger}  Y_{N}  Y_{N}^{\dagger}}\Big) \nonumber \\ 
&-\frac{27}{4} \mbox{Tr}\Big({Y_u  Y_{u}^{\dagger}  Y_u  Y_{u}^{\dagger}}\Big) \Big) \Big)+ 675 g_{2}^{2}  -960 \lambda_1 \Big)\Bigg] \, .
\end{align*}
\endgroup

\end{document}